\documentclass [12pt,draftclsnofoot, onecolumn]{IEEEtran}
\usepackage{cite,graphicx,amsmath,amssymb}
\usepackage{amsfonts,amssymb}
\usepackage{subfigure}
\usepackage{fancyhdr}
\usepackage{mdwmath}
\usepackage{mdwtab}
\usepackage{balance}
\usepackage{xcolor}
\usepackage{bm}
\usepackage{amsthm}
\usepackage{algorithm}
\usepackage{algorithmic}
\usepackage{multirow}
\usepackage{flafter}
\usepackage{mathrsfs}
\usepackage{url}
\usepackage{mathrsfs}
\usepackage{setspace}
\usepackage{graphics}
\usepackage{makecell}
\usepackage{diagbox}
\usepackage{hyperref}
\usepackage{epstopdf}

\usepackage{array}
\newcolumntype{P}[1]{>{\centering\arraybackslash}p{#1}}

\theoremstyle{definition}

\newtheorem{corollary}{Corollary}

\begin{document}

\title{ Distributed Intelligence in Wireless Networks }
%


\author{
Xiaolan~Liu,~\IEEEmembership{Member IEEE,} Jiadong~Yu,~\IEEEmembership{Member IEEE,}
Yuanwei~Liu,~\IEEEmembership{Senior Member IEEE,}
Yue~Gao,~\IEEEmembership{Senior Member IEEE,}
Toktam~Mahmoodi,~\IEEEmembership{Senior Member IEEE,}
Sangarapillai Lambotharan,  \IEEEmembership{Senior Member IEEE,}
Danny H. K. Tsang, \IEEEmembership{Fellow IEEE}

\thanks {\protect \onehalfspacing { X. Liu is with the Institute for Digital Technologies, Loughborough University, London E20 3BS, U.K. (Email: xiaolan.liu@lboro.ac.uk).}}

\thanks {\protect \onehalfspacing J. Yu is with the Internet of Things Thrust, The Hong Kong University of Science and Technology (Guangzhou), Guangzhou, Guangdong 511400, China (Email: jiadongyu@ust.hk).}

\thanks {\protect\onehalfspacing Y. Liu is with the School of Electronic Engineering and Computer Science, Queen Mary University of London, London, E1 4NS, U.K. (Email: yuanwei.liu@qmul.ac.uk).}

\thanks {\protect\onehalfspacing Y. Gao is with the School of Computer Science, Fudan University, Shanghai 200438, China (Email: gao\_yue@fudan.edu.cn).}

\thanks {\protect\onehalfspacing T. Mahmoodi is with the Department of Engineering, King's College London, London, WC2R 2LS, U.K. (Email: toktam.mahmoodi@kcl.ac.uk).}

\thanks {\protect\onehalfspacing S. Lambotharan is with the Wolfson School of Mechanical, Electrical and Manufacturing Engineering, Loughborough University, Loughborough LE11 3TU, U.K. (Email: s.lambotharan@lboro.ac.uk).}

\thanks {\protect\onehalfspacing D.H.K. Tsang is with the Internet of Things Thrust, The Hong Kong University of Science and Technology (Guangzhou), Guangzhou, Guangdong 511400, China, and also with the Department of Electronic and Computer Engineering, The Hong Kong University of Science and Technology, Clear Water Bay, Hong Kong SAR, China (Email: eetsang@ust.hk).
}

%


}

\maketitle

\begin{abstract}

The cloud-based solutions are becoming inefficient due to considerably large time delays, high power consumption, security and privacy concerns caused by billions of connected wireless devices and typically zillions bytes of data they produce at the network edge.
A blend of edge computing and Artificial Intelligence (AI) techniques could optimally shift the resourceful computation servers closer to the network edge, which provides the support for advanced AI applications (e.g., video/audio surveillance and personal recommendation system) by enabling intelligent decision making on computing at the point of data generation as and when it is needed, and distributed Machine Learning (ML) with its potential to avoid the transmission of large dataset and possible compromise of privacy that may exist in cloud-based centralized learning.
Therefore, AI is envisioned to become native and ubiquitous in future communication and networking systems.
In this paper, we conduct a comprehensive overview of recent advances in distributed intelligence in wireless networks under the umbrella of native-AI wireless networks, with a focus on the basic concepts of native-AI wireless networks, on the AI-enabled edge computing, on the design of distributed learning architectures for heterogeneous networks, on the communication-efficient technologies to support distributed learning, and on the AI-empowered end-to-end communications. We highlight the advantages of hybrid distributed learning architectures compared to the state-of-art distributed learning techniques.
We summarize the challenges of
existing research contributions in distributed intelligence in wireless networks and identify the potential future opportunities.


\end{abstract}


\section{Introduction}

\subsection{Native-AI Wireless Networks}

Recently, Artificial Intelligence (AI) techniques have been successfully deployed in several areas, including computer vision, Natural Language Processing (NLP), and smart decision making.
In wireless communications, AI is expected to become an indispensable part of the future wireless networks, such as in regard to the sixth Generation (6G) networks and beyond.
On one hand, the wireless network performance will need to be improved to support and enable AI services for the connected users and things. On the other hand, AI algorithms have huge potential to offer intelligence-based solutions for solving challenging problems in wireless communications which cannot be solved by traditional analytical approaches. Therefore,
with the advancement in wireless networks and their software defined capabilities,
AI will become native and ubiquitous in wireless networks \cite{letaief2019roadmap}.
The native AI wireless networks provide a new vision of traditional wireless networks by deeply integrating native AI-based
intelligent architecture into every aspect of the wireless networks, such as communications technologies, information technologies and data technologies.
It redefines the device-pipe-cloud, supports distributed AI services,
and truly enables pervasive intelligence.
The current research on native AI wireless networks mainly focuses on the three areas \cite{letaief2019roadmap,wu2021toward}, AI-based network optimization and management, AI-empowered wireless communication and AI-enabled distributed data processing.

\begin{figure}[t]
  \centering
  \includegraphics[width=4.5in]{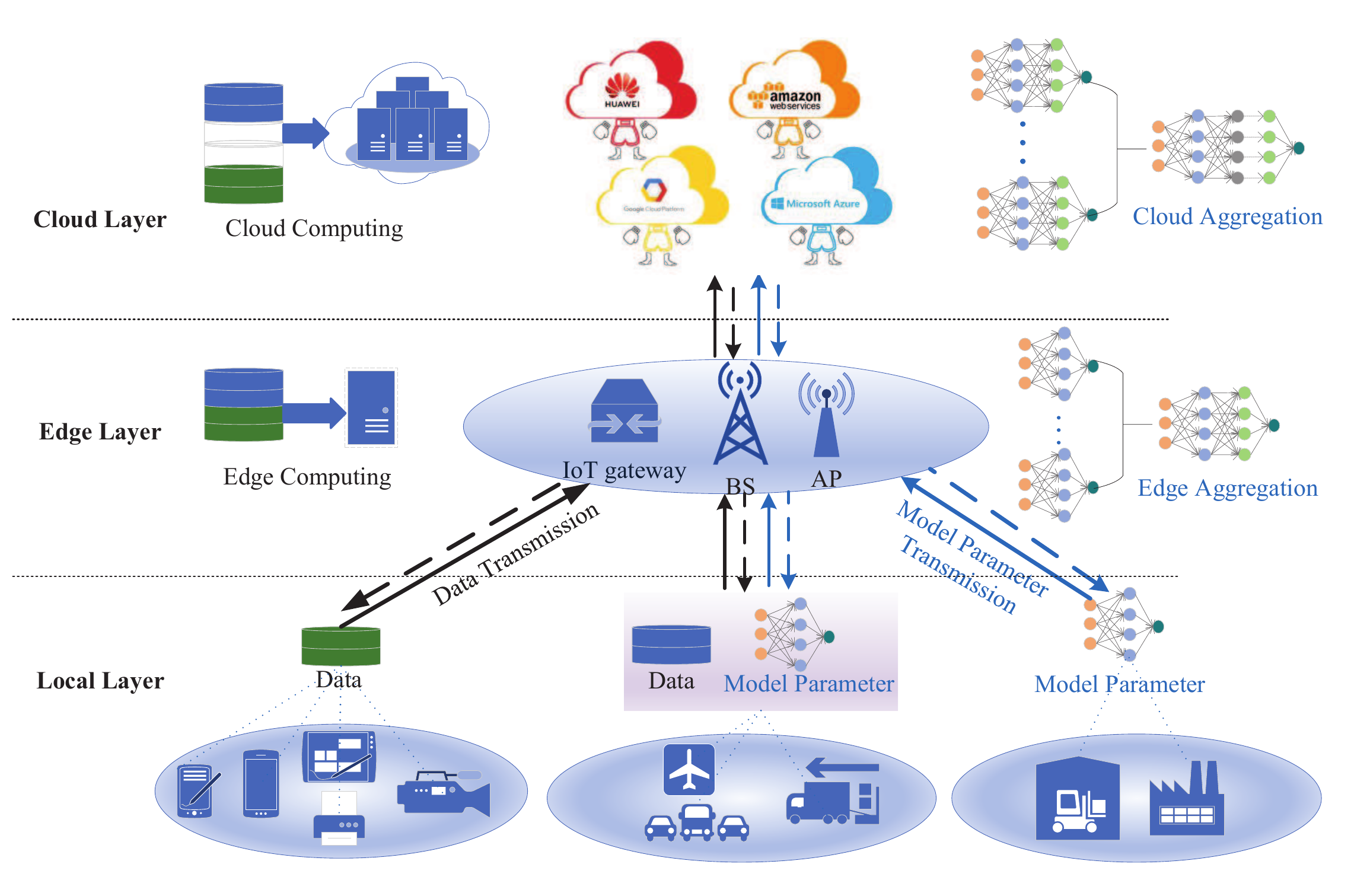}\\
  \caption{The illustration of distributed intelligence in wireless networks. }\label{DI_intelligence}
\end{figure}

\subsection{AI-enabled Distributed Data Processing}

Currently, there are hundreds billions of wireless devices (e.g., 13.8 billions Internet of Things (IoT) devices \cite{Knud2020}, 6.37 billions smart phones \cite{Radicati2021,GSMA2021}  and  5 millions drones \cite{Levitate2020} etc.) around the world,
and the number is expected to increase faster in the next decade. The massive devices are equipped with increasingly advanced sensors, computing and communication capabilities, and they are geographically distributed in different smart-x environment, e.g., smart homes, smart cities and smart agriculture, with the capability to undertake various crowdsensing tasks \cite{Mach2017}, to extract features from a large amount of data and make decisions using Machine Learning (ML) algorithms, especially Deep Learning (DL) (e.g., Convolutional Neural Network (CNN) and deep neural network (DNN)). In the upcoming 6G communications, each network element will natively integrate communication, computing, and sensing capabilities, facilitating the evolution from centralized intelligence in the cloud to ubiquitous intelligence on deep edges. As shown in Fig. \ref{DI_intelligence}, 6G will employ a deep edge architecture to enable massive machine learning in a distributed and collaborated manner.
This three-layer architecture integrates the cloud platform (e.g., cloud servers), the edge devices (e.g., WiFi router, Base Stations (BSs), IoT gateway or micro-datacenter), local devices (e.g., smart phones, IoT devices and vehicles), and advanced wireless communications, which can support
AI-enabled applications at the edge of wireless networks \cite{Taleb2017}.
The use of edge layer pushes the computational resources geographically closer to the local devices compared to the cloud platform, and thus the physical proximity between the computational servers and information-generation sources promises the advantages of edge computing, including low latency, high energy efficiency, proper privacy protection, and reduced bandwidth consumption.

From Fig. \ref{DI_intelligence}, to support AI services at the edge, the ML models can be learned by either centralized or distributed ML model training. For the conventional centralized learning, the massive data generated at local layer can be directly transmitted to edge layer or cloud layer for learning ML models.
Recently, an emerging distributed ML architecture built on deep edge intelligence has been shown to have the potential to meet the large-scale intelligence requirements of future society and manufacturing. This means the ML models can be learned through the collaboration of distributed devices and a centralized parameter server.
From Fig. \ref{DI_intelligence}, the model aggregation can be performed at the edge layer by deploying the parameter server at BS, WiFi router or IoT gateway. Also, a hierarchical distributed learning architecture can be applied by performing model aggregation at the cloud layer for complex ML model training and inference. For simplicity,
we will consider the BS as the parameter server located at the edge layer to illustrate the architecture of edge computing and distributed learning in wireless networks.

The conventional centralized learning approach of offloading raw data  to the edge incurs huge cost (e.g., large time delay, energy consumption and wide bandwidth) due to the transmission of large dataset, and reveals privacy and security concerns.
To address these challenges, distributed learning was proposed to let users keep their private data locally and share only model parameters or smashed data instead of raw data to a central server \cite{verbraeken2020survey}.
Federated Learning (FL) as a popular distributed learning technique was first proposed to provide communication-efficient distributed ML model training, in which the users perform local model training on their own private datasets, and then share their local model updates instead of raw data to a central server where a model parameter aggregation is performed to update the global model \cite{mcmahan2017}. Considering the diversity of users with different computational capabilities and resources (e.g., the size of local datasets, different data distributions and wireless channel qualities), FL is not always efficient since it requires that all the users are capable of computing gradients but this may not be possible for some users. Moreover, the users have to offload the local updates of the full ML model and this causes large communication overhead for users when the ML model is complex. Fortunately, the users with weak computational capability can choose centralized learning by migrating the training task to the server or a distributed learning approach that only runs partial ML model locally with the rest running at the server, and it is known as Split Learning (SL).
To this end, a hybrid learning architecture could be more energy-efficient for heterogeneous wireless networks to benefit from different learning approaches, such as Hybrid Centralized and Federated Learning (HCFL) \cite{Elbir2020}, and Hybrid Split and Federated Learning (HSFL) \cite{Liu2021,Liu2022_energy}.
Moreover, in distributed learning, the contributions of different users to the global model update are different, and thus scheduling the users with more contributions, i.e., large local model updates and good channel qualities, for participating in model training is  important.
Since the participation consumes users' energy and could possibly reveal their privacy, not all the users are actively willing to contribute to global model training without sufficient compensation. The incentive mechanism was stuided to encourage users to join the  model training by introducing rewards and payment for them based on Shapley value, Stackelberg game, auction and contract theory \cite{Zhan2021}.
On the other hand, due to the diversity of users, each user may not complete their local computation at the same time. Those users completing local model updates slower than others are called stragglers. When performing model aggregation, the stragglers will cause an adverse effect to the
convergence of the global model \cite{Lee2021}.
Therefore, asynchronous distributed learning was introduced to address the harmful effects from the stragglers by adopting dynamic learning rates and
using a regularized loss function \cite{Xu2021}.

\subsection{AI-based Network Optimization and Management}

Recently, a wide range of new applications, like mobile payment, mobile games and eXtended Reality (XR) services (including Virtual Reality (VR), Augmented Reality (AR) and Mixed Reality (MR)), has been deployed in wireless networks.
This requires an intelligent use of communications, computing, control, and storage resources from the network edge to the core, and across multiple radio technologies and network platforms. Therefore, to meet diverse service requirements, the existing technologies, such as Software Defined Networks (SDN), Network Functions Virtualization (NFV), and network slicing will need to be further improved relying on AI-based methods. Last but not least, the volume and variety of data generated in wireless networks are growing significantly. This requires data-driven algorithms, such as the ML algorithms, to extract insights from the massive data and it opens up great opportunities for intelligent network planing to achieve real-time additivity to dynamic network environments. Therefore, AI will be an indispensable tool to facilitate intelligent learning, reasoning, and decision making
in 6G wireless networks.

AI techniques are powerful for the quick analysis of big data and extracting insights from the data, which has achieved sustained success in many research areas, including automatic control in robotics, image processing in computer vision, speech recognition and natural language processing.
In Fig. \ref{DI_intelligence}, the introduction of edge layer provides distributed computing resources for the implementation of AI techniques in analyzing the big data generated at local layer. Benefiting from distributed learning architecture, the complex AI models can be trained and inferred  efficiently at the network edge. The big data analytics accomplished by AI techniques, including four different types, namely descriptive analytics, diagnostic analytics, predictive analytics, and prescriptive analytics, can support both AI services deployed at the network edge and network performance improvement in wireless networks.

\begin{figure}[t!]
  \centering
  \includegraphics[width=4in]{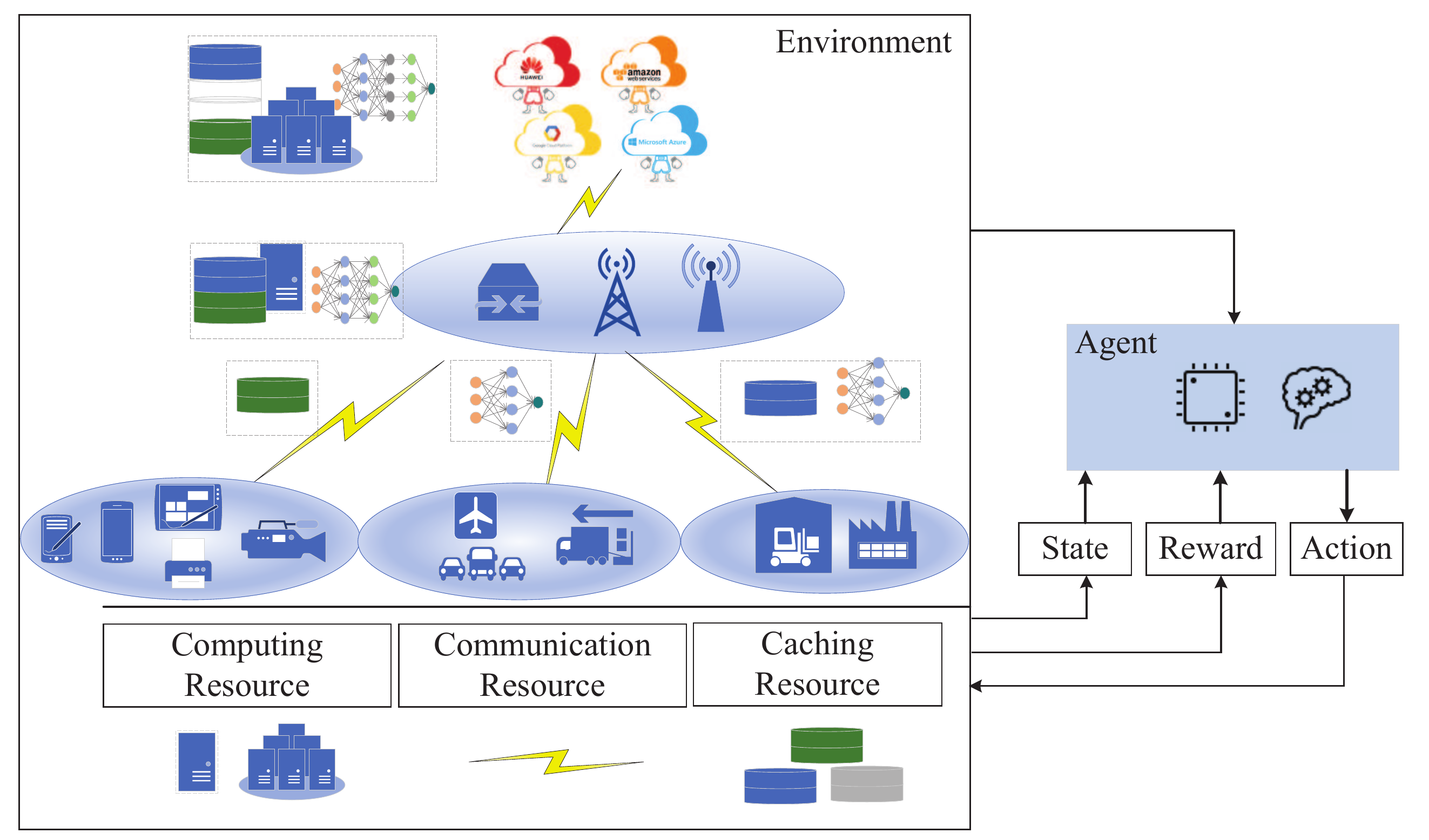}\\
  \caption{\small{AI techniques provide intelligent solutions for wireless communications.}}
  \label{AI_solutions}
\end{figure}

Due to the increase of network scale, density, and heterogeneity, it is hard or even impossible to model such dynamic wireless system with traditional optimization approaches. The conventional network optimization assumes the objective function to be available in nice algebraic forms, and allows an optimizer to evaluate a solution by simple calculation \cite{letaief2019roadmap}. However, the mapping between a decision and its effect on the physical system is cost prohibitive to define and may not be analytically available. Recent advances in AI technologies, such as statistical learning, Reinforcement Learning (RL) and DL algorithms, can solve the formulated complicated network optimization problems in future wireless networks since they can find the asymptotically optimal solutions iteratively using the Stochastic Gradient Descent (SGD) methods. Specifically,  the RL techniques including Deep RL (DRL), Multi-Armed Bandit (MAB) theory and multi-agent RL algorithms can establish a feedback loop between the decision maker and the physical system, so that the decision maker can iteratively refine its action based on the system’s feedback to reach the optimality eventually. As shown in Fig. \ref{AI_solutions}, RL techniques have been broadly applied to address several emerging issues in communication and networking, including resource allocation, wireless caching, computation offloading and user scheduling etc.

\begin{figure}[t!]
  \centering
  \includegraphics[width=4in]{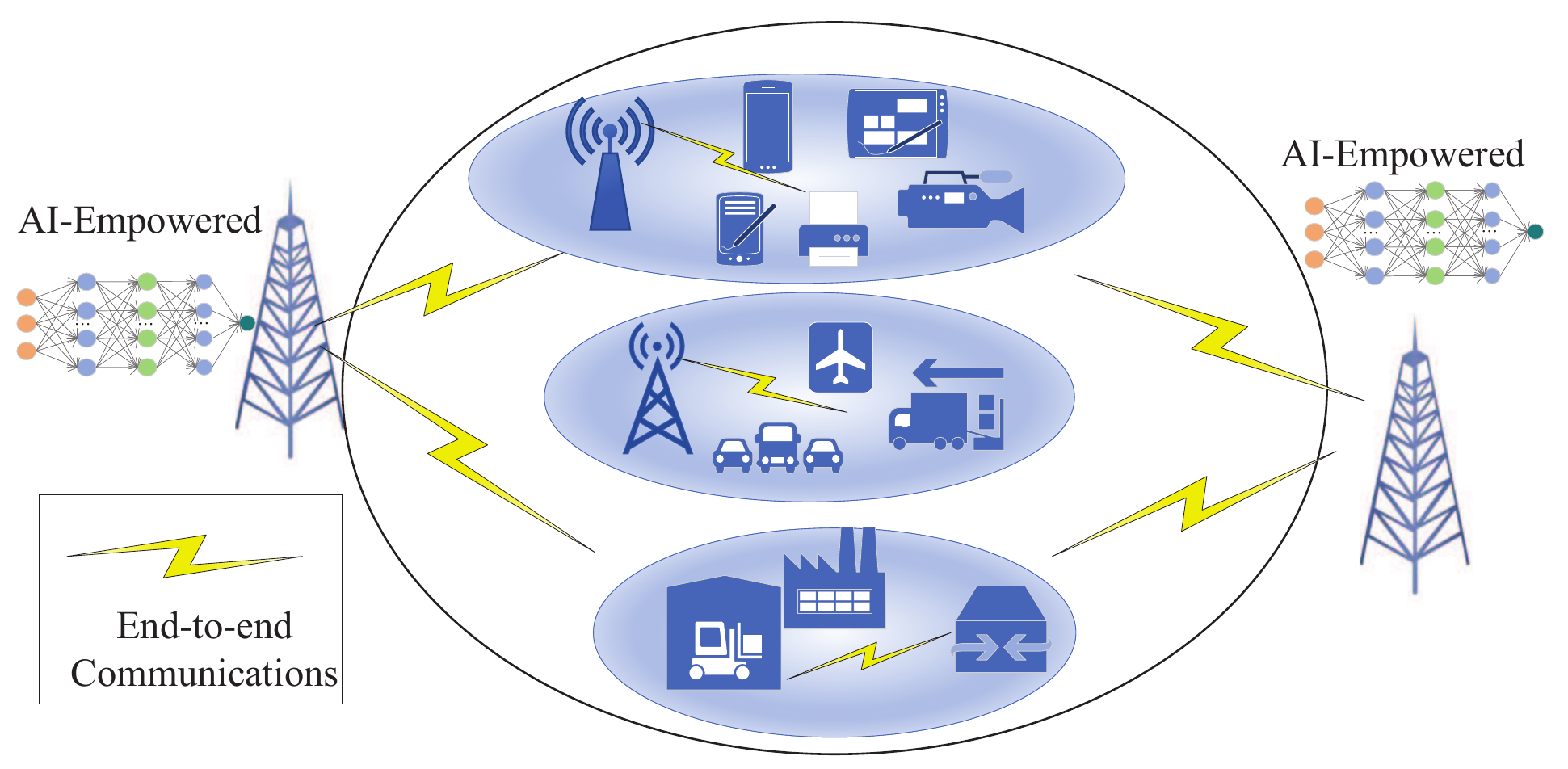}\\
  \caption{\small{AI-empowered end-to-end communications.}}
  \label{AI_wireless}
\end{figure}

\subsection{AI-empowered Wireless Communications}

As discussed above, to support AI services in wireless networks, the data generated by the end users need to be offloaded to the central server that uses centralized learning algorithms to unlock their potential, or kept locally by relying on distributed learning and offloading local model parameters. In both scenarios, the requirements on communication channel quality are high since offloading raw data needs large bandwidth and high data rate communication while offloading model parameters needs ultra reliable and low latency communication links. The current 4G/5G is restricted to support AI services with high requirements on Key Performance Indicators (KPI), and thus ML techniques have been studied in wireless communications to improve those KPI metrics \cite{Kato2020}. The potentials of ML techniques have been widely studied in block-based communication systems to support different locally optimized objectives, such as signal compression, modulation, channel coding and so on. To realize global optimization in communication systems, ML for end-to-end systems has been proposed to further enhance the communication efficiency\cite{8052521}.

With the development of wireless communications that involve emerging advanced technologies, the complicated environment bring unprecedented challenges to the communication system modelling. In conventional chain-shape block-based communication systems, DL has been deployed in different independent modules for multiple purposes with significantly improved performance \cite{8382166}, such as
interference alignment \cite{wang2021joint}, jamming resistance, modulation classification \cite{o2018over}, physical coding \cite{zhang2021countermeasures} and so on. However, existing approaches with separate blocks can not holistically capture the comprehensive aspects of the real-world system. Therefore, the potential of the AI-empowered end-to-end communication systems to support future network has been discussed \cite{letaief2019roadmap}. As shown in Fig. \ref{AI_wireless}, the envisioned intelligent end-to-end system can realize self-optimization communication with the help of advanced sensing, data collection, and AI technologies.

\begin{table}[t]
\centering
\caption{Comparison between Related Existing Survey}
\label{table1}
\resizebox{\columnwidth}{!}{%
\begin{tabular}{|ll|ll|llllll|l|}
\hline
\multicolumn{2}{|l|}{\textbf{Year of publication}}                                                                                                           & \multicolumn{1}{l|}{2021}                                    & 2021                                       & \multicolumn{1}{l|}{2020}                            & \multicolumn{1}{l|}{2020}                            & \multicolumn{1}{l|}{2021}                            & \multicolumn{1}{l|}{2021}                            & \multicolumn{1}{l|}{2021}                            & 2021                            &                                                                                                   \\ \hline
\multicolumn{2}{|l|}{\textbf{Existing survey}}                                                                                                                         & \multicolumn{1}{l|}{\cite{verbraeken2020survey}} & \cite{liu2021distributed} & \multicolumn{1}{l|}{\cite{9060868}} & \multicolumn{1}{l|}{\cite{9205981}} & \multicolumn{1}{l|}{\cite{9530714}} & \multicolumn{1}{l|}{\cite{9446488}} & \multicolumn{1}{l|}{\cite{Park2021}} & \cite{9562559} & Our work                                                                                          \\ \hline


\multicolumn{2}{|l|}{\diagbox{\textbf{Key points}}{\textbf{Main topic}}}                                                                                        & \multicolumn{2}{l|}{\textbf{FL/DML}}                                                & \multicolumn{6}{l|}{\textbf{FL/DML for wireless communications}}                                                                                                                                                                                                                                                   & \textbf{\begin{tabular}[c]{@{}l@{}}Distributed intelligence \\ in wireless networks\end{tabular}} \\ \hline
\multicolumn{1}{|l|}{\multirow{3}{*}{\textbf{\begin{tabular}[c]{@{}l@{}} Main ML\\ techniques\end{tabular}}}}                               & DL                   & \multicolumn{1}{l|}{\checkmark}               &                                            & \multicolumn{1}{l|}{\checkmark}       & \multicolumn{1}{l|}{\checkmark}       & \multicolumn{1}{l|}{}                                & \multicolumn{1}{l|}{}                                & \multicolumn{1}{l|}{}                                & \checkmark       & \checkmark                                                                         \\ \cline{2-11}
\multicolumn{1}{|l|}{}                                                                                                                & RL                   & \multicolumn{1}{l|}{\checkmark}               &                                            & \multicolumn{1}{l|}{}                                & \multicolumn{1}{l|}{}                                & \multicolumn{1}{l|}{}                                & \multicolumn{1}{l|}{}                                & \multicolumn{1}{l|}{}                                & \checkmark       & \checkmark                                                                         \\ \cline{2-11}
\multicolumn{1}{|l|}{}                                                                                                                & MARL                 & \multicolumn{1}{l|}{}                                        &                                            & \multicolumn{1}{l|}{}                                & \multicolumn{1}{l|}{}                                & \multicolumn{1}{l|}{}                                & \multicolumn{1}{l|}{}                                & \multicolumn{1}{l|}{}                                & \checkmark       & \checkmark                                                                         \\ \hline
\multicolumn{1}{|l|}{\multirow{2}{*}{\textbf{\begin{tabular}[c]{@{}l@{}}Distributed \\ learning\end{tabular}}}}                       & FL                   & \multicolumn{1}{l|}{\checkmark}               & \checkmark                  & \multicolumn{1}{l|}{\checkmark}       & \multicolumn{1}{l|}{\checkmark}       & \multicolumn{1}{l|}{\checkmark}       & \multicolumn{1}{l|}{\checkmark}       & \multicolumn{1}{l|}{\checkmark}       & \checkmark       & \checkmark                                                                       \\ \cline{2-11}
\multicolumn{1}{|l|}{}                                                                                                                & Beyond FL            & \multicolumn{1}{l|}{}                                        &                                            & \multicolumn{1}{l|}{}                                & \multicolumn{1}{l|}{\checkmark}       & \multicolumn{1}{l|}{}                                & \multicolumn{1}{l|}{\checkmark}       & \multicolumn{1}{l|}{\checkmark}       & \checkmark      & \checkmark                                                                         \\ \hline
\multicolumn{1}{|l|}{\multirow{2}{*}{\textbf{\begin{tabular}[c]{@{}l@{}}ML for\\ communications\end{tabular}}}}                       & Block-based system   & \multicolumn{1}{l|}{}                                        &                                            & \multicolumn{1}{l|}{}                                & \multicolumn{1}{l|}{\checkmark}       & \multicolumn{1}{l|}{}                                & \multicolumn{1}{l|}{}                                & \multicolumn{1}{l|}{}                                &                                 & \checkmark                                                                         \\ \cline{2-11}
\multicolumn{1}{|l|}{}                                                                                                                & End-to-end system    & \multicolumn{1}{l|}{}                                        &                                            & \multicolumn{1}{l|}{}                                & \multicolumn{1}{l|}{}                                & \multicolumn{1}{l|}{}                                & \multicolumn{1}{l|}{}                                & \multicolumn{1}{l|}{}                                &                                 & \checkmark                                                                        \\ \hline
\multicolumn{1}{|l|}{\multirow{5}{*}{\textbf{\begin{tabular}[c]{@{}l@{}}Communications\\ for\\ distribtted\\ learning\end{tabular}}}} & Aggregation          & \multicolumn{1}{l|}{\checkmark}               & \checkmark                  & \multicolumn{1}{l|}{\checkmark}       & \multicolumn{1}{l|}{\checkmark}       & \multicolumn{1}{l|}{\checkmark}       & \multicolumn{1}{l|}{}                                & \multicolumn{1}{l|}{\checkmark}       & \checkmark       & \checkmark                                                                         \\ \cline{2-11}
\multicolumn{1}{|l|}{}                                                                                                                & Compression          & \multicolumn{1}{l|}{}                                        &                                            & \multicolumn{1}{l|}{\checkmark}       & \multicolumn{1}{l|}{\checkmark}       & \multicolumn{1}{l|}{\checkmark}       & \multicolumn{1}{l|}{}                                & \multicolumn{1}{l|}{\checkmark}       & \checkmark       & \checkmark                                                                         \\ \cline{2-11}
\multicolumn{1}{|l|}{}                                                                                                                & Security and privacy & \multicolumn{1}{l|}{\checkmark}               & \checkmark                  & \multicolumn{1}{l|}{\checkmark}       & \multicolumn{1}{l|}{\checkmark}       & \multicolumn{1}{l|}{\checkmark}       & \multicolumn{1}{l|}{\checkmark}       & \multicolumn{1}{l|}{}                                &                                 & \checkmark                                                                         \\ \cline{2-11}
\multicolumn{1}{|l|}{}                                                                                                                & Asychronous          & \multicolumn{1}{l|}{\checkmark}               &                                            & \multicolumn{1}{l|}{}                                & \multicolumn{1}{l|}{\checkmark}       & \multicolumn{1}{l|}{}                                & \multicolumn{1}{l|}{}                                & \multicolumn{1}{l|}{}                                &                                 & \checkmark                                                                    \\ \cline{2-11}
\multicolumn{1}{|l|}{}     & User scheduling      & \multicolumn{1}{l|}{}                                        &                                            & \multicolumn{1}{l|}{}                                & \multicolumn{1}{l|}{}                                & \multicolumn{1}{l|}{}                                & \multicolumn{1}{l|}{}                                & \multicolumn{1}{l|}{\checkmark}       &                                 & \checkmark               \\ \hline
\multicolumn{1}{|l|}{\multirow{3}{*}{\textbf{Others}}}                                                                                & Edge computing       & \multicolumn{1}{l|}{}                                        &                                            & \multicolumn{1}{l|}{\checkmark}       & \multicolumn{1}{l|}{}                                & \multicolumn{1}{l|}{\checkmark}       & \multicolumn{1}{l|}{}                                & \multicolumn{1}{l|}{}                                & \checkmark       & \checkmark                                                                         \\ \cline{2-11}
\multicolumn{1}{|l|}{}                                                                                                                & Incentive mechanism  & \multicolumn{1}{l|}{}                                        &                                            & \multicolumn{1}{l|}{\checkmark}       & \multicolumn{1}{l|}{\checkmark}       & \multicolumn{1}{l|}{}                                & \multicolumn{1}{l|}{}                                & \multicolumn{1}{l|}{}                                &                                 & \checkmark                                                                         \\ \cline{2-11}
\multicolumn{1}{|l|}{}                                                                                                                & Architectures        & \multicolumn{1}{l|}{}                                        &                                            & \multicolumn{1}{l|}{}                                & \multicolumn{1}{l|}{}                                & \multicolumn{1}{l|}{}                                & \multicolumn{1}{l|}{}                                & \multicolumn{1}{l|}{\checkmark}       &                                 & \checkmark  \\ \hline
\end{tabular}
}
\end{table}

\subsection{Motivation and Contributions}
The aforementioned research works have laid the basic foundation for understanding the development of applying AI techniques in wireless communications. There are some surveys and tutorials that have tried to address this interdisciplinary problem of AI and wireless communications from the aspects of edge intelligence \cite{chen2019deep,Xu2020}, and distributed ML in wireless communications\cite{verbraeken2020survey,liu2021distributed,9060868,9205981,9530714,9446488,Park2021,9562559}, but their focus is different from our work. Particularly, The authors in \cite{chen2019deep} focused on an overview of the deep learning applications at the network edge. In \cite{Xu2020}, the authors mainly identified edge intelligence from edge caching, edge training, edge inference, and edge offloading.
In \cite{verbraeken2020survey,liu2021distributed}, the authors mainly reviewed the fundamental concepts and techniques of distributed ML, with the focus on FL algorithms. In \cite{9060868}, the authors presented several applications using FL algorithms in mobile edge networks and further introduced the implementation challenges of FL algorithms.
Similar works have been discussed in \cite{9205981,9530714}, the authors illustrated
the basic principles behind implementing FL in supporting efficient and intelligent wireless communications. Apart from FL, the authors in \cite{9446488,Park2021,9562559} explored a broad aspect of distributed ML in wireless communications. Specifically, the latest applications of distributed ML in wireless networks and the practical challenges of which, as well as privacy and security concerns were reviewed in \cite{9446488}.
In \cite{Park2021}, the authors presented the communication-efficient techniques and DML frameworks based on a few selected use cases. Furthermore, the use of
communication techniques for the efficient deployment of distributed learning algorithms in wireless networks has been provided in \cite{9562559}, in which an overview of several emerging distributed learning paradigms, including FL, distributed inference and federated distillation were presented.

Although the aforementioned research contributions present either edge intelligence or distributed ML in wireless networks, the analytical of different ML techniques for edge computing, some distributed learning architectures and the use of DL techniques for end-to-end communication have not been covered.
Besides, a clear illustration of the development of AI algorithms in wireless communications from different aspects is missing. Motivated by the aforementioned inspirations, we develop this survey paper with the goal of comprehensively investigating the major issues, challenges and opportunities of distributed intelligence in wireless networks that falls under the umbrella of native AI wireless networks, with the focus on intelligent data processing, network management optimization, and communication performance improvement.
Table \ref{table1} illustrates the comparisons of this
survey with the existing relevant surveys and tutorials.

To highlight the significance of our contributions, this survey is starting with the introduce of native-AI wireless networks, which provides the readers with the clear concepts of  ML techniques assisting wireless communications and wireless networks supporting AI services at the network edge. We continue present the ML techniques for edge computing, followed by distributed learning architectures and communication-efficient technologies for distributed learning. We then address the use of DL techniques to improve wireless communication performance. Finally, we identify the existing challenges and potential opportunities for achieving distributed intelligence in wireless networks.
The main contributions of this survey are stated as  follows:

\begin{itemize}
    \item [1)] We present a comprehensive survey on the recent advances and on the state-of-art in deploying distributed intelligence in wireless networks. The basic concepts of edge computing and distributed learning techniques are introduced and key advantages are summarized. Moreover, the research challenges and potential opportunities are also discussed.
    \item [2)] We investigate different ML techniques for optimizing computation offloading and resource management in edge computing networks. RL techniques, including DRL, multi-agent RL and Federated RL (FRL), and other learning techniques, such as DL and imitation learning, are reviewed. Furthermore, we summarize the application scenarios and complexity of those ML techniques and traditional optimization methods. The challenges of exploiting the existing ML techniques for complicated edge computing have been identified and the potential solutions are also underlined.
    \item [3)] We review the state-of-art hybrid learning architectures and asynchronous distributed learning for heterogeneous networks. We also demonstrate that the HSFL framework achieves better learning performance in the wireless networks with diverse users. Moreover, we investigate the motivation of wireless users for joining in global model update by reviewing the design of incentive mechanism schemes. Besides, we identify the challenges of traditional communication technologies, followed by the review of the state-of-art communication-efficient technologies, user scheduling, over-the-air communication and gradient compression.
    \item [4)] We identify the potentials of DL in wireless communications with a review of investigating DL techniques to optimize the traditional communication blocks and redesign end-to-end communication structure. Moreover, DL for current advanced communication technologies are also reviewed. A range of challenges and potential opportunities of using
    \item [5)] We provide the future opportunities and challenges to improve network efficiency, to cope with diverse users, and to prevent privacy leakage and security concerns for distributed intelligence in wireless networks.
\end{itemize}

\begin{figure}[!ht]
  \centering
  \includegraphics[width=6in]{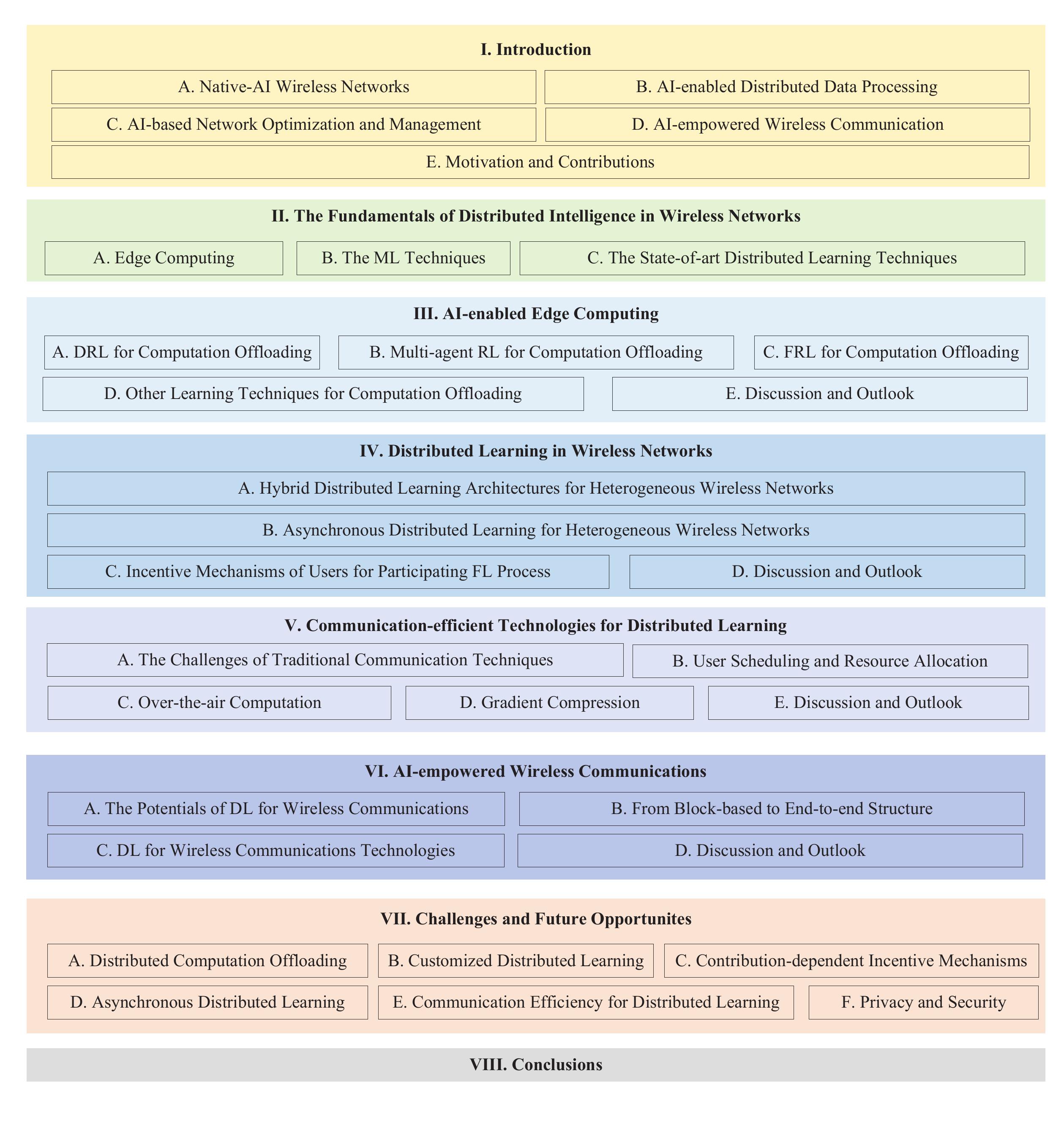}\\
  \caption{The structure of this survey.}\label{surveystructure}
\end{figure}

\subsection{Organization}

The rest of this paper is organized as follows. Sections II provides the fundamentals of edge computing, ML algorithms and distributed learning techniques. Section III discusses different AI techniques enabled edge computing. In Section IV, distributed learning in wireless networks is discussed. Followed by the communication-efficient technologies for distributed learning are illustrated in Section V.
Section VI reviews investigating DL techniques to optimize the traditional communication blocks and end-to-end communications. The open issues and future opportunities are discussed in Section VI. The structure of this survey paper is depicted in Fig. \ref{surveystructure}.

\section{The Fundamentals of Distributed Intelligence in Wireless Networks}

\subsection{Edge Computing}
Edge/fog computing was proposed to pave the way for the evolution of new
era applications and services, which follows a wireless distributed computing framework and is promising to handle data processing for the explosive growth of data generated from the massive wireless devices (e.g., mobile phones, sensors, drones,etc.).
As discussed in \cite{li2017scalable,Li2016a}, the conventional wireless distributed computing networks performed data processing by partitioning the dataset into N files and distributing them to the users to exploit the decentralized computational resources from them.
Edge computing is able to process the massive amount of data generated from geographically distributed users through pushing the computational resources from the central to the network edge which is near the data source. Recently, the increasingly advanced applications, such as mobile payment, smart healthcare, mobile games, and XR applications \cite{Qualcomm2020,Tara2017}, put higher requirements on the resource capacity of smart devices. Cloud computing, was first put forward by Google \cite{Google2008}, which was gradually introduced into the mobile environment to provide computational resources for users with highly demanding applications.  However, the explosive growth of wireless devices (e.g., IoT devices, mobile phones, drones etc.) has brought higher requirements on the transmission bandwidth, latency, energy consumption, application performance,
and reliability \cite{Cisco2020}. In this context, cloud computing is hard to meet the high-performance requirements of users due to limited bandwidth, large latency and high energy consumption.
Instead of replacing cloud computing, edge computing is introduced as a complementary paradigm, so that these challenges can be addressed to let the computational resources  be accessible ubiquitously through deploying a large number of edge nodes in a distributed manner at the network edge. As shown in Fig. \ref{DI_intelligence}, an edge layer is added to the conventional cloud-local wireless networks, more distributed computational resources are deployed at the edge layer to provide edge computing.

The first edge computing concept cloudlet \cite{Satyanarayanan2011} was proposed to bring the  computational or storage resource closer to the users. Cloudlet is a small-scale datacenter or a cluster of computers designed to quickly provide cloud computing services to mobile devices, such as smartphones, tablets and wearable devices, within close geographical proximity.
To support the big data processing for advanced applications with  billions of connected devices at the network edge, a more general concept of edge computing, fog computing with a focus on IoT applications, was introduced by Cisco as it can offer:  a) low latency and location awareness due to proximity of the computational devices to the edge of the network,
b) wide-spread geographical distribution when compared to
the cloud computing, c) interconnection of a large number of end devices
(e.g., wireless sensors), and d) support of streaming and
real-time applications \cite{bonomi2012fog}. However, the aforementioned edge computing concepts are not integrated into an architecture of the mobile network, which causes that the Quality of Service (QoS) and Quality of Experience (QoE) for mobile users can be hardly guaranteed. Therefore, MEC network was proposed to place computation capabilities and service
environments at the edge of cellular networks \cite{hu2015mobile}. By deploying edge server at the cellular BSs, the mobile users can support advanced applications and services flexibly and quickly. The European Telecommunications Standards Institute (ETSI) Industry Specification Group (ISG) further extends its name of MEC to Multi-access Edge Computing (MEC) to embrace the challenges of more wireless communication technologies, such as
Wi-Fi \cite{MEC2019}.

With the distributed architecture, edge computing is able to deal with enormous traffic originated from geographically distributed local devices compared to the centralized cloud computing. The importance and benefits of edge computing are discussed in detail from a set of factors as follows \cite{Baktir2017}.

\begin{itemize}
    \item \textbf{Providing real-time QoS:} the IoT devices and wearable devices are designed for delay-sensitive use cases, and most of them demand high QoS requirements due to the mobile and interactive environment. For instance, the healthcare data generated by the body worn sensors needs to be processed immediately in case of an emergency \cite{Abdellatif2019}. As another example, the AR and VR experiences rely on the graphics rendering on the edge/cloud to augment latency-sensitive on-device head tracking, controller tracking, hand tracking and motion tracking \cite{Siriwardhana2021}.
    The legacy cloud servers cannot support these applications because of the large delay of accessing them through the Wireless Area Network (WAN).  Edge computing could provide the solution by deploying the edge servers closer to the users, which reduces the overall latency through high Local Area Network (LAN) bandwidth and decreased number of hops.

    \item \textbf{Decreasing energy consumption:} the limited battery capacity is still a challenge for mobile phones and especially for most IoT devices, and thus reducing energy consumption is always an important goal in wireless networks. Computation offloading has been demonstrated to be an effective method to reduce the total energy consumption by offloading the intensive computational tasks to edge or cloud \cite{Li2014Exploring,Nir2014,Tao2017}. It is also stated that offloading tasks to the edge servers results in lower energy consumption compared to offloading them to the cloud platforms. Certainly, executing tasks locally at the device causes the highest energy consumption.

    \item \textbf{Reducing network congestion:}  the limited bandwidth of the core network makes it vulnerable to the network congestion. In 2020, tens millions of devices are generating 2.5 quintillion bytes of data per day, and this rate is expected to increase \cite{cloud10key}. The conventional approach is to transmit the data through the core network to the cloud servers for processing, which causes heavy burden on core network.  Edge computing prevents this by keeping the traffic at the edge servers and also optimizes the utilization of the limited bandwidth.

    \item \textbf{Scalability:}
    the number of mobile users is expected to increase to 10.3  billions and the number of IoT devices will reach 30.9 billions by 2025 \cite{andalibi2021making}, which creates a significant scalability problem. The conventional cloud cannot provide scalable environment for the data and applications due to highly possible network congestion caused by the data transmission of tens of millions of end devices. With edge computing, if one edge server becomes congested and fails to satisfy the incoming requests, the corresponding service can be transferred to another edge server nearby and let the computing service to be handled there.

\end{itemize}

As mentioned above, edge computing has similar working mechanism as cloud computing, but distributes the computational resources closer to the local devices. Instead of offloading intensive computational tasks to the remote cloud, the end devices recur to the edge servers in the vicinity for computational resources; generally, there are several nearby edge servers that can be accessed by each end device. However, the edge servers have limited power and computational resources compared to the cloud server which is assumed to be super powerful, which makes the computation offloading problem more complicated due to the need of considering edge server selection and resource management \cite{Mach2017}. In cloud computing, the key point of computation offloading is to decide
whether to offload or not, how
much and what should be offloaded. In edge computing, in addition to those points,  we need to address where and how to offload, and how much resources should be allocated. Recently, researchers have studied the joint computation offloading and resource management problem with the goal of minimizing energy consumption and execution delay \cite{Sardellitti2015,You2017}. They formulated the joint problem as a combinatorial optimization problem with non-linear constraints and proposed the computation offloading algorithms based on convex optimization \cite{Mao2017,Chen2018,Tran2019}, Lyapunov optimization \cite{Mao2016,Mao2017a,Liu2017} and game theory \cite{Zhang2018,Pham2018}. Moreover, the design of computation offloading scheme can be modeled as the process of making decisions on offloading and resource allocation by interacting with the dynamic environment, which is then investigated by exploiting the RL algorithms in many research works \cite{Li2018,Chen2019}.

\subsection{The ML Techniques}

\subsubsection{The DL Algorithms}

The basic data-driven deep learning based algorithm adopts a fully connected feed-forward neural network with multiple hidden layers to extract the data representation \cite{hornik1989multilayer}. This multi-layer neural network can be established by supervised learning, unsupervised learning and RL. Without the knowledge of the mathematical model, deep learning can learn from the large amount of labelled data and the hyper-parameters can be tuned based on the domain knowledge for superior insight extraction. Hence, deep learning has been widely applied to the fields that the mathematical description cannot be easily obtained. There are mainly three different kinds of deep learning architectures: DNN, CNN, and Recurrent Neural Network (RNN) \cite{schmidhuber2015deep}. We will briefly introduce DNN and CNN below.

\noindent\textbf{a. DNN}

\begin{figure}[!t]
\centering
\includegraphics[width=6in]{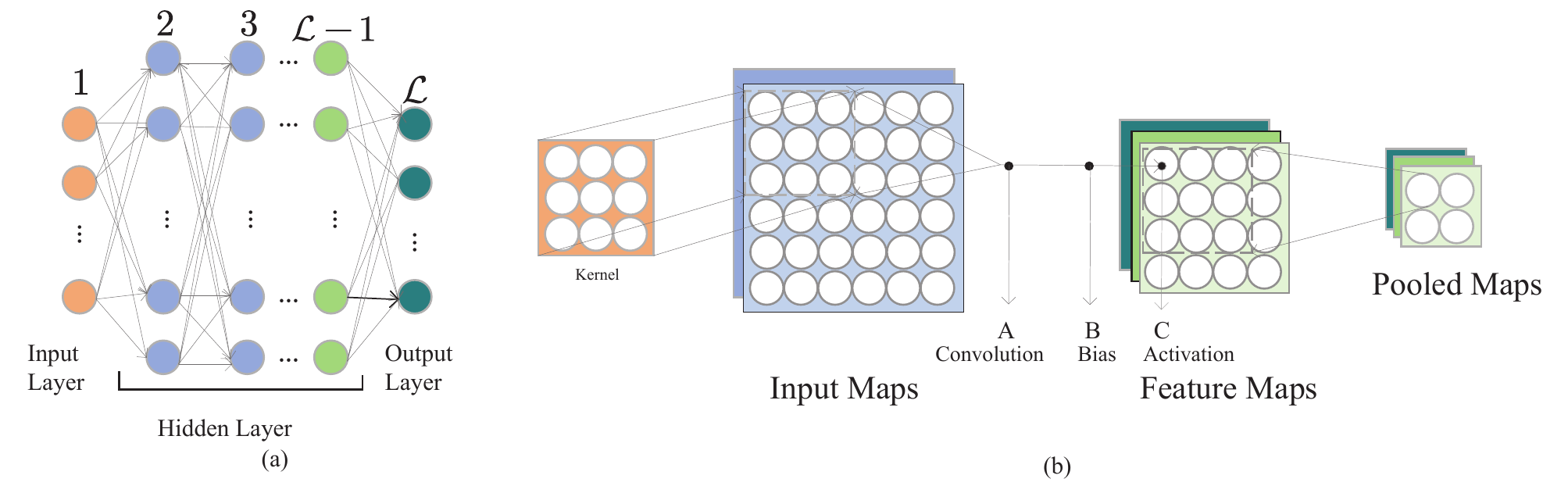}
\caption{The structure of different DL algorithms, (a) DNN, (b) CNN.}
\label{dnn}
\end{figure}

Generally, DNN is a deeper version of Artificial Neural Networks (ANNs) with multiple layers (more than three hidden layers). The structure of the DNN is shown in Fig.\ref{dnn} (a) \cite{schmidhuber2015deep}. In DNN, each layer consists of multiple neurons, each of which has an output that is a non-linear function, like Sigmoid function or ReLU function.
To express the DNN propagation principle, we use $\boldsymbol{i}_{l}$ to represent the input of the $l^{th}$ layer neurons. $o_{l,n_e}$ represents the output of the $n_e^{th}$ neuron at $l^{th}$ layer. $\boldsymbol{W}_{l}^\text{(DNN)}$ and $\boldsymbol{b}_l^\text{(DNN)}$ denote the weight matrix and the bias vector of the $l^{th}$ layer. Hence, each neuron's output can be expressed as
\begin{equation}
\begin{aligned}
o_{l,n_e}=f_{l,n_e}\left({b}_{l,n_e}^{\text{(DNN)}}+\boldsymbol{W}_{l,n_e}^{\text{(DNN)}^T}\boldsymbol{i}_{l}\right),
\end{aligned}
\end{equation}

with $f_{l,n_e}$ as the activation function for the $n_e^{th}$ neuron at the $l^{th}$ layer, and $(\cdot)^{T}$ denotes the transpose.

During the training phase of constructing DNN, the parameter set $\boldsymbol{\theta}_l=(\boldsymbol{W}_l^\text{(DNN)},\boldsymbol{b}_l^\text{(DNN)})$  represents the weights and biases of the DNN model at the $l^{th}$ layer, which can be obtained through back propagation gradient to recursively minimize the loss function until convergence. Conventionally, the gradient descent method is to find the local minimum by taking steps proportional to the gradient of the function, this can be represented by
\begin{align}
 \boldsymbol{\theta}^{(\tau+1)}=\boldsymbol{\theta}^{(\tau)}-\eta\nabla
 \text{Loss}\left({\boldsymbol{\theta}^{(\tau)}}\right),
\end{align}
where $\boldsymbol{\theta}^{(\tau)}$ represents the model parameter set at time slot $\tau$, and $\text{Loss}({\boldsymbol\theta})$ is the loss function with current parameter set, $\eta$ is the learning rate.

Different from gradient descent which calculates the gradient by taking the whole dataset into account, SGD has been proposed to handle much larger dataset in practical scenarios through calculating the model updates based on the mini-batch of data. This can be formulated as
\begin{equation}
\begin{aligned}
&\text{Loss}\left({\boldsymbol{\theta}}\right)=\sum_{d=1}^{D}\text{Loss}_{d}\left({\boldsymbol{\theta}}\right),\\
&\boldsymbol{\theta}^{(\tau+1)}=\boldsymbol{\theta}^{(\tau)}-\eta\nabla
\text{Loss}_{d}\left({\boldsymbol{\theta}^{(\tau)}}\right)
\end{aligned}
\end{equation}
where $D$ is the number of mini-batches of the whole dataset. It has been proved that SGD has higher probability of avoiding from local minimum and data redundancy \cite{6638950}.

\noindent \textbf{b. CNN}

Compared to DNN, CNN puts additional convolutional and pooling layers before feeding the data into the neural network\cite{krizhevsky2017imagenet}. It has been widely utilized to deal with computer vision  and signal compression problems. In Fig. \ref{dnn} (b), the structure of CNN with two-dimensional (2-D) kernel is plotted. There are three main volumes: input maps, feature maps, and pooled maps.
The convolution between a 2-D kernel $w_{m,n,k}^\text{CNN}$ and the $k^{th}$ map at $(x,y)$ spatial location is the sum of products of the weights of the kernel and the elements of the map that are spatially coincident with the kernel \cite{8496892}. Specifically, $m$ and $n$ are variables that indicate the kernel height and width, respectively. At point $A$ in Fig. \ref{dnn} (b), the summation of the overall $K$ depth of the input volume at spatial coordinate $(x,y)$
can be written as
\begin{align}
conv_{x,y}=\sum_{k}\sum_{m,n}{w}_{m,n,k}^\text{(CNN)}{v}_{m,n,k},
\end{align}
where ${v}_{m,n,k}$ is the value of the spatially corresponding elements on the input maps. Then a scalar bias $b^\text{(CNN)}_{x,y}$ is added at point $B$ in Fig. \ref{dnn} (b) as
\begin{align}
    z_{x,y}=conv_{x,y}+b^\text{(CNN)}_{x,y}.
\end{align}
Therefore, the feature map can be expressed as
\begin{align}
    a_{x,y}=f(z_{x,y}),
\end{align}
with $f(\cdot)$ as the activation function. Based on aforementioned steps, the complete feature map can be generated. Next, in the pooling layer, the neurons in the feature maps are then grouped together for average pooling or maximum pooling.
During the training stage, the weights of the 2D-kernel and the bias of each feature map are learned through minimizing the output error and then performing back propagation.

\begin{figure}[t]
  \centering
  \includegraphics[width=3.5in]{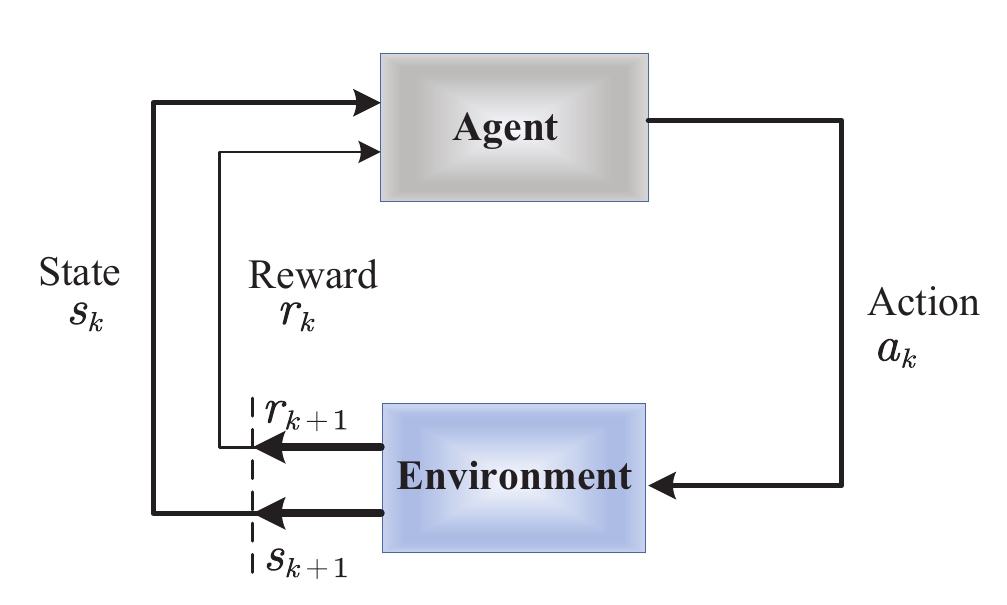}\\
  \caption{\small{The framework of the RL technique.}}
  \label{RL}
\end{figure}

\subsubsection{RL Techniques}
As illustrated in Fig. \ref{RL}, the goal of RL techniques is to create an intelligent agent in the environment that can learn efficient policies to maximize the long-term rewards by taking controllable actions, where the process of the agent taking actions and changing state through interacting with the environment can be modeled as a Markov Decision Process (MDP).
The DRL approach is the combination of deep learning and RL techniques, but it focuses more on RL and aims to solve decision-making problems. The role of deep learning is used to explore the powerful representation ability of DNN to represent a large number of
states and approximate the action values to estimate the
quality of the action in the given states, so that the DRL is able to solve the explosion of state-action space or continuous state-action space problems. The typical application scenario of DRL is to solve various scheduling problems, such as decision-making problems in games, rate selection of video transmission and resource allocation in wireless communications. There are two main DRL approaches introduced as follows.

\noindent\textbf{a. Value-based DRL}

As a representative of value-based
DRL, Deep Q-Network (DQN) was proposed to approximate action values using DNN, which breaks the curse of
high-dimensional input data and successfully maps it to actions \cite{Mnih2015}. However, the  non-linear approximator, DNN, makes DQN unstable due to the  correlations that exist in the sequence of observations.   Hence, the experience replay is used to remove the correlations by using a random sample of prior actions instead of the most recent action to proceed.
Besides, the Double-DQN algorithm that can reduce the observed overestimated action values was studied in \cite{Hasselt2015}, and the  Dueling-DQN proposed in \cite{Wang2015Dueling} can learn which states are (or are not) valuable without having to learn the effect of each action at each state.

\noindent \textbf{b. Policy Gradient-based DRL}

Policy gradient is a policy-based RL algorithm, which relies upon optimizing parametrized policies with respect to the long-term cumulative reward by gradient ascent. Policy gradient algorithms typically proceed by sampling the stochastic policy and adjusting the policy parameters in the direction of greater cumulative reward.
Instead, the Deterministic Policy Gradient (DPG) algorithm was considered in \cite{Silver2014Deterministic}, and it is demonstrated that it has a significant performance advantage over stochastic policy gradients.
By combining DQN and DPG, the Deep Deterministic Policy Gradient (DDPG) algorithm was proposed by using the DNN to parameterize the policy that is then optimized by the policy gradient method \cite{Lillicrap2015}.
Besides, there are a few other state-of-art policy-based DRL algorithms, such as Asynchronous Advantage Actor-Critic (A3C) that enables parallel actor-learners to train the neural network \cite{Mnih2016},  Trust Region Policy Optimization (TRPO) that is effective for optimizing large non-linear policies like neural networks \cite{Schulman2015},
and  Proximate Policy Optimization (PPO) that improves TRPO with simpler implementation \cite{Schulman2017}.
Specially, all of them rely on an \emph{Actor-Critic} (AC) framework, in which
both the \emph{Critic} and \emph{Actor} functions are parameterized with DNN, known as critic network and actor network. The critic network is used to estimate the value function of the state-action pair, while the actor network is in charge of policy updating in the direction suggested by the critic network.

\begin{figure}[t]
  \centering
  \includegraphics[width=3.5in]{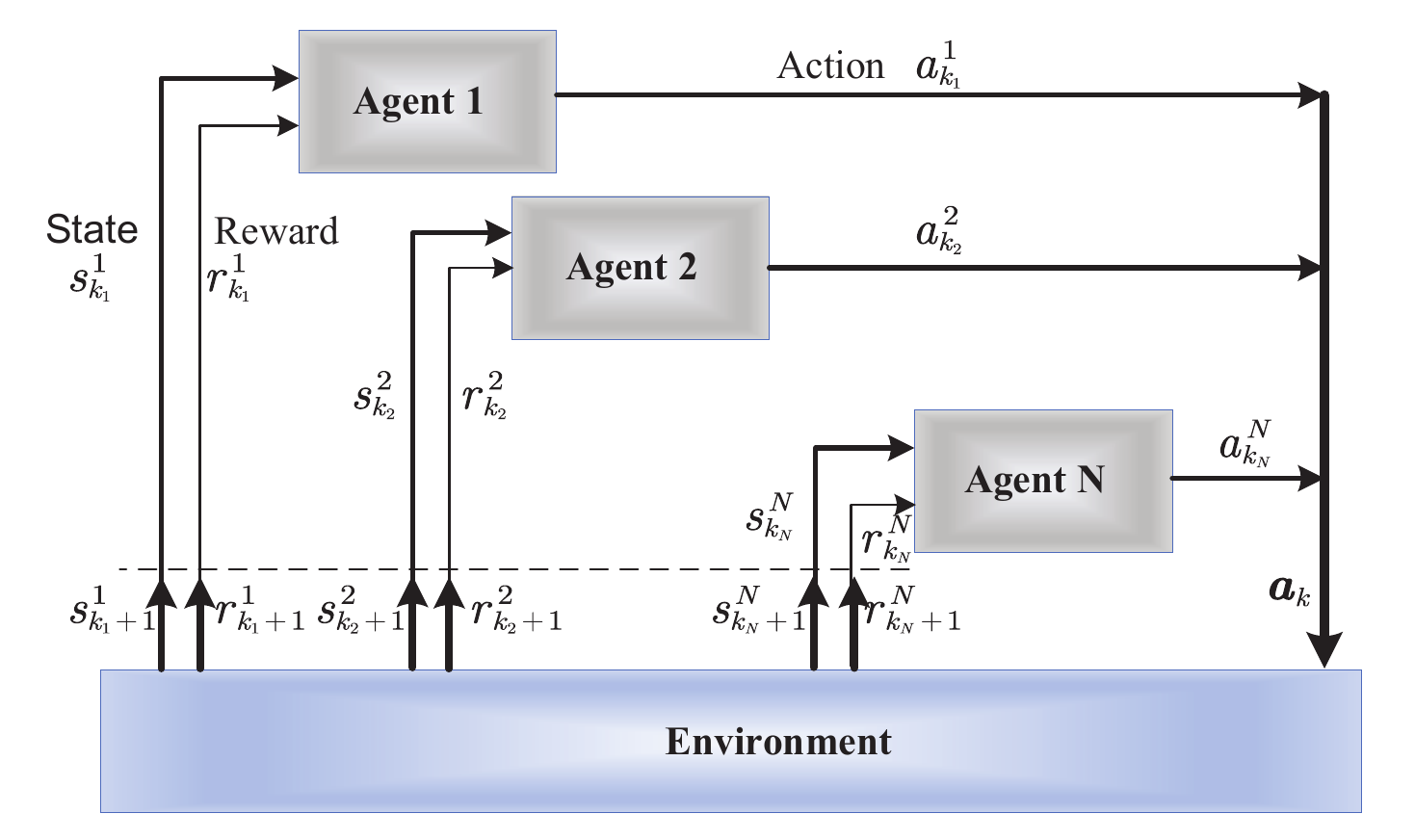}\\
  \caption{\small{The framework of multi-agent RL framework.}}
  \label{MARL_1}
\end{figure}

\subsubsection{Multi-agent RL Techniques}
In the aforementioned RL algorithms, a RL agent is modeled to perform sequential decision-making by interacting with the environment, which is always formulated as an MDP problem.
However, most of the successful RL applications, e.g., the games
of Go and Poker, robotics, and autonomous driving, involve the participation of more
than one single agent, which naturally fall into the realm of multi-agent RL \cite{Zhang2019Multi}. The research on multi-agent RL can be traced back to 1990s \cite{Tan1993,Littman1994}, and most recently it re-emerges due to the advances in single-agent RL techniques.
Specifically, multi-agent RL can address the sequential decision-making problem of multiple agents that operate in a common environment, each of which aims to optimize its own long-term return by interacting with the environment and other agents \cite{Zhang2019Multi}.

Markov game, also known as stochastic game, is a general extension of MDP in multi-agent scenario to include multiple adaptive agents with an interacting or competing goal, and the framework of Markov game has long been used to develop multi-agent RL algorithms originated from \cite{Littman1994}.
Different from the RL algorithm that the environment changes its state only based on the action of one agent, both the evolution of the system and the reward received by each agent depend  on the joint action of all agents in the multi-agent RL algorithm as shown in Fig. \ref{MARL_1}.
The multi-agent RL algorithms are categorized into three groups according to the types of multi-agent tasks that they address.
\begin{itemize}
  \item \emph{Cooperative setting} - a fully cooperative setting is the case that all the agents collaborate to optimize a common long-term return.
  \item \emph{Competitive setting}- in a fully competitive setting, the return of agents usually sum up to zero, which is typically modeled as a zero-sum Markov game.
  \item \emph{Mixed setting}- a mixed setting is usually modeled as a general-sum game, where no restriction is imposed on the goal and the relationship among the agents.
\end{itemize}

\begin{figure}[t]
  \centering
  \includegraphics[width=5.5in]{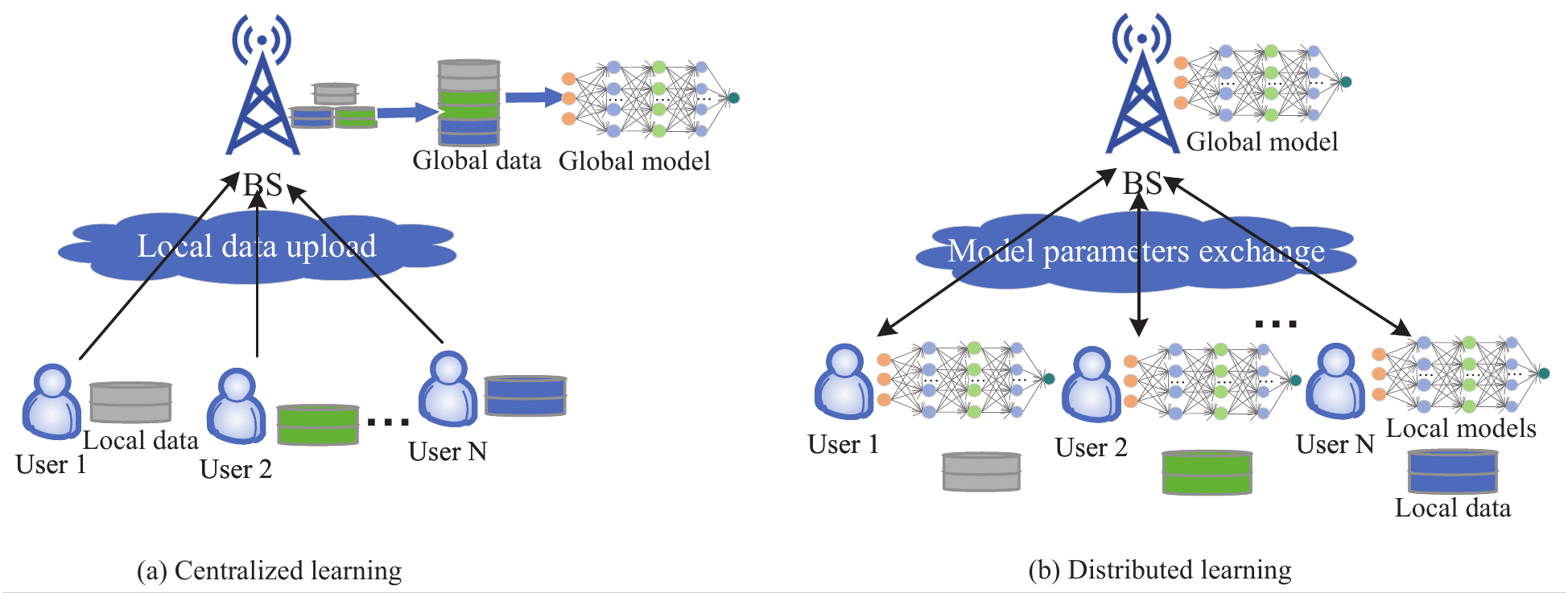}\\
  \caption{The architecture of  (a) centralized learning and (b) distributed learning.}\label{distributed_learning}
\end{figure}

\subsection{The State-of-art Distributed Learning Techniques}
The traditional centralized ML algorithms typically gather the distributed raw data generated at different devices/organizations to a single central server/cluster with shared data storage as shown in Fig. \ref{distributed_learning} (a).
The centralized approach faces the challenges of large computational power and long training time, and most importantly, serious data privacy and security concerns. When the training data becomes huge, e.g., terabyte of data, or is inherently distributed to be stored and processed on individual machines, the model training
process can be carried out by exploiting distributed resources (e.g., computational resource, power and data) over the end devices, which is distributed ML.

Distributed ML has been investigated since 2000 on deploying the structure of distributed computing to speed up the training process so as to reduce the training time. Multiple parallelization techniques have been introduced into distributed ML, such as MapReduce and Hadoop framework relying on  distributed file system \cite{2007a}, Apache Spark saving expensive reads from the disk \cite{Shanahan2015} and Parameter Server with relaxing the stringent requirement of synchronization \cite{li2013parameter}, which could address the large-scale data challenges. More details about the popular architectures of the conventional distributed ML can be found in \cite{Hu2021}.
In wireless networks, due to the naturally distributed characteristics of the data generated over the wireless devices,
a new concept of distributed ML, simply called distributed learning in this article, appears to train a global ML model by keeping the dataset locally at user devices, which exploits the distributed computational resources of the wireless devices, saves the communication cost and protects the users' privacy.
The architecture of the distributed learning is shown in Fig. \ref{distributed_learning} (b).
In the following, we will introduce two state-of-the-art distributed learning algorithms, FL and SL algorithms.

\begin{figure}[t]
  \centering
  \includegraphics[width=3.5in]{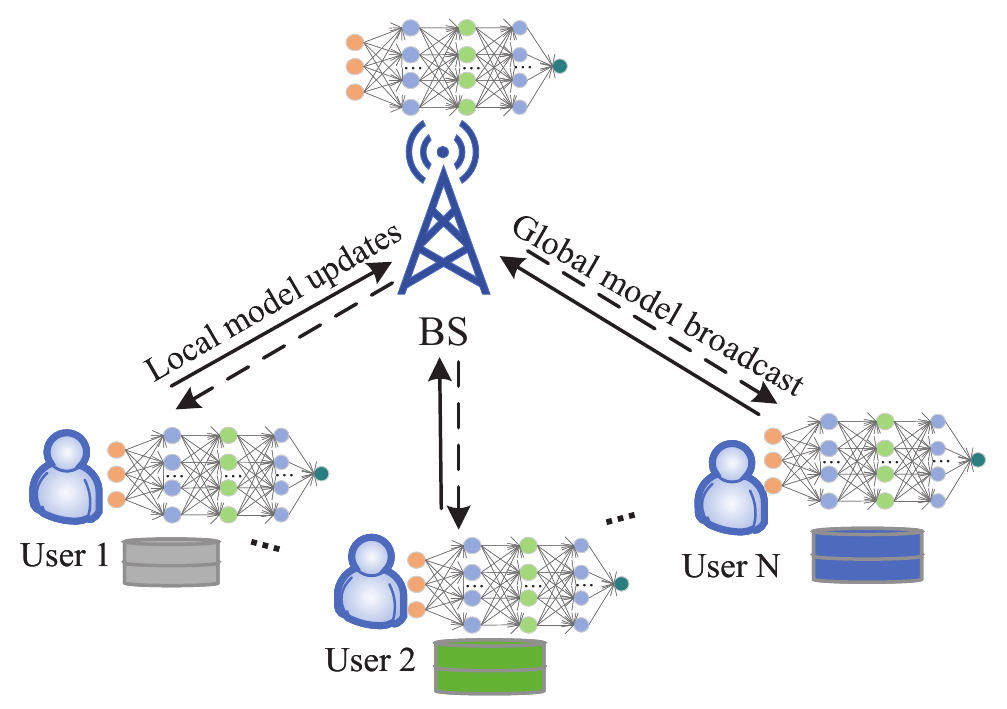}\\
  \caption{The learning procedure of FL technique.}\label{FL}
\end{figure}

\subsubsection{The FL Techniques}
FL is a special type of distributed learning, where multiple users collaboratively train a global model while keeping the raw data distributed at local users without being moved to a single server or data center. The architecture of FL is shown in Fig. \ref{FL}.
FL is flexible and reliable to train ML models in heterogeneous system. This is because of its unique characteristics: a) it does not require the direct raw data transmission from the distributed users, b) it exploits the distributed computational resources from multiple regions and organizations, c) it generally takes advantage of encryption or other defense techniques to ensure
the data privacy and security.

FL was first proposed in \cite{mcmahan2017} to handle the machine learning model training with decentralized data from mobile devices, which was demonstrated to be robust for unbalanced and non-Independent and Identically Distributed (non-IID) data distributions. As a distributed learning scheme, FL brings the learning task to the edge level instead of performing model training at a central entity, which enlightens a series of studies on FL over different areas. For instance, FL has been successfully applied to Google's predictive keyboards \cite{yang2018applied}.  In FL, the communication overhead of FL is proportional to the number of model parameters and it is significantly reduced especially when the users hold the local datesets that are much larger than the model size, which also avoids the transmission of large raw dataset. However, FL struggles with supporting distributed model training for DNNs with large and complex model parameters over capacity-limited wireless channels.


\begin{figure}[t]
\centering
\hspace{-25mm}
\subfigure[The structure of splitNN]{
\includegraphics[scale=0.5]{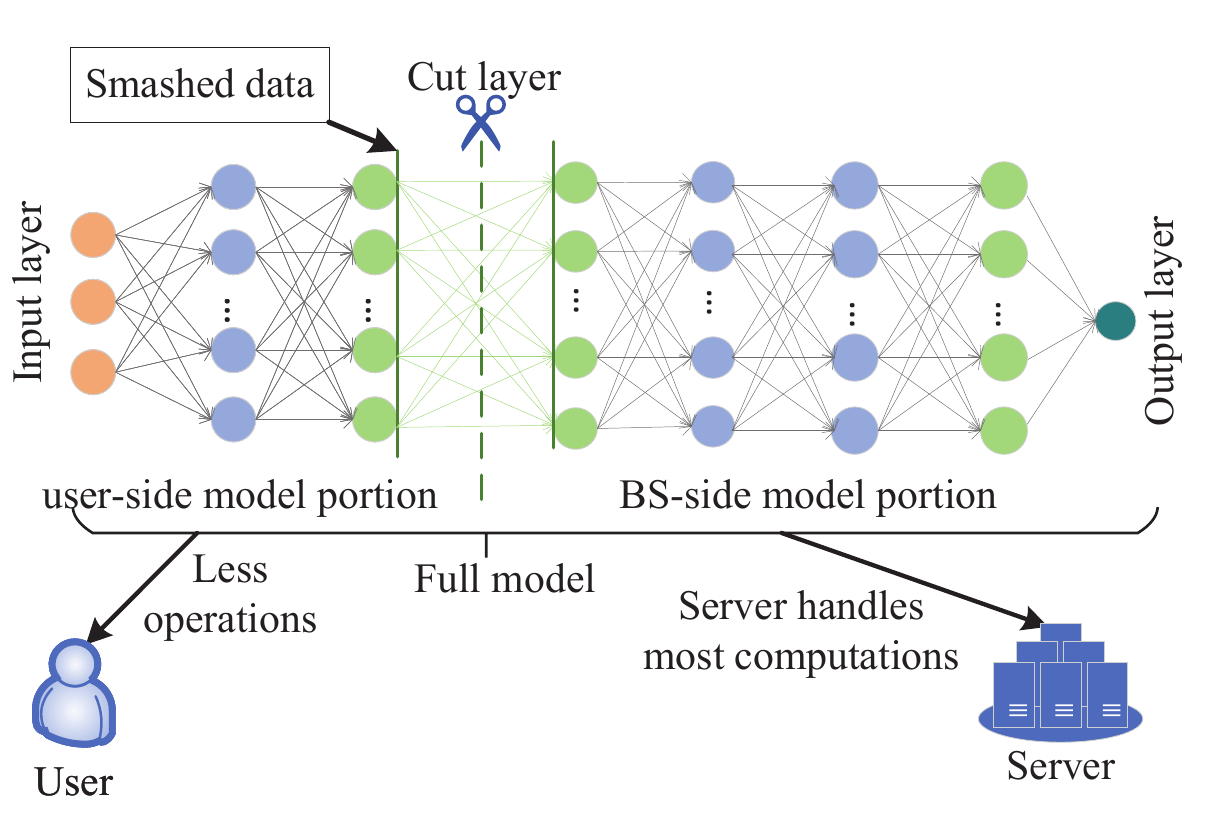}
}
\subfigure[The learning procedure of SL technique]{
\includegraphics[scale=0.5]{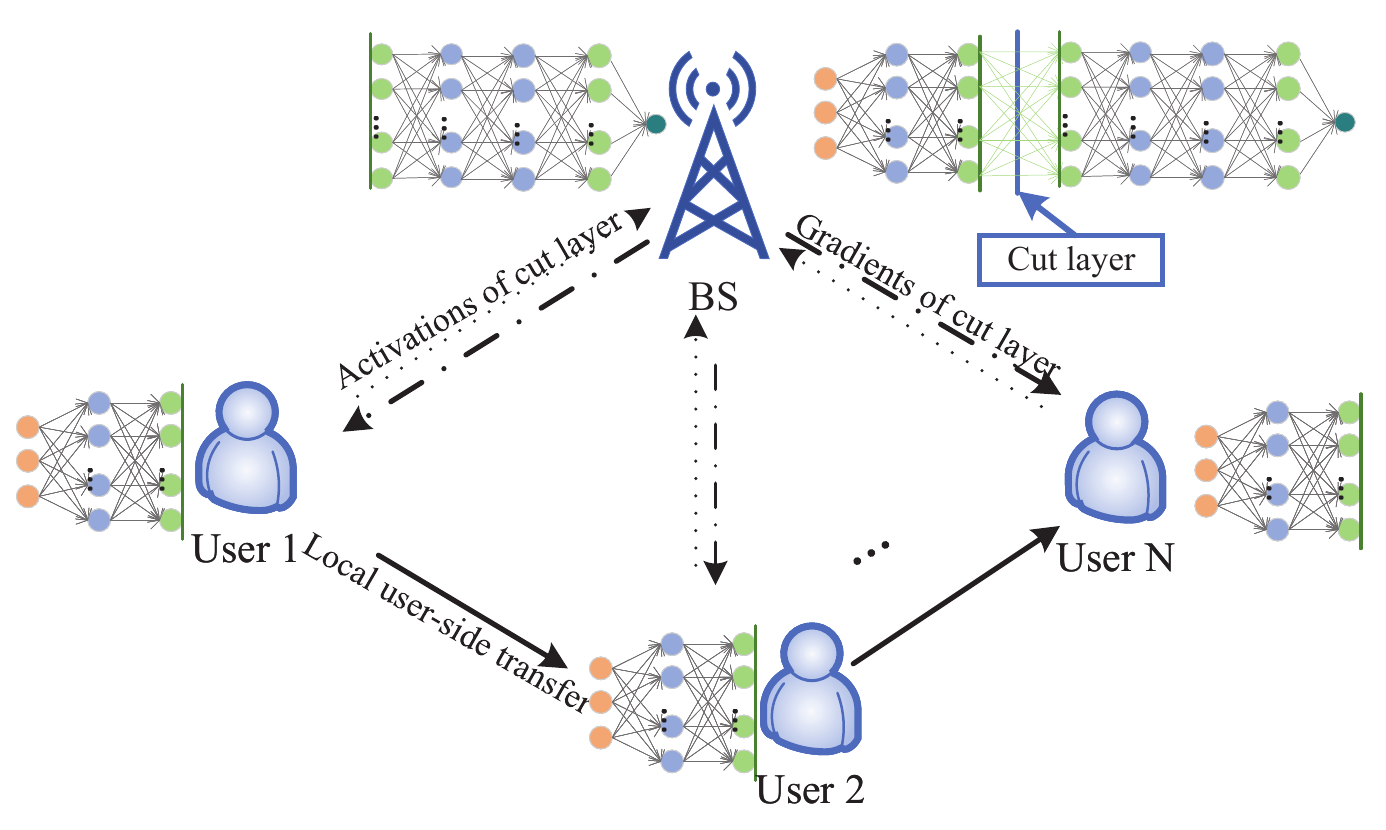}
}
\caption{The learning mechanisms of SL techniques.
}
\label{SL}
\end{figure}

\subsubsection{The SL Techniques}
SL, also known as split neural network (SplitNN), was proposed to address the problem of training a DNN model over multiple data entities. In SL, a DNN is split into multiple sub-networks (e.g., each sub-network includes a few NN layers) by the cut layer, each of which is trained on a different entity \cite{Gupta2018} as shown in Fig. \ref{SL} (a). Similar to FL, SL also provides a solution to train a DNN model while keeping the raw data locally at the distributed users, whereas the users only train a sub-network of the DNN model instead of the full model and the other sub-network is trained by a more powerful parameter server. The architecture of SL is shown in Fig. \ref{SL} (b).

Take the sequential SL as an example, all the users collaborate with the parameter server to train a full ML model sequentially. The parameter server distributes a lower sub-network to the users and itself holds the upper sub-network, and then the training of the full model is carried forward on the user's local dataset by transferring the output of the cut layer to the parameter server.  Next, the parameter server calculates the loss values and the gradients, and then updates its upper sub-network and sends  the gradients of the cut layer back to the user for updating its  lower sub-network. At last, the user returns the updated sub-network to the server, and then the training process of the next user will start.  In this case, only the outputs of the cut layer are shared between users and the parameter server, no raw data is shared so that the user privacy and security are protected. SL was first proposed to be applied in medical applications \cite{Vepakomma2018,vepakomma2019reducing}, where a model is trained with the sensitive health data from different hospitals.

The communication cost of SL mainly depends on two parts: the model size of the first few NN layers prior to the split and the size of the activations that is up to the size of the dataset owned by the user. Therefore, SL requires much lower communication bandwidth when training over the dataset distributed over a large number of users but is being relatively larger in settings with a smaller number of users \cite{vepakomma2018no}.
\subsection{Discussion and Outlook}

Given the increasing research contribution on the interdiscipline
of AI techniques and wireless communication at the network edge, its advantages are becoming obvious, especially at decision-making, network management and AI services support. The ML techniques have become a powerful tool for data analytics and intelligent decision-making, which is able to extract the insights from data and hence provides intelligent network solutions and advanced applications at the network edge. This is particularly essential in large-scale wireless networks supporting a large number of distributed BSs and users. At the time of writing the native-AI enabled wireless networks is still in its infancy. To investigate the proper ML technique for the specific application scenario is challenging. Hence, further research is required to identify the cooperation of different ML techniques in next generation wireless networks.

Moreover, an important issue needs to be addressed is to improve distributed learning performance  i.e., less convergence latency and higher energy-efficiency, under dynamic environment. Specifically, the design of new distributed learning architectures based on FL and SL needs more attention when considering the heterogeneity of wireless networks.

\section{AI-enabled Edge Computing}
Edge computing has been proposed as a promising solution to handle data processing of a large volume of security-critical and time-sensitive data \cite{Yousefpour2018,bittencourt2018internet}. With the distributed deployment of edge devices, edge computing can shift computational and caching capabilities from distant and centralized cloud to the network edge.
This enables AI-based data analytics to be performed in a distributed manner, and thus to support ubiquitous AI services. However, the edge devices are typically resource-constrained and have heterogeneous computation capabilities, thereby causing critical challenges on
resource management and wireless caching \cite{hong2019resource,nguyen2019market}. In addition, with the increasingly powerful chips integrated at the local devices, they are able to handle some simple computational tasks. Thus, deciding which task should be offloaded to the edge, how much power is used to transmit the data, and when and where (i.e., in multiple edge devices) the task is offloaded, is necessary and full of challenges. Recently, many researchers \cite{Sardellitti2015,You2017} have put much attention to this problem from the aspect of optimizing computation offloading scheme and resource allocation, with the proposed algorithms based on convex optimization, Lyapunov optimization, game theory and ML techniques.

Computation offloading plays an important role in edge computing, and it provides a paradigm of appropriately allocating computation resources between different layers (e.g., wireless networks normally consist of three-layer architecture that include local, edge and cloud layers) \cite{Mach2017}. Efficient edge computing relies on the edge device or the end devices making optimal decisions on computation offloading and resource allocation.
Conventional centralized computation offloading methods require complete and accurate network information, so that the edge device can make optimal decisions on which users offload their data while others execute data processing locally, and achieve optimal resource allocation based on the obtained prior network information \cite{Mao2016,Mao2017,Liu2017}. The joint computation
offloading and resource allocation problems are often modeled
as combinatorial optimization problems with non-linear constraints that are difficult to optimize efficiently using traditional
optimization methods.
Therefore, in \cite{Liu2016,Xu2017}, the authors leveraged the RL techniques to extract valuable knowledge from the environment and then to make adaptive decisions, and hence they offered distributed
computation offloading policy and optimal resource allocation for the end users without the need of a priori knowledge of network statistics.

The above studies focus on centralized intelligent
approaches for computation offloading, which model the sophisticated global optimization
problem as a single-agent RL problem that
requires a central agent to collect the global state information of
the environment to make the global decisions for the entire system. This becomes challenging when the number of users increases. Moreover, in edge computing enabled wireless networks, the computation offloading problem involves the interaction among multiple decentralized users, wherein each user is considered as an intelligent agent and can make its decisions individually based on its local observation of the environment. Since the single-agent RL only learns a decision-making rule for one user without considering the influences of the existences of other users on its behaviors, the multi-agent RL is investigated to solve the decision-making problems with more than one agent coexisting in a shared environment \cite{weiss1999multiagent,Busoniu2008}.
Next, we present an overview of using RL techniques to optimize computation offloading scheme and resource allocation solutions from single-agent RL to multi-agent RL algorithm for computation offloading, and also the approach of introducing FL into RL technique as FRL is discussed to address the multi-user computation offloading problem. Moreover, other machine learning techniques, such as DL technique and imitation learning, have also been investigated to learn computation offloading strategies \cite{Yu2017,Yu2020Intelligent}. The comparisons of different machine learning techniques and traditional mathematical algorithms for computation offloading are summarized in Table \ref{offloading_ML}.

\begin{table}[t]
\centering
\caption{Comparisons of ML techniques and traditional method based offloading algorithms}
\label{offloading_ML}
\resizebox{\columnwidth}{!}{
\begin{tabular}{|m{2cm}|m{6cm}|m{4cm}|m{5cm}|m{1cm}|}
\hline
\textbf{Offloading algorithms } & \textbf{Characteristics}                                                       & \textbf{Dataset~ }                                & \textbf{Complexity }                                         & \textbf{Ref. }  \\
\hline
Traditional                     & Game theory, Lyapunov optimization                                             & Not need dataset                                  & High algorithm complexity, especially multiuser multi-server & \cite{Sardellitti2015,You2017}                \\
\hline
DRL                             & Consider computation offloading as MDP                                         & Collect dataset while training                    & High time complexity                                         & \cite{Li2018,Chen2019}                \\
\hline
Multi-agent RL                  & Consider multi-user computation offloading as Markov Game,
  parallel training & Each user collects its own dataset while training & Depends on the complexity of each DRL agent                  & \cite{Liu2020,Nasir2019}                 \\
\hline
DL                              & Formulate computation offloading as multi-label classification
  problem       & Generate dataset before training                  & Depends on offline training complexity                       & \cite{Yu2017,Ali2019}                \\
\hline
Imitation Learning              & Mimic expert behaviours                                                        & Generate expert dataset                           & Depends on offline training complexity                       &\cite{Yu2020Intelligent,Wang2021multiagent}               \\
\hline
\end{tabular}}
\end{table}

\subsection{DRL for Computation Offloading}
To address the problem of optimal computation offloading and efficient resource allocation, the conventional method formulates this joint problem as either a convex optimization or mixed-integer problem, but this finite-time optimization has the drawback that the computation offloading parameters are considered to be irrelevant under different system states. In this case, the long-term performance over dynamic system states changing is not maintained \cite{You2017,Chen2021}.
MDP is an effective mathematical tool to model the impact of user's actions in a dynamic environment, and it allows seeking the optimal action for achieving a particular long-term goal. To this end, the optimization of computation offloading policy under dynamic environment can be modeled by MDP.
When modeling the computation offloading problem as an MDP problem, a state transition probability matrix that describes the system dynamics needs to be constructed to obtain the optimal offloading policy. However, the system dynamics are hard to measure or model in most real-world scenarios, and thus obtaining the state transition probability matrix is intractable, especially when the state and action spaces are large \cite{Zhan2020}.

RL techniques have been used as promising solutions to tackle this challenge based on trial-and-error rule, where the RL agent, i.e., the user, can adjust its policy to achieve a best long-term goal according to the future reward feedback from the environment without prior knowledge of system models. In \cite{Alam2016,Xu2017,Ranadheera2017}, the authors investigated the dynamic computation offloading process and developed RL algorithms to learn the optimal offloading mechanism with the goal of minimizing latency and choosing the energy efficient edge server. Besides, DRL algorithm has been proved to be more effective for enabling RL to handle large state spaces by leveraging the powerful DNNs to approximate state-action values, which is envisioned to solve complex sequential decision-making problems. Therefore, DRL is particularly suitable for solving the computation offloading problems in dynamic environment. First, DRL can target the optimization of long-term offloading performance, which outperforms the one-shot and greedy application of the approaches studied in static environments. Second, the optimal offloading policies can be obtained without any prior information of the system dynamics (e.g., wireless channel or task arrival characteristics) by using the DRL techniques. Third, thanks to the powerful representation capability of the DNN, the optimal offloading policy can be adequately approximated even in complicated problems with vast state and/or action spaces \cite{Zhan2020}.

Recently, many researchers have investigated the DRL techniques to learn the optimal offloading mechanisms and at the same time optimize resource allocation \cite{Li2018,Chen2019}. In \cite{Chen2019}, the authors first proposed a DQN-based algorithm to learn the optimal computation offloading policy, in which the high dimensional state spaces were handled. In \cite{Ning2019}, a DRL algorithm was implemented to learn optimal decisions on resource allocation for vehicular edge computing networks, where the DQN is improved by applying dropout regularization and double DQN. Besides, the joint computation offloading and resource allocation problem has also been formulated and solved with DRL algorithms in recent studies \cite{Li2018,Yan2020,Huang2020a}. The authors in \cite{Li2018} jointly optimized the offloading decision and computational resource allocation to minimize the sum cost of the MEC system. In \cite{Yan2020}, a DRL framework was proposed to jointly optimize the offloading decisions and resource allocation with the goal of minimizing the weighted sum of users' energy consumption and task execution latency. In \cite{Huang2020a}, a DRL-based online offloading framework was proposed for wireless-powered MEC network to obtain optimal task offloading decisions and wireless resource allocations under the time-varying wireless channel conditions.
The authors in \cite{Chen2019b} investigated the optimal task offloading policy, computation and communication resource allocation, by the proposed intelligent resource allocation framework based on a multitask DRL algorithm. Moreover, in \cite{Ho2020}, the authors proposed the DRL-based algorithm for joint edge server selection, optimization of offloading decision and handover in a multi-access edge wireless network. Specifically, in \cite{Bi2021}, the authors combined the advantages of Lyapunov optimization and DRL algorithms, and proposed a novel online stable offloading framework that achieves making joint action of binary computation offloading and resource allocation in each short time frame without the assumption of knowing the future realizations of random channel conditions and data arrivals. In \cite{Wang2021b}, the Meta-RL (MRL) algorithm was proposed to address computation offloading problems, so new users can learn their offloading policies fast based on their local data and meta policies. Additionally, the MRL training in the MEC system can leverage resources from both the MEC server and the users.

However, the above research works rely on centralized decision making at the server, which limits the scalability of most RL-based algorithms due to the huge decision space and the overwhelming information collection from the MEC system. Moreover, implementing DRL into computation offloading optimization problems needs numerous interactions with training environments to obtain experiences with large quantities and high diversity, which causes huge costs due to the trial-and-error process (also known as exploration costs). Thus, the huge training cost lies in training a high-performance DRL agent for the MEC system, which is often unaffordable for a single MEC environment. To address this challenge, the authors in \cite{Qiu2021} proposed a distributed and collective DRL-based algorithm to adaptively learn the offloading and channel allocation decisions. Based on exploring the domain of distributed DRL training \cite{Ong2015}, the proposed algorithm assimilates experiences and knowledge from multiple MEC environments to obtain a collective DRL agents with high performance by adopting the experience-sharing scheme between the master agent and distributed agents, and thus the cost of the trial-and-error process is spread over the distributed system.

The aforementioned centralized offloading algorithm is restricted by the increasing scale of the network and is inability to observe the local environments, and it also causes huge cost for the edge server.
The distributed offloading is explored from the following two cases: a) a distributed DRL algorithm that enables each user to make its offloading decision without knowing the task models and offloading decisions of other users, which still relies on the broadcast information from the edge server in each time slot \cite{Tang2020}, b) a distributed DRL training is proposed to train a collective DRL agents by assimilating experiences from multiple MEC environment\cite{Qiu2021}, which not only exploits the distributed computational resources of multiple MEC servers, but also obtains more diverse training data. However, both of the distributed offloading approaches
does not discuss the scenario where multiple users make offloading decisions together while sharing the computational and communication resources. This is more practical in real world since when one user is making its own offloading decisions, other users are also making their decisions and thus affect the decision making of the user. Therefore, the distributed offloading considering the competitive behaviours among the users needs to be studied.

\subsection{Multi-agent RL for Computation Offloading}

When there are multiple users making decisions simultaneously in an edge computing system, each user's decisions are affected by the other users' decisions, so the degree of cooperation plays an vital role in the design of computation offloading policies. In this multi-agent system, each user is regarded as an agent that can only observe its local environment information. At each time point of observation, the edge computing system is in a system state, each agent takes an action according to the computation offloading policy of all the users in a vector form, and then the system responds to their actions by moving to a new system state according to the probability distribution and sending the rewards to each agent.
This multi-agent edge computing system models more practical computation offloading scenarios where more than one agent make decisions together to achieve goals, which may be cooperative or in conflict with each other. Compared to the single-agent computation offloading problem that falls under a category of single-agent RL and can be solved by the popular RL algorithms, like Q-learning and DQN, this multi-agent computation offloading problem falls under the category of multi-agent RL \cite{Xu2021b}.

Due to the simultaneous learning of multiple agents, it is challenging to  solve the formulated multi-agent RL computation offloading framework. Recently, researchers have put much more attention to investigate this problem by proposing the algorithms to solve it in a centralized or decentralized manner.
In the centralized approach, a central trainer collects the reward information from the individual agents, and dispatches the actions to them.
In \cite{Amiri2018}, the BSs are considered as the agents to execute RL independently for obtaining the Q-values, and then the Q-values are shared with new BSs as cooperative learning. With a similar information sharing mechanism, a distributively executed dynamic power allocation scheme was developed by using deep Q-learning, which is suitable for large-scale networks\cite{Nasir2019}.
This is based on a distributed framework with a centralized training assumption, in which the BS trains a single DQN using the transitions collected from all agents, while each agent has the same copy of the DQN parameters for decentralized execution.
However, this approach becomes impractical as the number of agents increases. Therefore, a decentralized framework where each agent independently learns its own strategy to maximize individual return was proposed, which is able to deal with large-scale networks. In \cite{Liu2020}, an independent learner based multi-agent Q-learning was proposed by considering the other users as part of the environment, in which each user is modeled as a RL agent observing its local environment information to independently learn a task offloading strategy that minimizes its energy consumption and task execution latency. While the Independent Q-learning (IQL) algorithm \cite{Tan1993} avoids the scalability problems of centralized algorithm and works well in practice as shown by empirical evidence \cite{Matignon2012}, it faces the challenges of non-stationary environment from the point of view of each agent as it contains other agents who are also learning themselves.

Therefore, the distributed multi-agent RL schemes with collaborations among users
were investigated to address this challenge. A distributed multi-agent DRL
scheme with collaborative exploration of the environment was proposed to solve the joint problem of computation offloading and resource allocation \cite{Huang2021}. The agents independently learn their individual strategies based on their local observations, and refine their learned strategies through a learning process driven by the specially designed reward function.
In \cite{Sacco2021}, the proposed decentralized multi-agent RL algorithm solves the computation offloading problem with the agents sharing their estimate of the value function with each other at the critic step. Unfortunately, the information sharing among users causes high communication overhead and is even infeasible due to the large-scale deployment of a beyond 5G network. To address those challenges, a distributed ML-agent RL framework without information sharing among users in the MEC system was studied in \cite{Chen2021,Zhang2021a}. In \cite{Chen2021}, each user independently learns its computation offloading policy by forming and updating conjectures on the behaviours of other users using the historical information retrieved from the BS. In \cite{Zhang2021a}, each Cloud Center (CC) is considered as an agent, where each CC determines the task offloading strategy independently by learning explicit models of other CCs as stationary distributions over their actions. Additionally,  RNN architecture was studied to improve the offloading strategy when the multi-agent RL algorithm is applied \cite{Gao2020,Chen2020a}. A Long Short-Term Memory (LSTM) network was introduced in the multi-agent DDPG network to accurately estimate the current state information of the MEC system in \cite{Gao2020}. The LSTM technique combined with the DQN could overcome the partial observability and the curse of high dimensionality in local network state space faced by each vehicle user pair for packet scheduling and resource management in vehicular networks \cite{Chen2020a}. In \cite{Zhan2020a}, the authors formulated the dynamic decentralized computation offloading game as a multi-agent Partially Observable Markov Decision Process (POMDP), and then they designed an algorithm that can achieve the optimal offloading strategy by combining policy gradient DRL-based approach with DNC. DNC is a special recurrent neural network and is capable of learning and remembering the past hidden states of inputs.  Moreover, the authors in \cite{Xu2021b} considered the practical challenges of deploying the previously mentioned deep multi-agent RL algorithms and studied applying them to solve task offloading with reward uncertainty.


\subsection{FRL for Computation Offloading}
FRL was first studied in \cite{Zhuo2019}, which built an MultiLayer Perceptron (MLP) as the shared value network to compute a global Q-network output with its own Q-network output and the encrypted output of Q-networks from other agents. RL has the problem of learning efficiency caused by low sample efficiency. Even though distributed RL can address this problem by sharing information (i.e., raw data, parameters or gradients) to the central server for model training, but there is a possibility of agent information leakage. In a multi-agent system, each user can only observe partial environment information which is not enough for the agent to make decisions. Furthermore, many RL algorithms requires pre-training in simulated environments as a prerequisite for application deployment, but the simulated environments cannot accurately reflect the environments of the real world. To this end, FL with the ability of aggregating information can integrate the information from different users and can bridge simulation-reality gap, and also it can provide privacy protection. Then, the idea of combining FL with RL, known as FRL, is generated to address the challenges that exist in RL. In FRL, the three dimensions of sample, feature and label in FL can be replaced by environment, state and action, respectively \cite{qi2021federated}.

As discussed in the previous sections, RL has been widely exploited to solve computation offloading problem, but the optimal solutions are obtained only with many assumptions of the environment. In the complex computation offloading environment, each user only knows its own information of waiting tasks and resources, and can receive notifications from the BS. Also, each user's decisions on offloading and resource allocation are affected by the others in the same edge system making decisions at the same time. The collected RL training data from one edge system may be not enough to reflect the complex offloading environment, and thus more diverse data from multiple edge systems are required to be integrated for obtaining complete environment information. Moreover, it is still challenging to implement the trained RL by the proposed algorithms in the aforementioned research work for solving practical computation offloading problems due to the time-varying wireless channels, limited resources and randomly generated computational tasks. As a result, FL has been introduced in RL-based computation offloading algorithms to address the above challenges with its ability of information aggregation.

In \cite{Wang2019}, the authors first applied the FL framework to train the DRL agents for intelligent joint resource management of communication and computation in MEC systems. With FL, the DRL agents are efficiently trained in a distributed fashion, and they can handle unbalanced and non-IID data and cope with the privacy issues.  In \cite{Shen2020}, the FL was used to conduct the training process of DRL agents for optimizing decision making about computation offloading and energy allocation in IoT edge computing networks. Moreover, in \cite{Huang2021,Hou2021}, the FL framework was introduced to train the multi-agent RL algorithm.  A distributed multi-Agent DDPG (MADDPG)-based joint hierarchical offloading and resource allocation algorithm was proposed to exploit the FL to train multi-agent deep RL model in Cybertwin networks, which solves the sensitive information leakage issue and relieves the computational pressure at the edge.
In \cite{Huang2021}, each Small Base Station (SBS) adopts independent learning algorithm while treating the agents as part of the environment in the formulated multi-agent DRL framework, and then the SBSs exchange their model parameters with each other. Finally, each SBS agent performs model aggregation of FL based on the collected model parameters from the other agents. In \cite{Prathiba2021}, an effective radio resource management based on federated Q-learning was proposed to optimize resource allocation for computation offloading in 6G-Vehicle-to-everything (V2X) communications, where the locally trained Q-tables are shared to the vehicle edge center pool for global aggregation.

\subsection{Other Learning Techniques for Computation Offloading}

\subsubsection{DL for Computation Offloading}
Different from most works discussed in the previous sections, DL has also been exploited to design dynamic offloading strategy. In \cite{Yu2017}, the computation offloading problem is formulated as a multi-label classification problem. To obtain the optimal offloading policy, an exhaustive strategy is used to search the optimal solution in an offline way, and then the obtained solution can be used to train a DNN with the composite state of the edge computing network as the input, and the offloading decision as the output. Likewise, the authors in \cite{Ali2019} proposed DL algorithm to avoid exhaustive decision-making process by training a DNN over the dataset generated by their mathematical model. By this means, once the DNN is trained, it can be used as a decision-maker for offloading specific components. To solve a heavy burden caused by massive training dataset in multi-user task offloading problem, a distributed DL-based computation offloading algorithm was proposed by training multiple parallel DNNs with the  channel gain as the input and the output is the offloading decision \cite{Wang2022}.

\subsubsection{Imitation Learning for Computation Offloading}
Another promising ML technique, imitation learning, has also been investigated to design offloading schedule \cite{Yu2020Intelligent,Wang2021multiagent}. In \cite{Yu2020Intelligent}, a novel deep imitation learning based offloading scheme has been proposed, the ML model is first trained from learning demonstrations in an offline manner based on behavioral cloning. Then, the near-optimal online offloading decisions can be made at a very fast speed with quick and easy deployment.
In \cite{Wang2021multiagent}, a multi-agent imitation learning based computation offloading algorithms was proposed, which allows multiple learning agents to imitate the behaviors of corresponding experts for good scheduling policies. They designed the expert policies by enabling the experts to obtain full observation of system states and then form the demonstrations including state-action pairs for the learning agent to mimic.

\subsection{Discussion and Outlook}

The aforementioned literature has investigated different ML techniques to design optimal computation offloading strategies with a focus on RL algorithms. Although multi-agent RL and FRL frameworks have been proposed to address multi-user computation offloading problem, to obtain the optimal offloading strategies needs to analyze the collaborations or competitions among users. The solution process is always formulated as a Markov game by finding a Nash Equilibrium (NE) to get the optimal strategy, but which is challenging to find the NE or even the NE is not existed for some complicated problems. Furthermore, the above studies provide good prospects for the application of FRL to edge computing, but there are still many
challenges to overcome, including the adaptive improvement of the algorithm, and the training time of the model from scratch.

Imitation learning is another potential technique to design offloading strategy, which can train an ML model from expert demonstrations in an offline manner and learn a near-optimal offloading policy to make fast online offloading decisions. Compared to RL techniques, it enables easier deployment for practical use. However, it is difficult to obtain the expert dataset and the learning adaptability is bad.

\section{Distributed Learning in Wireless Networks}

To support a wide range of emerging AI-services in wireless networks,
the conventional approach is to collect all the raw data from the users and then train the ML models in a centralized fashion as shown in Fig. \ref{DI_intelligence}.
However, the centralized learning approach is restricted by the limited bandwidth, energy consumption, privacy and security concerns. Therefore,  distributed learning, including FL and SL algorithms, have been proposed to allow the parameter server and wireless users to collaboratively train the ML models by only exchanging model parameters instead of raw data as shown in Fig. \ref{DI_intelligence}. Benefiting from the distributed learning architecture, a broad range of advanced AI applications can be deployed at the network edge.  In this section,

\subsection{Hybrid Distributed Learning Architectures for Heterogeneous Wireless Networks}

As a representative of distributed learning, FL has been studied a lot recently in wireless networks to improve communication efficiency for training ML models in a distributed manner, which is deployed at the network edge to exploit computational capabilities of the end users.
The FL architecture requires that all the users are capable of gradient computation, which is hard to be satisfied considering the diversity of the users in terms of computational capacities. Besides, the local dataset, energy and communication resources are also diverse over the users. It is demonstrated that SL is more communication-efficient than FL in the scenarios of large model size and small local dataset \cite{Singh2019}, and the communication overhead of which depends on the user's local dataset size.
To this end, a new distributed learning architecture relying on hybrid learning technique could be a better solution, which uses the idea to benefit from different learning algorithms according to the users' unique characteristics. For instance, an HFCL framework was proposed to let the users incapable of sufficient computational power deploy CL while the rest use FL \cite{Elbir2020}.
In \cite{Thapa2020}, a Split Federated Learning (SFL) was proposed to combine the parallel model training mechanism of FL and network splitting structure of SL, which is beneficial for resource-constrained environment where full model training and deployment are
not feasible at the local users.
In our recent work, we further propose an HSFL architecture in which the users with small data size and weak computational capability are allowed to choose SL while the others use FL \cite{Liu2021}. In the following, two hybrid distributed learning architectures are introduced.

\begin{figure}[t]
  \centering
  \includegraphics[width=3.5in]{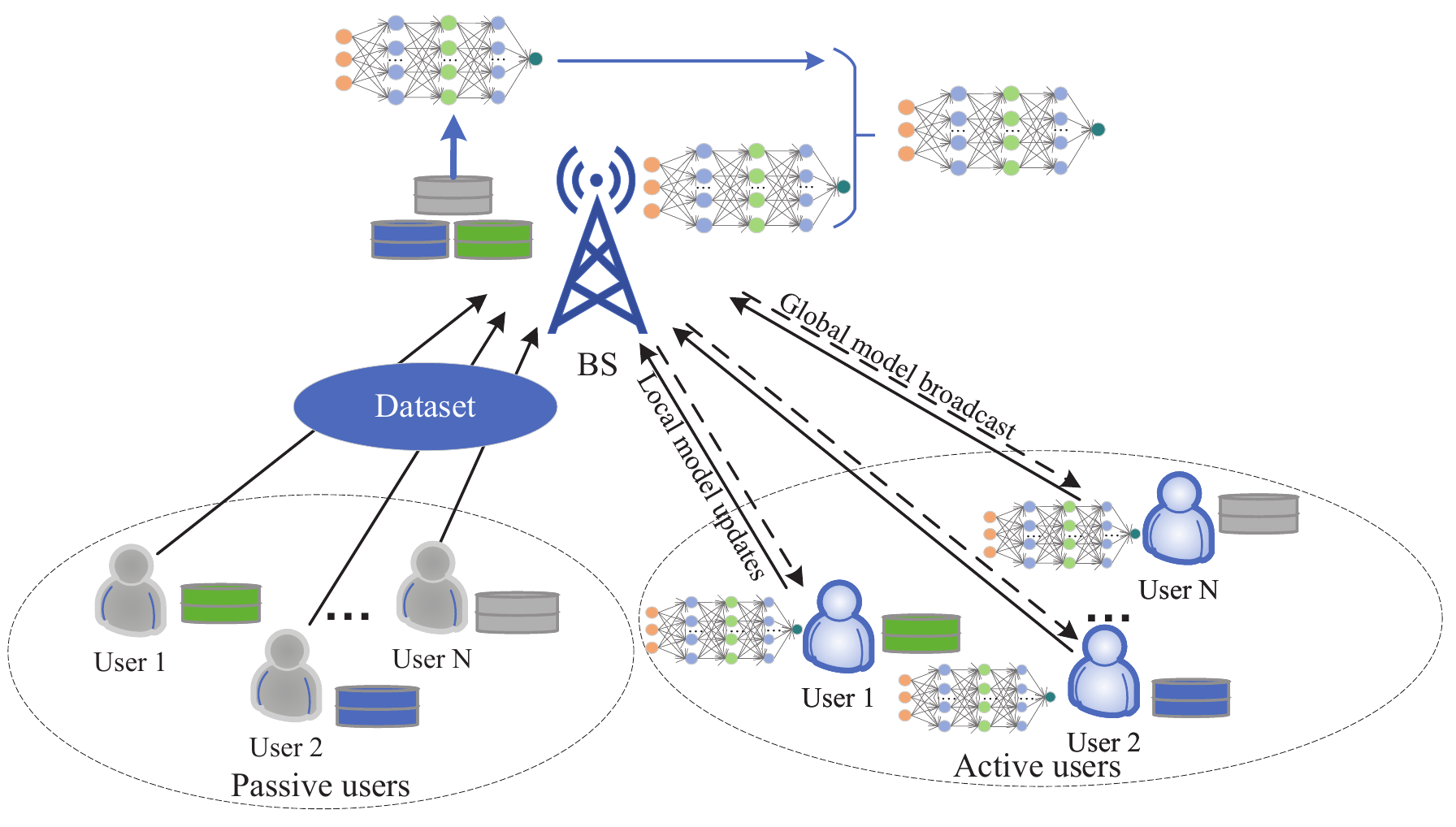}\\
  \caption{The learning mechanism of HFCL framework. }\label{HFCL}
\end{figure}

\begin{figure}[t]
  \centering
  \includegraphics[width=3.5in]{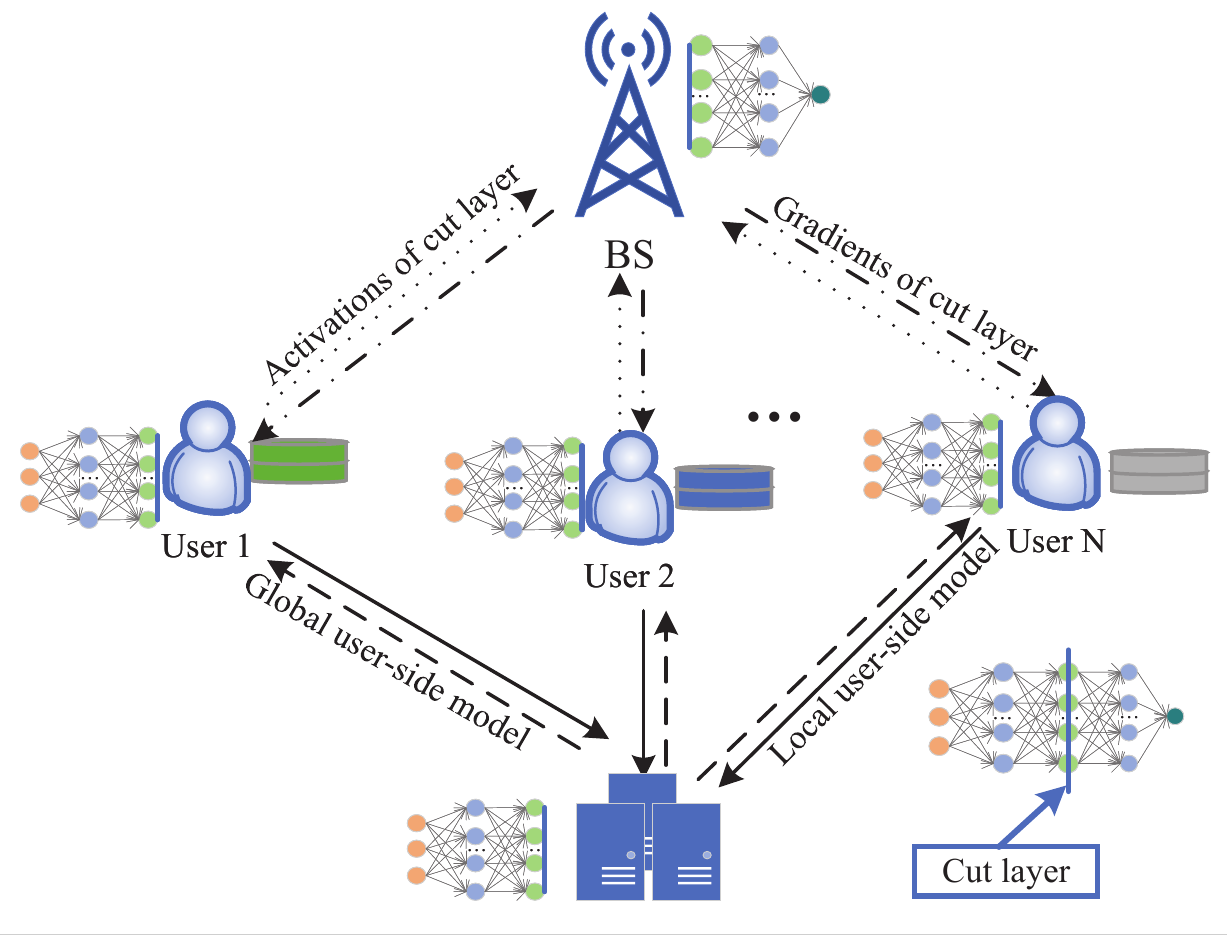}\\
  \caption{The learning mechanism of the SFL algorithm. }\label{splitfed}
\end{figure}

\subsubsection{HFCL Algorithm}
FL algorithms bring the learning tasks to the edge level, wherein the users are required to be computationally powerful since they have to train the full ML model. However, this requirement may not always be satisfied due to diverse computational capabilities of the users.
In \cite{Elbir2020,Elbir2021}, the authors proposed an HFCL framework to train ML models efficiently exploiting the computational capabilities of the users, which is achieved by only letting the users that have enough computational resource employ FL while the other users resort to CL by transmitting their local raw dataset to the BS. The learning mechanism of the HFCL framework is illustrated in Fig. \ref{HFCL}

In HFCL, the users are grouped into active and passive user sets depending on their computational capabilities to either perform CL or FL, respectively.
In this case, the passive users transmit their local datasets to the parameter server which then uses them to train the ML model. On the other hand, the active users upload the gradient information calculated locally based on their private datasets. Next, the parameter server performs model aggregation with the computed gradients from users and the parameter server itself and then sends the updated model parameters back to the active users. The HFCL faces the challenge that the active users need to wait for the passive users completing their data transmission at the beginning of the model training, followed by the model aggregation at the parameter server before they can update their local model parameters.
To address this problem, the authors proposed a sequential dataset
transmission approach where the local datasets of the passive users  are divided into smaller blocks, so that both active and passive users can perform gradients and data transmission during the same communication round.

\subsubsection{HSFL Algorithm}

In \cite{Singh2019}, the authors demonstrated that the FL is more communication-efficient and computation-efficient when the users have large local datasets and the model size is small, otherwise SL is more efficient \cite{Singh2019}.
Moreover, the user-side computational cost in SL is significantly reduced compared to FL because of the network splitting structure. The disadvantages of FL are that each user needs to train a full ML model but some resource-constrained users cannot afford that, and that both the server and users have full access to the local and global models which causes privacy concerns from the model's privacy perspective. On the other side, the disadvantages of SL are that only one user engages with the server at one time while the others stay idle, causing a significant increase in training period.
To address these issues in FL and SL,
SFL was proposed to exploit the advantages of FL and SL \cite{Thapa2020}. The architecture of SFL is presented in Fig. \ref{splitfed}.

In Fig. \ref{splitfed}, the full ML model is divided into two parts by the cut layer, and one is user-side model residing at the users and the other one is the BS-side model residing at the BS.
All the users carry out forward propagation through the user-side model with their local datasets in parallel, and pass the activations of the cut layer to the BS. The BS then conducts forward propagation and back propagation on the BS-side model with the received activations from each user separately in parallel. Then, it computes the gradients of the cut layer and sends them back to the respective users for calculating the gradients of user-side models. Afterwards, the BS updates its BS-side model using
\emph{FedAvg}, and the user-side model updates are sent to a fed server for model aggregation using \emph{FedAvg}.
With the parallel training architecture, the SFL shortens the training time in SL and achieves a similar performance to SL in terms of test accuracy and communication
efficiency. By using the network splitting structure, it has better communication efficiency than FL when the users have small local datasets, and it has better model privacy than FL because the users/BS cannot access to the BS-side/user-side model except for some smashed data of the cut layer.

\begin{figure}[t]
  \centering
  \includegraphics[width=3.8in]{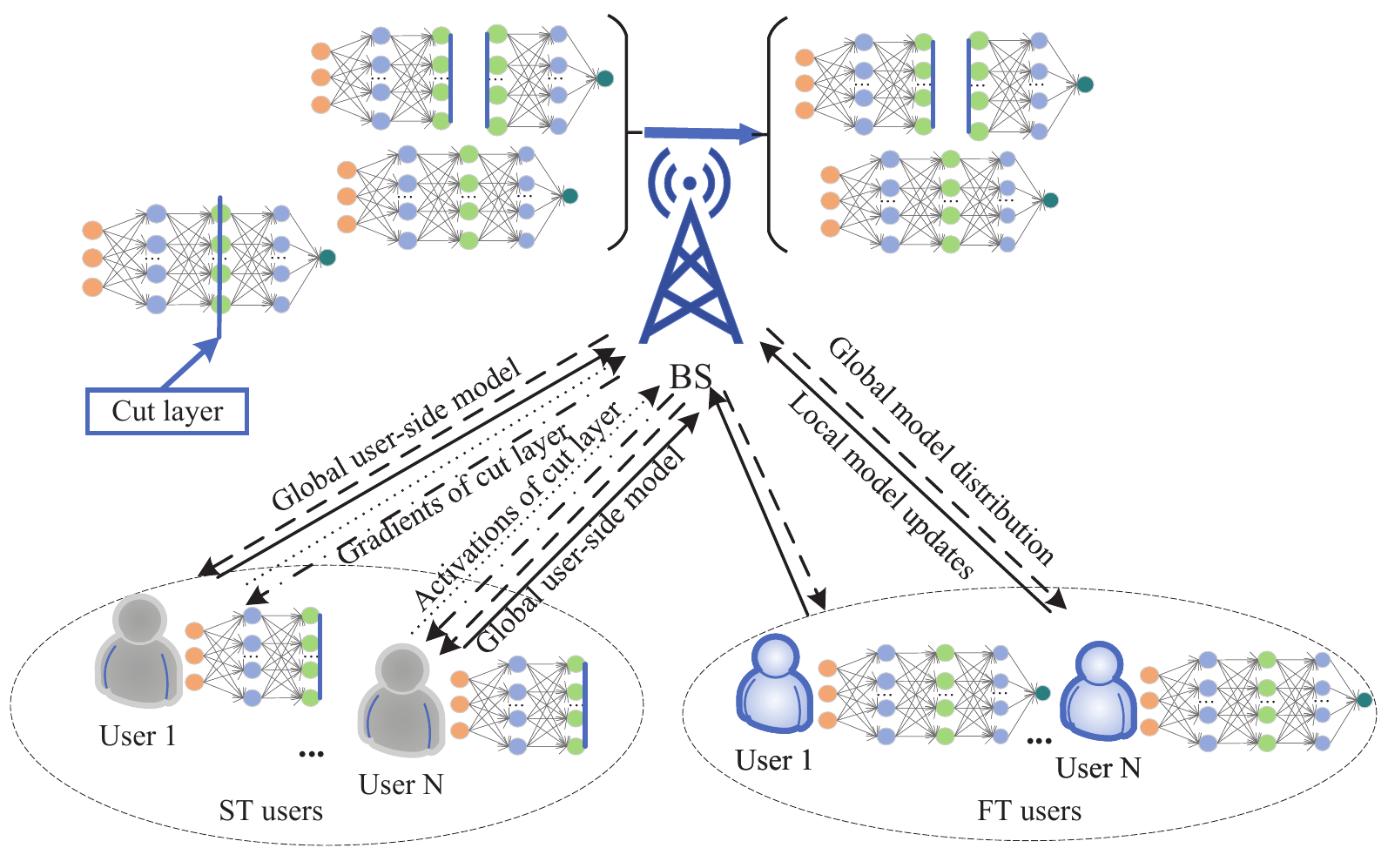}\\
  \caption{The learning mechanism of the HSFL framework}\label{HSFL}
\end{figure}

However, the SFL algorithm still experiences high communication overhead as in SL when the users have large local datasets since the training dataset could be highly imbalanced and distributed over the users. Fortunately, this
issue can be compensated by letting some users use FL which is more communication efficient than SL in this scenario. Based on the works in \cite{Singh2019,Thapa2020}, we propose
an HSFL framework that also aims to seek the advantages of FL and SL, and it can eliminate the drawbacks of SFL \cite{Liu2021}. The HSFL adopts the same  parallel model training mechanism as FL and the same network splitting structure as SL, but it has a different architecture from SFL.
The illustration of the architecture for HSFL is shown in Fig. \ref{HSFL}.
In HSFL, the users are allowed to choose either FL or SL method according to their own unique characteristics, such as the users with small datasets and powerful computational capabilities would prefer FL in which the users run a full ML model locally, and the users with large dataset and weak computational capabilities would resort to SL wherein the users only run a part of the full ML model locally while the server runs the remaining part of the ML model.

During the training process, the BS initializes the architectures and weights of the global full ML model, and also divides a copy of it into two sub-models as the global user-side model and the global BS-side model.
The users choosing SL method receives
the global user-side model, while the users choosing FL receives the global full model, and then they respectively compute their local gradients with their local datasets in parallel.
Specifically, the users choosing SL
follows the same rule as SL to train the full ML model by engaging with the BS. Afterwards, the users compute their local gradients and send them to the BS which then performs model aggregation with the received local model updates and the updates of its own BS-side model. Later, it sends the updated global full model back to the users choosing FL and the updated global user-side model to the users choosing SL, respectively. The challenge of HSFL is how to decide which learning method, i.e., SL or FL, for each user. We further design a metric to measure the characteristics of each user, called diversity index, which is defined as the weighted sum of four parameters, including the computational capability, dataset size, dataset diversity and user diversity.
Considering the scenario of deploying  HSFL in wireless networks, we formulate the learning method selection and user selection problem as a Multiple-Choice
Knapsack Problem (MCKP) and propose an energy-efficient user scheduling algorithm [19] to select a subset of users in each communication round and schedule each user with either SL or FL method.

\begin{figure}[t]
  \centering
  \includegraphics[width=3.5in]{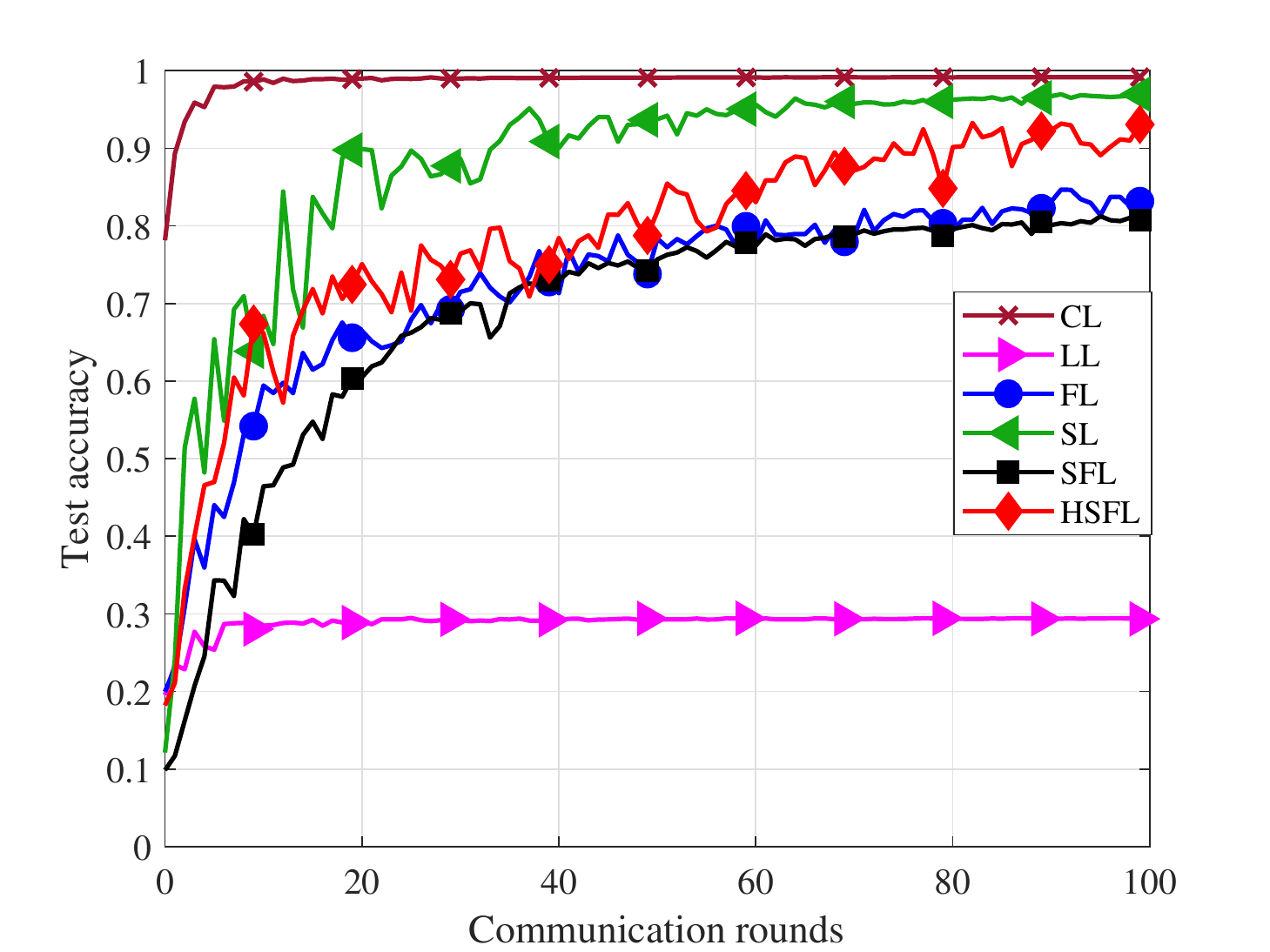}\\
  \caption{The learning accuracy performance of different distributed learning architectures.}\label{accuracy}
\end{figure}

As discussed above, the state-of-art distributed learning architectures have their unique characteristics and can be efficiently used in specific application scenarios. In Table \ref{distributed_table}, we summarize the comparisons of different distributed learning architectures. Moreover, the convergence performance of implementing different distributed learning architectures in wireless networks with the best channel user scheduling scheme under non-IID data is shown in Fig. \ref{accuracy}.

\begin{table}[t]
\caption{Comparisons of different distributed learning architectures}
\label{distributed_table}
\centering
\resizebox{\columnwidth}{!}{
\begin{tabular}{|m{3cm}|m{5cm}|m{3cm}|m{3cm}|m{3cm}|m{1cm}|}
\hline
\textbf{Distributed learning architecture } & \textbf{Characteristics }                                                                               & \textbf{Training mechanism }                                                       & \textbf{Communication overhead }                              & \textbf{Computing ability requirements }              & \textbf{Ref. }  \\
\hline
FL                                          & A full model at local for all the users                                                                 & Parallel                                                                           & Model parameters~                                             & High at local                                         & \cite{mcmahan2017}               \\
\hline
SL                                          & A sub-model at local, another at the BS for all the users                                               & Sequential                                                                         & Smashed data (activations and gradients of the cut layer)     & Low at local                                          & \cite{Gupta2018}               \\
\hline
SFL                                         & A sub-model at local, another at the BS for all the users                                               & Parallel                                                                           & Smashed data                                                  & Low at local                                          & \cite{Thapa2020}               \\
\hline
HFCL                                        & Central users: a full model at BS; Federated users: a full model at local                               & Central users: Training together at BS; Federated users: Parallel                  & Central users: Raw dataset; Federated users: Model parameters & Very low for central users; High for federated users~ & \cite{Elbir2020}               \\
\hline
HSFL                                        & Split users: a sub-model at local while another sub-model at BS; Federated users: a full model at local & Split users: Parallel by collaborating with BS; Federated users: Parallel at local & Split users: Smashed data;~Federated users: Model parameters~ & Low for split users; High for federated users         & \cite{Liu2021,Liu2022_energy}               \\
\hline
\end{tabular}}
\end{table}

\subsection{Asynchronous Distributed Learning for Heterogeneous Wireless Networks}
Due to the heterogeneity of the users with different computational capacities and energy resources, the local computation at each user does not complete at the same time. Moreover, the wireless network is always heterogeneous in terms of each user accessing diverse specturm resources and  suffering dynamic wireless channel for the transmission of model parameters. In most studies over distributed learning,  the model aggregation at the parameter server is assumed to be synchronous. However, in this case, the server has to wait for the stragglers, i.e., the slow users, before performing model aggregation. Thus, research on asynchronous optimization methods has gained significant attention to solve the straggler problems in distributed learning.

\subsubsection{The Issues of Classical FL in Wireless Networks}

Recently, FL has received significant achievements to train a global model on datasets partitioned across a number of users, which exploits the large amount of training data from diverse users and provides privacy preservation for them. However, when applying the classical FL to resource-constrained users, a few issues are emerged as follows:

\begin{itemize}
\item  \textbf{Heterogeneity:} The heterogeneity of users in terms of different computation capacities, dataset and wireless channel conditions causes different completion time of local gradient computation, so the aggregation server has to wait for the slow users.

\item  \textbf{Unreliability:} The selected users may go offline unexpectedly due to their unreliability, which also causes the aggregation server to wait for the local gradients from the unreliable users.

\item \textbf{Low round efficiency:}
Due to the heterogeneity of user diversity (different computational abilities and channel conditions of users) and data diversity (training dataset size and distribution over users), the users who finish local gradient updates early have to wait for those straggler users in each training round.

\item \textbf{Low resource utilization:}
Due to limited spectrum resource and inefficient user scheduling algorithms, some competent user may be rarely selected.

\item  \textbf{Security and privacy concerns:} There are several attacks that can compromise the
security of classic FL, including poisoning and backdoor attacks. The privacy concern comes from the possible data leakage during the training process.

\end{itemize}

\begin{figure}[t]
  \centering
  \includegraphics[width=5in]{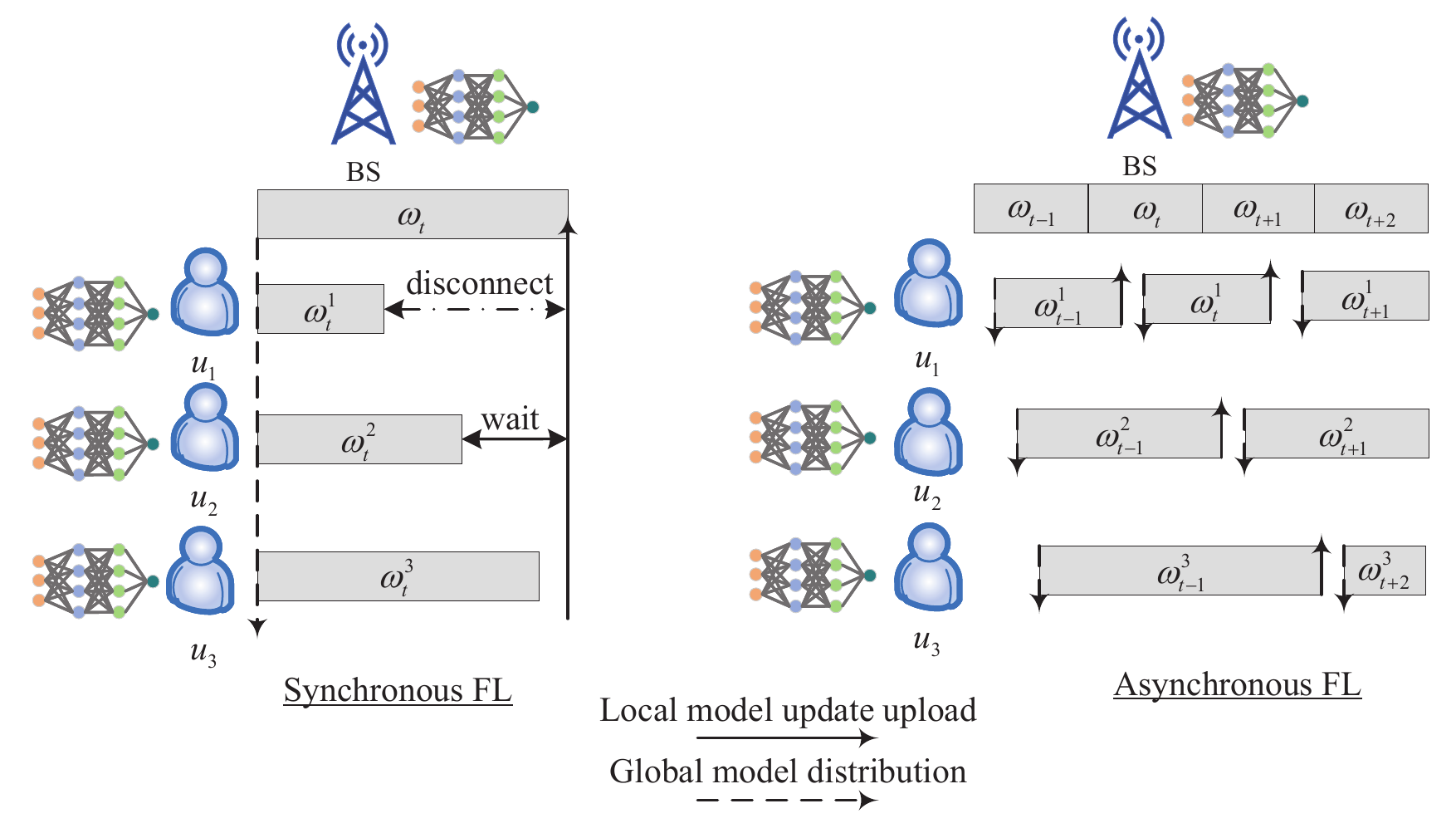}\\
  \caption{The training procedures of synchronous vs. asynchronous FL.}\label{synch_asynch}
\end{figure}

To overcome the above-identified challenges, asynchronous training has been widely studied in traditional distributed SGD, known as asynchronous stochastic gradient descent, for stragglers and heterogeneous latency \cite{dean2012large,li2014scaling,Zheng2016}. The authors in \cite{dean2012large} first developed an asynchronous stochastic gradient descent  procedure, Downpour SGD, to train large-scale ML models distributively.  Downpour SGD builds multiple replicas of a single DistBelief model, and divides the training data into a number of subsets and then runs a copy of the model on each of these subsets. It leverages the concept of a centralized sharded parameter server, through which the models can exchange their updates. This approach is asynchronous in the aspects that the model replicas run independently of each other and the parameter server shards also run independently of one another. Compared to synchronous SGD where one user failure will delay the entire training process, Downpour SGD is more robust to user failures since the other model replicas continue the training processing even if one user in a model replica fails. Asynchronous FL was first studied in \cite{Xie2019} by taking advantage of asynchronous training and combining it with federated optimization. In asynchronous FL, the BS server can perform model aggregation once it receives any local model updates from the users without the need of waiting for the lagging users shown in Fig. \ref{synch_asynch}. Due to the asynchrony of completing local model updates by the users, the local model updates uploaded in the same round may contain different fresh information and possess varying degrees of staleness because the local models are trained by using the global model versions received from different time periods. Moreover, the diversity of channel conditions causes the transmission of local model updates from different users asynchronous. Therefore, it is essential to design effective and efficient asynchronous FL algorithm in wireless networks that could deal with the staleness in the system appropriately with restricted communication resources.
The results in \cite{Zhang2021} demonstrated that the asynchronous FL could significantly improve the prediction performance of local edge models and it proves high efficiency when deploying ML or DL components to heterogeneous real-world embedded systems.

\subsubsection{Asynchronous FL in Wireless Networks}


The contradiction between the limited wireless resources and the explosive growth of the number of users is gradually intensifying nowadays, making it unrealistic to deploy a strictly synchronous FL system composed of a massive number of users with great heterogeneity over the wireless networks \cite{Wang2021}.
On one hand, a massive number of users trying to upload model parameters simultaneously will bring high communication overhead and cause congestion in the network. On the other hand, the BS can only perform model aggregation until all the local model updates from all the users are received, but some users with poor communication conditions and weak computational capabilities can greatly lag the training process,  leading to extremely low training efficiency.
Thus, an asynchronous FL could be much more scalable and applicable in wireless networks. In this case, the
local model updates trained from the same global model can be transmitted in different time slots, which can greatly reduce the instantaneous communication load. Additionally, the BS can perform global model update whenever it receives the local model updates without having to wait for the updates from all the users, which significantly improves the overall training efficiency.

\begin{table}[t]
\centering
\caption{The architectures of asynchronous FL }
\label{asyn_architecture}
\resizebox{\columnwidth}{!}{
\begin{tabular}{|m{4cm}|m{5cm}|m{5cm}|m{3cm}|m{1cm}|}
\hline
\textbf{The architecture of asynchronous FL } & \textbf{Working principles }                                               & \textbf{Challenges }                                  & \textbf{Advantages }                         & \textbf{Ref. }  \\
\hline
Classic Asynchronous FL                       & Aggregation once receives the local updates                                & High communication overhead, aggregation server crash & No need to wait                              & \cite{Xie2019}               \\
\hline
Semi-asynchronous FL                          & Aggregation after a specific period of time                                & Performance degradation by slow users                 & Alleviate the effects of the straggler users & \cite{Stripelis2021}               \\
\hline
Clustered asynchronous FL                     & Users are clustered into groups, Inner-group and inter-group
  aggregation & The clustering algorithm and metrics                  & Increase training efficiency                 & \cite{Xu2021Asyn}               \\
\hline
\end{tabular}}
\end{table}

\subsubsection{Hierarchical Architecture of Asynchronous FL}

In \cite{Xie2019}, a \emph{FedAsync} algorithm which combines a function of staleness with asynchronous update protocol was developed. However, the users have to transmit a large amount of data to the server, which may cause the server to crash. Moreover, the stale local updates from the stragglers can decrease the accuracy of the global model to a certain extent. To this end, researchers have developed two schemes to address these challenges, semi-asynchronous FL and cluster FL \cite{Xu2021Asyn}. In table \ref{asyn_architecture}, we summarize the characteristics of different asynchronous FL architectures.

The semi-asynchronous FL combines the classic FL and asynchronous  FL, in which the aggregation server
caches some local updates that arrive early and aggregates them after a specific period of time, which then can alleviate the effects of the straggler users. A data expansion method was used to alleviate the straggler phenomenon in \cite{Hao2020}, in which a semi-asynchronous communication method was proposed to speed up convergence for FL. In \cite{Stripelis2021}, a new energy-efficient semi-synchronous FL was proposed, which aggregates the local updates at a specific time interval determined by the slowest user.

Cluster FL is an effective approach to increase the training efficiency with reduce the transmission data from local users by grouping together users with similar performance, functionalities, or datasets \cite{Xu2021Asyn}.
To reduce the network congestion caused by a massive number of users simultaneous uploading local model updates in edge computing networks, the authors in \cite{Wang2021a} proposed a cluster-based FL mechanism. This mechanism divides users into different clusters, where users in one cluster transmit their local model updates to the cluster head for synchronous model aggregation while all cluster heads communicate with the edge server for global aggregation in an asynchronous way. A cluster-based asynchronous FL framework adopting appropriate time-weighted inter-cluster aggregation strategy was proposed in \cite{Sun2021}, which eliminated the straggler effect and improved the learning efficiency.

\subsubsection{User Scheduling and Resource Allocation for Asynchronous FL}

Recent studies have put considerable attention on device scheduling and resource allocation for asynchronous FL \cite{Lee2021 ,Hu2021,Yang2021}. In \cite{Lee2021}, the transmission scheduling scheme considering time-varying channels over multiple rounds, and stochastic data arrivals of the edge devices with asynchronous FL was first studied. The authors developed an asynchronous learning-aware transmission scheduling (ALS) algorithm for the scenario with the perfect statistical information about the system uncertainties, and further proposed a Bayesian ALS algorithm to learn the system uncertainties without requiring any priori information or with requiring only partially observable information. Furthermore, three device scheduling schemes, namely random, significance-based and frequency-based scheduling, were investigated for the heterogeneous wireless networks
by adopting the asynchronous FL framework with periodic
aggregation \cite{Hu2021}.
A RL-based device selection, UAVs placement and resource management algorithm was developed for deploying the asynchronous FL framework in multi-UAV-enabled networks \cite{Yang2021}, in which it also demonstrated that the proposed asynchronous online FL is particularly useful for streaming data with heterogeneous devices having different computing capabilities and communication conditions \cite{Chen2019}.

\subsubsection{Security and Privacy in Asynchronous FL Framework}

To ensure the security required by FL, blockchain network is introduced into FL framework to replace the classic central server to aggregate the global model, which avoids the real-world issues
such as interruption by abnormal local user training failure, dedicated attacks, and etc. Researchers studied blockchain enabled asynchronous FL framework by exploiting the decentralized property of blockchain network and the fast convergence performance of asynchronous FL strategy \cite{liu2021blockchain,feng2021blockchain}, this framework improved training efficiency and prevented poisoning attacks. In \cite{Cui2021}, the authors studied the blockchain enabled asynchronous FL framework
to mitigate the threats of poisoning attacks against IoT anomaly
detection models, and then devised a novel Generative Adversarial Network (GAN)-driven differentially private algorithm by injecting controllable noise into local model parameters.

\subsection{Incentive Mechanisms of Users for Participating FL Process}

\begin{figure}[t]
  \centering
  \includegraphics[width=3.5in]{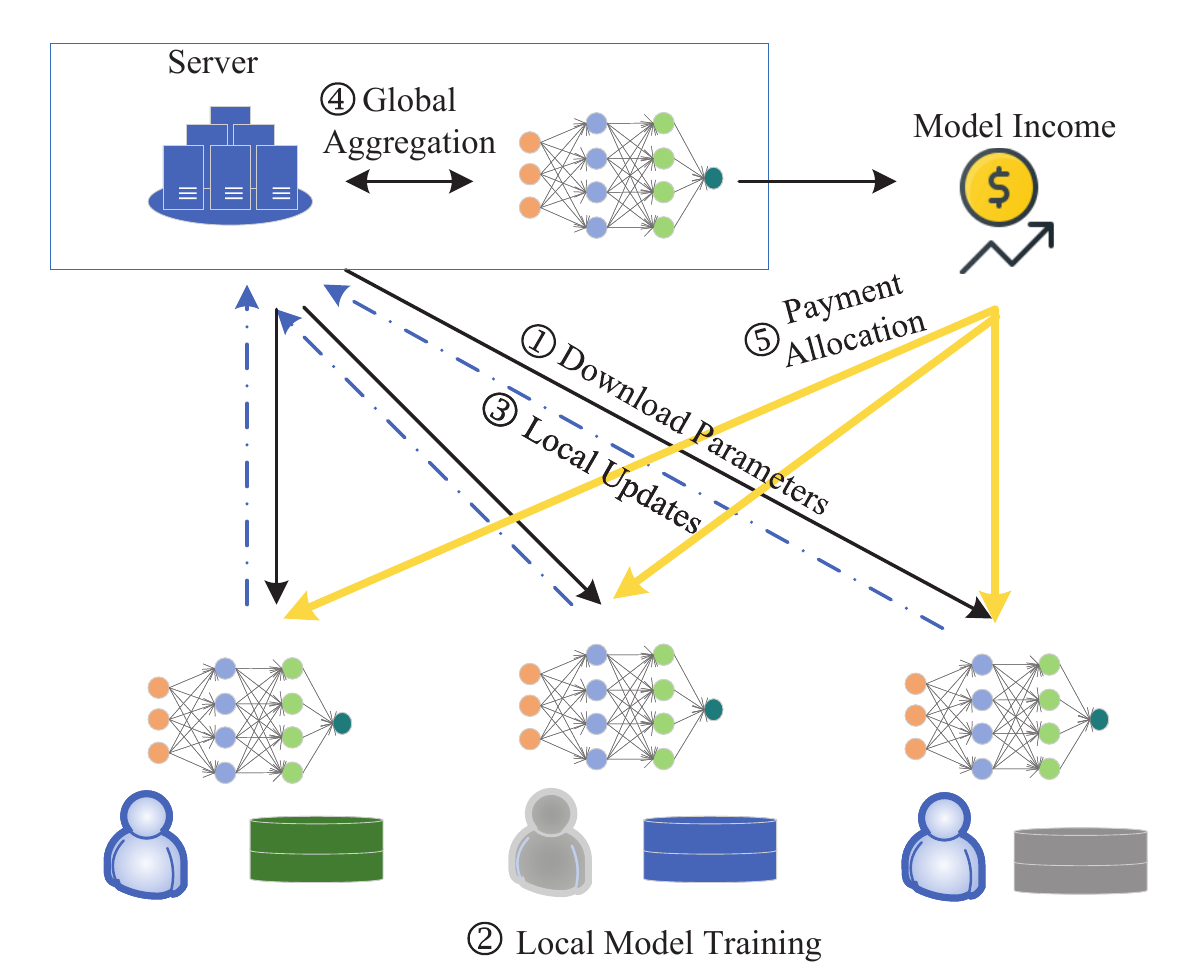}\\
  \caption{The architecture of incentive mechanism.}\label{incentive}
\end{figure}

Generally, to implement distributed learning architectures, all the users are assumed to voluntarily participate in global model aggregation without requiring any returns.
However, in practice, the participants may be reluctant to participate in this federation process without receiving compensation since training ML models
is resource-consuming \cite{Zeng2021}. In \cite{Liu2020a}, the incentive mechanism in FL was first studied by providing an incentive-compatible scoring system for building a payment system.  Fig. \ref{incentive} shows the architecture of incentive mechanism in FL, in which the users might be mobile devices, edge nodes, and IoT devices in cross-device FL or giant companies in cross-silo
FL. They provide various types of resources instead of only data, all of which are key factors to the training performance. After global model aggregation, the server will pay each user according to their individual contributions to the FL process.
In \cite{Zhan2021,Zeng2021}, the authors
did comprehensive surveys of incentive mechanisms for FL in recent research works. They identified the challenges of incentive mechanism design for FL,  and then summarized a taxonomy of existing incentive mechanisms for FL in terms of main techniques, such as Stackelberg game \cite{Pandey2019,Khan2019}, auction \cite{Zeng2020FMore,Jiao2019}, contract theory \cite{Kang2019,Lim2020}, Shapley value \cite{Wang2020incentive}, RL \cite{Zhan2020Incentive}, and blockchain \cite{Liu2020Fed}.
The Stackelberg game, auction and contract theory are mainly employed to perform user selection and payment allocation for  incentivizing users to participate in
FL process, while the Shapley value is used for fair assessment of FL user contribution.  Both RL and blockchain are introduced to improve the performance and robustness of the incentive schemes.

\subsubsection{The Reluctance of Users}
First, the FL process consumes resources including computational power, bandwidth, and private data, from participants, some of which might be constrained in scenarios like mobile networks and MEC systems.
Moreover, the privacy and security concerns are raised because the FL server can infer the important information of the training data \cite{Zhan2021}. In \cite{Yu2020}, the authors showed that many participants gain no benefit
from FL because the federated model is less accurate on their data than the models they can train
locally on their own, which removes their main incentive to join the FL process.
To this end, without proper incentives,
the users tend to opt out of the participation, may contribute either uninformative or outdated information, or even contribute malicious model information.

\subsubsection{Incentive Mechanism for FL in Heterogeneous Networks}

When deploying FL in wireless edge networks, the users, like mobile phones, IoT devices and drones, are always heterogeneous with different computational capacities, training data size,  power and communication resources, and this heterogeneity might degrade the performance of FL. Hence, the incentive mechanism design for FL in wireless edge networks should encourage more high-quality users to participate in the FL process so as to  eventually improve the convergence performance of FL \cite{Zeng2020FMore}. In \cite{Kang2019}, reputation is applied as the metric to select reliable users for participating in FL and is calculated in a decentralized manner through the consortium blockchain; the incentive mechanism using contract theory was proposed to stimulate high-reputation workers with high-quality data to join in model training.
Besides, the proposed
scheme should not introduce much computational cost and communication overhead since these resources are
constrained at some users.
In \cite{Khan2019}, a Stackelberg
game-based incentive mechanism was proposed to select a set of IoT devices willing to join the model training process while minimizing the overall training costs, i.e., computational and communication costs. Taking into account the non-IID data and the wireless channel constraints, an auction mechanism was designed to realize the trading between the FL
server and the users for pricing and task allocation \cite{Jiao2019}.
In addition,
a multi-dimensional contract-matching based incentive mechanism was designed to address the incentive mismatches and information asymmetry between the UAVs and the FL server \cite{Lim2020}. Due to the special challenges of unshared decisions and difficulties of contribution evaluation for FL in IoT applications, DRL algorithm was exploited to learn system states from historical training records and adjust the strategies of the parameter server and edge nodes according to the environmental changes \cite{Zhan2020Incentive}.

\subsection{Discussion and Outlook}
Table \ref{distributed_table} summarizes all the above-mentioned distributed learning architectures. We can observe that each distributed learning technique has unique advantage in term of communication efficiency and computation efficiency, while the HSFL technique can achieve a trade-off between them. However, to schedule each user with the proper learning method efficiently is still in its fancy.

Asynchronous distributed learning is an effective approach to address the straggler issue
appeared in classical distributed learning techniques due to diverse computational capacities of the users. However, the study of asynchronous distributed learning still requires more attention from both academia and industrial partners since its more suitable for practical use, particularly special attention is needed when deploying in wireless networks due to unreliable wireless communication links.
Moreover, it is also necessary to design asynchronous learning scheme for the hybrid distributed learning architectures since the
users with different learning method complete their local learning in different pace. The potential solution is to group the users completing at similar pace into one cluster, and then perform inner-cluster synchronous aggregation and inter-cluster asynchronous aggregation. In this case, the clustering algorithm, especially the clustering index, needs careful design.

Another important issue to be addressed is the design of incentive mechanisms for the users to join in model aggregation. The users want to get involved model aggregation only if they can receive either economical compensation or local model improvement. Hence, an effective reward mechanism needs to be designed to motivate more high-quality users to participate in model aggregation. Particularly, the metrics that can measure the quality of the users are required, such as reputation \cite{Kang2019}, training costs \cite{Khan2019}, and dataset quality.

\section{Communication-efficient Technologies for Distributed Learning}
In practice, as the trained ML model can become outdated with time, continually updating the model in the time-varying environment is essential.
Recently, distributed learning has replaced the conventional centralized learning as an
effective technique to provide ML model training in a distributed manner by exploiting the computational resources from distributed wireless users. Instead of sharing raw data between the users and the BS as in centralized learning, only model parameters need to be exchanged in distributed learning.
When deploying distributed learning in wireless networks, it relies on
reliable and high-throughput wireless channels to support the real-time transmission of model parameters.
In this way, ML can be treated as the data source transmitted from end-to-end rather than as the enabler at the end of the system. There are three main reasons that distributed learning has been broadly considered: end users have been empowered with strong computational abilities \cite{9187796}, a huge amount of data distributed at end users which can provide valuable information, and the awareness of data privacy \cite{8957702}. However, the dynamics of the wireless communication environment strongly affects the performance of ML model training. These challenges drive the researchers to focus on designing more efficient wireless communication techniques for distributed learning.

\subsection{The Challenges of Traditional Communication Technologies }

When deploying distributed learning in wireless networks, the qualities of wireless channels determines the convergence performance of ML model training and there are some critical bottleneck problems that need to be addressed when using the current wireless communication techniques to support distributed learning:
\begin{itemize}
    \item \textbf{Communication resource limitation:} Since a huge number of wireless users need to communicate with the central BS back and forth to exchange model parameters for the learning process in a distributed manner, it is urgently necessary to design the optimal resource management solution to tackle the limited communication bandwidth and transmission power \cite{8865093}. Although more frequency bands, such as mmWave, have been widely introduced to support massive connectivity, the vulnerable signal propagation still restrains the reliability of communication  due to different kinds of channel fading. Therefore, optimizing the
    design of distributed learning algorithms to reduce the communication overhead is important. Recent research studies address this optimization problem by reducing either the communication rounds or the transmitted gradients in each round  \cite{9205981}.

    \item \textbf{Communication conditions:} Apart from limited communication resources, the wireless channel conditions directly affect whether the BS can receive or decode the local model updates from the users. Specifically, the dynamic fluctuation of the communication channel may strongly distort the transmitted informationand results in reducing decoding accuracy at the receiver side \cite{8865093}. Moreover, the reliability and the robustness of the dynamic communication channel are the guarantee of successful information exchange to support distributed learning framework. Hence, it is necessary to develop reliable communication techniques to achieve robust and low-latency communications for the implementation of distributed learning  in wireless networks.

    \item \textbf{Computational resource limitation:} Generally, the wireless users are powered by the capacity-limited batteries, and they have diverse computational capabilities, such as IoT devices equiped with small CPUs, drones and mobile phones lacking in GPUs. Distributed learning requires the users to undertake some model training tasks, which consumes computational, energy, memory/storage resources of the users.  Therefore, it is indispensable to design simple and energy-efficient ML models to simplify the computation process \cite{8865093}, or to improve the distributed learning algorithm to efficiently exploit the diverse computational resources from users.

    \item \textbf{Dynamic network:} In real-world scenarios, the end devices can be both static and mobile. In such time-varying wireless networks, communications can be interrupted, connectivity can be blocked, and data can be outdated. Therefore, distributed learning is facing extreme challenges that are caused by environmental dynamics. Hence, designing more stable training schemes that consider asynchronous collaboration, prediction, or other mechanisms that are suitable to the dynamic networks is one of the emerging issues to be solved \cite{Park2021}.

    \item \textbf{Privacy and security concerns:} Although one of the purposes to utilize distributed learning in wireless communication is to preserve data privacy, the model gradients information or outputs transmitted through wireless communication links can still be disclosed and reversely traced, which means the privacy is only partially preserved \cite{10.1145/2810103.2813677}. This kind of privacy concern is named gradient leakage attacks \cite{8737416} and membership inference \cite{deepleakage}. Similarly, security concern comes out when adversaries launch attacks on heterogeneous devices in the network and cause distributed learning faults\cite{9205981}.

\end{itemize}

Therefore, the transmission of model parameters has higher requirements on reliable and low-latency wireless communication links since the convergence performance the ML model is decided by the performance of wireless communication. The traditional communication technologies, resource allocation and data transmission methods need to be improved by considering the convergence of ML model as the performance metrics. In the following, we will investigate the state-of-art techniques, including over-the-Air Computation (AirComp), gradient compression, and user scheduling and resource allocation, to improve the performance of wireless communication for supporting distributed learning.

\subsection{User Scheduling and Resource Allocation}

\subsubsection{The Reasons for User Scheduling}
The emerging paradigm of FL is to provide a decentralized ML model training approach for a wireless edge network with a large number of resource-constrained mobile devices collecting the training data from their environment \cite{mcmahan2017}.
Recently, to obtain a high-performance model with low-latency training process, many research works \cite{Xia2021,Amiri2020b,Ren2020,Cui2021a,Vu2020,Yang2020a,Huang2020} have been conducted to investigate user scheduling schemes by addressing the following challenges:
\begin{itemize}
    \item \textbf{Dynamic channel condition:} The dynamic wireless environment cannot always guarantee good channel qualities, the spectrum resources are limited;
    \item \textbf{Heterogeneous computational  resources:} The available computational resources of individual users vary over time because of other possible task execution. The heterogeneity of users with different computational capabilities causes straggler effects;
    \item \textbf{Heterogeneous data distribution:} The statistical heterogeneity of different data distributions (i.e., IID, non-IID and imbalanced data) over users leads to the drift of local model updates, and it results in different local updates that are of dissimilar significance to the model convergence \cite{cho2020client}.
\end{itemize}
To address the above challenges, recent research works have proposed different metrics to optimize user scheduling for model aggregation.

\subsubsection{The Metrics of User Scheduling Scheme}

An intuitive design of user scheduling scheme is to aggregate as many local model updates as possible from users since the whole dataset is distributed over the users. This can be achieved by optimizing the following metrics.

\noindent\textbf{a. Metric-The Number of Users }

In \cite{Yang2020}, three user scheduling policies, including random scheduling, round robin, and proportional fair, were studied in terms of FL convergence rate for wireless networks. The analyses revealed that there exists a trade-off between the number of scheduled users and sub-channel bandwidth in the optimization of FL convergence rate under a fixed amount of available spectrum.
In \cite{Nishio2019,Yang2018}, the authors have studied this user selection problem with the goal of maximizing the number of selected users for each round under constrained resources.

\noindent\textbf{b. Metric-Channel Conditions }

In FL, the convergence performance of the ML model heavily relies on the transmission of model parameters, so the channel conditions are necessary to be considered for user scheduling.
In \cite{Yang2020}, the user scheduling scheme, proportional fair, in terms of channel conditions is studied. The authors in \cite{Amiri2020b} investigated federated learning over wireless fading channels and schedule one user for transmission based on the channel conditions in the proposed user scheduling scheme. Moreover, they generalize it as best channel scheduling scheme by selecting several users with the best channel gains \cite{Amiri2020}.

\noindent\textbf{c. Metric-The Importance of Local Model Updates }

Due to the non-IID and imbalanced data distribution over the users, each user is of different significance to the global model update. The authors in \cite{kang2020reliable} proposed a reliable
UE selection scheme by considering the reliability of the dataset owned by users. In \cite{Amiri2020b,Ren2020,Ma2021}, the authors studied a novel user scheduling scheme by considering both the channel conditions and the importance of the local model updates calculated at the users for implementing FL at wireless edge. In \cite{Ren2020}, the scheduling policy is derived in closed form to achieve the optimal trade-off between channel quality and the importance of local model update when scheduling one user in each round. The authors demonstrated that the channel based scheduling shows the lowest
testing accuracy performance while the model update based user
scheduling has the best performance for AirComp FL. A trade-off performance is achieved by considering them together \cite{Ma2021}.

\noindent\textbf{d. Metric-Age of Update }

The aforementioned user scheduling schemes are focused on either exploiting the limited spectral resources or investigating the diversity of local datasets to maximize
the number of updates collectible by the BS in each round of
global communication but ignore the staleness of these updates.
In \cite{Yang2020a}, a new metric,  Age-of-Update (AoU), was proposed to measure the staleness of local model updates in each round, and then a user scheduling algorithm that considering both the straggler effect and the communication quality was developed to minimize the AoU. This scheme aims to keep the collection of all the local updates as fresh as possible while considering the fairness among all the users. The authors in  \cite{yin2020joint} considered AoU as the performance metric of the user fairness to optimize the user selection policy, transmission power and CPU-cycle frequency.

\subsubsection{The Optimization of User Scheduling and Resource Allocation}

Due to resource limitations in  wireless networks, including the limited communication resource and the scare energy resource at the users, the joint user scheduling and resource allocation problem has been studied by a series of works. A joint learning, user scheduling and resource allocation problem was formulated to optimize the uplink Resource Block (RB) allocation and transmit power allocation so as to decrease the packet error rates of each user and improve the FL performance in wireless networks \cite{Chen2019a}.
An optimization problem that jointly designs the
power allocation and user scheduling scheme for the UAV swarm network was formulated to reduce the FL convergence round \cite{Zeng2020}. The authors in \cite{Ma2020} investigated the optimal user scheduling policy and power allocation scheme with Non-Orthogonal Multiple Access (NOMA)-based FL uplink communication during the entire learning process, and thus the aggregation latency was reduced. In \cite{Shi2021}, a joint bandwidth allocation and user scheduling problem was formulated to optimize the convergence rate of latency constrained wireless FL. The energy-efficient radio resource management strategy was investigated for bandwidth allocation and user scheduling in Federated Edge Learning (FEEL) network, which can effectively reduce the sum of energy consumption of devices while providing a guarantee on learning speed \cite{Zeng2020a}. The developed optimal bandwidth allocation scheme suggests allocating more bandwidth to the devices with worse channel conditions or weaker computational capabilities in individual learning round. To consider the long-term effect of FL, the authors in \cite{Xu2021} brought a long-term perspective to client selection and resource allocation problem, they identified the varying significance of learning rounds, and how this would affect the resource allocation to optimize learning performance for FL in wireless networks.

\subsubsection{MAB-based Optimization for User Scheduling and Resource Allocation}

The above studies investigate user scheduling scheme based on the assumption of the availability of prior information regarding the Channel State Information
(CSI) and the knowledge of the available computational resources of each user. However, in practice, it is costly or even impossible to obtain these dynamic environment information, especially for a large-scale network. Hence, a more practical scenario without knowing the
prior information needs to be considered, and MAB tool with the ability of estimating the statistical information based on the trail-and-error rule
has been exploited to design online scheduling schemes \cite{Yoshida2020,Xia2020,Cho2020,Xu2021a}. The aim of the MAB problem is to determine the arm so as to maximize the total rewards obtained in sequential decisions. In \cite{Yoshida2020}, an MAB algorithm was proposed to estimate which users are expected to have rich and available computational power and high throughput when designing the user selection strategy. Moreover, the proposed MAB-based client selection algorithm can perform exploration by selecting the users that are selected less often, and exploitation by selecting the users with rich resources, to achieve efficient user scheduling. The authors in \cite{Xia2020} discussed the client selection problem in both ideal and non-ideal scenarios (ideal scenario: always has available clients, IID and balanced dataset; non-ideal scenario: non-IID and imbalanced properties of local datasets, and dynamic availability of clients) by formulating it as an MAB problem and further proposing an Upper Confidence Bound  (UCB)-based algorithm to strike  a balance between the exploration and exploitation actions. Different from \cite{Yoshida2020,Xia2020} that reduce the time consumed per round
with fixed number of training rounds, the authors in \cite{Xu2021a} used the MAB tool to reformulate the client scheduling problem, but aiming to reduce the number of training rounds and the time consumed per round simultaneously. In \cite{Cho2020}, the client selection problem was investigated aiming to achieve faster convergence by adopting the MAB algorithms to find a balance between selecting users with larger local loss (i.e., exploitation) and ensuring user diversity in selection (i.e., exploration).

\subsection{Over-the-air Computation}
In FL, the global model aggregation procedure consists of the transmission of local model updates from each user, followed by the computation of their weighted average at a central server.
To realize efficient uplink model aggregation in FL, an analog AirComp was proposed as a communication and computation co-design approach by exploiting the additive
nature of the wireless multiple access
channels \cite{Amiri2020,Sery2019}. With AirComp, the users transmit their model updates via analog signalling, i.e., without converting to discrete coded symbols which need to be decoded at the server-side. Through joint user selection and beamforming design at the central server \cite{Yang2018}, the scheduled users then simultaneously transmit in the same communication link such that their signals overlap at the server. Given the perfect CSI at the users and accurate transmission
timing, the signal overlapped from the devices to the server over-the-air naturally produces the arithmetic sum of the local
model updates. To deal with residual channel gain and synchronization in AirComp, the authors in \cite{Shao2021} referred to it as a misaligned AirComp and devised a sum-product maximum likelihood estimator to estimate
the arithmetic sum of the transmitted symbols; the beamforming techniques were employed at the server to alleviate the destructive effects of the interference and noise terms due to the lack of CSI at the users and perfect CSI at the server \cite{Amiri2019a,Amiri2020a}.

Towards developing more efficient AirComp schemes, a broadband analog aggregation scheme (BAA) was proposed to support the transmission of high-dimensional updates and which dramatically reduces the communication latency \cite{Zhu2020}. Additionally, the authors extended the BAA to FEEL in which transmitters are limited to Quadrature Amplitude Modulation (QAM) and designed a one-bit broadband digital aggregation scheme for the current digital wireless system by featuring digital modulation of local gradients \cite{Zhu2020a}. Moreover, to address the channel noise caused by analog AirComp, the authors in \cite{Sery2020} developed an AirComp FL algorithm by introducing pre-coding at the users and scaling at the server.
In \cite{Sun2019}, an online energy-aware dynamic user scheduling policy was proposed to deal with non-IID data in FEEL by introducing data redundancy.

\subsection{Gradient Compression}

Synchronous SGD has been widely used for distributed training to enhance the efficiency of large-scale distributed learning \cite{7239545}. Although the overall computation time can be significantly reduced by adding more computational nodes and performing data parallelization, the gradient updates are still costly \cite{li2014communication}. To efficiently scale up distributed training, it is crucial to overcome the communication barrier when deploying the bandwidth-consuming parallelizing SGD. Therefore, reducing the communication data size to save the transmission spectrum bandwidth in the wireless network has been extensively studied.

One way to reduce the size of the transmitted gradients is to quantize them to low-precision values \cite{seide20141,alistarh2016qsgd,wen2017terngrad,zhou2016dorefa}. In \cite{seide20141},  1-bit SGD with error feedback was proposed for speech-scale DNNs. Quantized SGD (QSGD) was proposed by considering the trade-off between communication bandwidth and convergence
time in \cite{alistarh2016qsgd}. To be specific, the proposed QSGD scheme has the adjustable number of bits sent per iteration with possibly higher variance. Similarly, the authors in \cite{wen2017terngrad} developed the ternary gradients. Apart from quantizing gradients, in \cite{zhou2016dorefa}, the authors also considered bit convolution kernels to accelerate both model training and inference. The other way to save the gradients data size is named as gradient sparsification \cite{strom2015scalable,aji2017sparse,dryden2016communication,chen2018adacomp}. The intuitive way is to eliminate the small-amplitude gradients below a pre-defined constant threshold and only send the remaining gradients \cite{strom2015scalable}. In \cite{aji2017sparse}, the authors proposed gradient dropping by sparsifying the gradients, and then they combined the layer normalization  to keep the convergence speed. However, the pre-defined threshold is difficult to select in practice.
To avoid inappropriate threshold selection, the authors in \cite{dryden2016communication} proposed a method that only chooses a fixed proportion of positive and negative gradients, respectively. In \cite{chen2018adacomp},  a more advanced technique that can automatically tune the compression rate based on local gradients activity was studied.

To greatly reduce the communication bandwidth, Deep Gradient Compression (DGC) was proposed in \cite{lin2017deep}, which not only employs gradient sparsification to reduce the bandwidth but also employs some other techniques, such as local gradient accumulation, momentum correction, local gradient clipping, momentum factor masking, and warm-up training, to guarantee the model convergence and improve learning accuracy,
Recently, a joint local compression and global aggregation approach called Analog distributed SGD (A-DSGD) was proposed to further address the bandwidth limitation in wireless communications \cite{Amiri2020}. In A-DSGD, the distributed local users first sparsify their gradients based on a compression ratio and then project them to a lower-dimensional space imposed by the available channel bandwidth. These projections are sent directly over the multiple access channels without employing any digital coding. This approach can significantly reduce the communication overhead and save transmission resources as multiple distributed devices can transmit compressed gradients to the centre server simultaneously through the same channel.

\begin{table}[t]
\centering
\caption{Communication-efficient Technologies for Distributed Learning}
\label{table:communicationefficient}
\resizebox{\columnwidth}{!}{
\begin{tabular}{|m{4cm}|m{7cm}|m{6cm}|m{1cm}|}
\hline
\textbf{ Technology }                                           & \textbf{ Purpose }                                                                                          & \textbf{ Benefits }                                                               & \textbf{ Ref. }  \\
\hline
Joint user scheduling and~resource allocation & Appropriately schedule the users for limited channel access, heterogeneity of users and data distributions & Improve the FL performance, reduce aggregation latency, reduce energy consumption &  \cite{Ma2020,Shi2021,Zeng2020a,Xu2021,Yoshida2020,Xia2020,Cho2020,Xu2021a}                \\
\hline
Over-the-air computation                                                          & Analogy model aggregation without converting to discrete symbols                                            & Reduce communication latency                                                      &\cite{Amiri2020,Sery2019,Shao2021,Amiri2019a,Amiri2020a,Zhu2020,Zhu2020a,Sery2020,Sun2019}                \\
\hline
Gradient compression                                            & Compress gradient updates before transmission                                                               & Reduce communication data size, save transmission spectrum
  bandwidth            &\cite{seide20141,alistarh2016qsgd,wen2017terngrad,zhou2016dorefa,strom2015scalable,aji2017sparse,dryden2016communication,chen2018adacomp,lin2017deep}                \\
\hline
\end{tabular}}
\end{table}

\subsection{Discussion and Outlook}
When deploying distributed learning in wireless networks, the transmission of model parameters is decided by the quality of the wireless communication links.
This raises higher standards for the classical communication technologies, and thus some novel technologies are required to improve them
for providing better communication performance with higher reliability and lower time latency.
Table \ref{table:communicationefficient} summarizes the state-of-art technologies, including  user scheduling and resource allocation, over-the-air computation, and the gradient compression, to assist the
the share of model parameters in distributed learning with better communication-efficiency. User scheduling and resource allocation is a good approach to deal with limited spectrum resource and diverse users with the goal of optimizing the convergence performance of distributed learning, but the complexity of user scheduling scheme itself may affect the convergence time and needs to be minimized.

Apart from optimizing resource allocation and scheduling schemes for efficient communications, AirComp is a promising technique that backtracks the analog communications for direct model aggregation without digital converting. Furthermore, reducing the size of local model updates at the information source before the transmission is another efficient technique that can not only save the spectrum bandwidth, but also reduce the transmission latency. However, most of the aforementioned techniques are mainly studied separately, it is an inevitable trend to find an effective way to conceive the synergistic architecture among these technologies.

\section{AI-empowered Wireless Communications}
In section III and IV, wireless networks with edge computing are able to provide native AI services by leveraging RL techniques to enable intelligent decision-making and optimal network management and implementing distributed learning architecture to exploit the distributed computational resources of the local and edge devices.
Due to the emerging AI applications deployed at the network edge, the requirements of communication such as near-instant millisecond latency, massive connectivity, and ubiquitous coverage have become urgently desired \cite{9349624}. To meet these demands, the promising technologies, such as millimetre Wave (mmWave) \cite{6932503}, massive MIMO \cite{6798744}, NOMA, and Intelligent Reflection Surfaces (IRSs) \cite{8910627} have been proposed to improve communication performance. However, it is still challenging to apply the conventional communication theories directly to the complex application scenarios. The recent advances in AI techniques have enabled training ML models to predict the different modules in the block structure-based communication systems (i.e., modulation scheme, channel condition, transmitted signal, etc). Moreover, the AI-enabled end-to-end communication can improve communication performance by assuring low communication latency. In this section, we will discuss about the DL for physical layer wireless communications techniques.



\begin{figure}[t]
    \centering
    \includegraphics[width=6in]{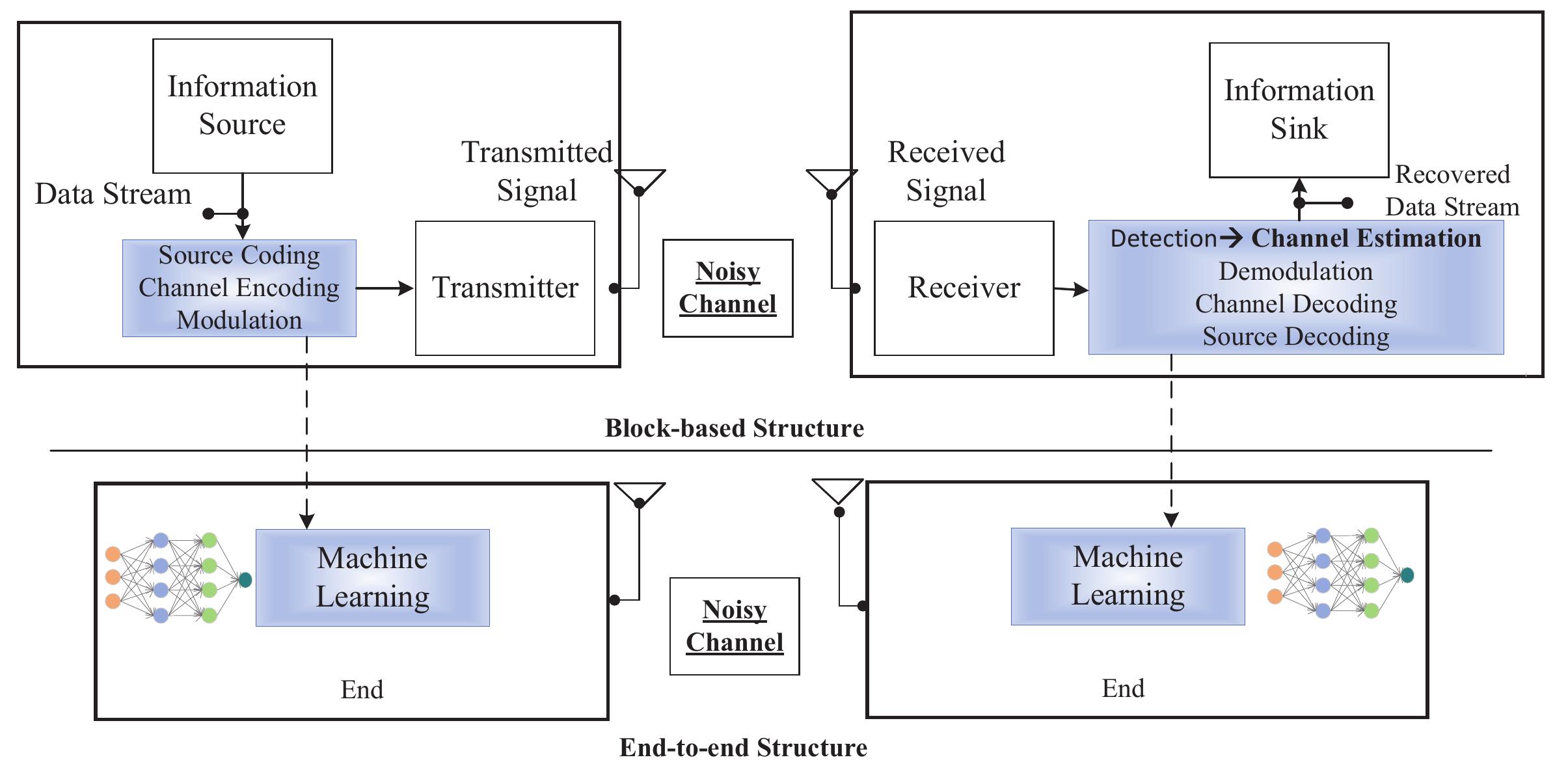}
    \caption{The block diagram of conventional communications block and the end-to-end structure.}
    \label{fig:blockdiagram}
\end{figure}

\subsection{The Potentials of DL for Wireless Communications}

To meet the demands of the large data stream and high-speed processing in emerging complex environments, there are some challenges that need to be tackled in wireless communications\cite{8786074}: complex channel modelling, local optimal in the block structure of the communication system, and efficient computation.
\subsubsection{DL solutions in wireless communications}
There are mainly three aspects potentials of DL motivates researchers to find intelligent solutions for the aforementioned limitations:
\begin{itemize}
    \item \textbf{Model-driven DL channel modelling:} Conventionally, the communication system heavily relies on mathematical models to characterize the dynamic wireless environment. However, in the real world, the environment can be very difficult to be modeled. Especially, the complex scenarios with unknown or unpredictable effects are not possible to be characterized or expressed by using mathematical models. Hence, DL is introduced to address this impossibility, which can act as a black box to replace the conventional mathematical model. Moreover, by combining it with domain knowledge, model-driven DL can assist complex modelling \cite{8715338}.
    \item \textbf{Replace the block structure:} The traditional communication structure consists of multiple blocks, such as encoding, decoding, modulation, demodulation and detection blocks, as shown in Fig. \ref{fig:blockdiagram}. Conventionally, researchers focus on optimizing specific blocks for different purposes and combining them to achieve global optimal performance through the entire system.
    However, this is not always guaranteed with such a simple combination of locally optimized blocks. Therefore,  DL has been exploited to obtain the global optimization of the entire end-to-end communication system  by replacing the separated optimization of each block \cite{8052521}.
    \item \textbf{Parallel efficient processing:} With the emergence of resource-constrained wireless devices, a large amount of data exchange causes a huge computational costs. It is even more challenging for real-time data processing in a complex environment with advanced technologies. One of the convincing reasons to apply DL to deal with such concern is that the trained DL methods can pass through parallel distributed memory architectures, such as the graphic processing unites and specialized computation chips that can demonstrate fast and energy-efficient computational ability \cite{7738524}.
\end{itemize}

\subsubsection{Partial and Complete Algorithm Replacement}
The implementation of DL in physical layer can be categorized into partial algorithm replacement and complete algorithm replacement \cite{8054694}. For partial algorithm replacement, inspired by the idea that unfold the inference iterations as layers in a deep network \cite{DBLP:journals/corr/HersheyRW14}, part of the existing algorithm can be replaced by the neural network layers. For example, \cite{7905837} considered the application of neural network architecture for sparse linear inversion in compressive sensing to assist recovering the sparse signal from the noisy measurements. Differently, for complete algorithm replacement, the DL algorithms can be treated as the black boxes that can be trained and applied for multiple purposes in the communication systems.

\subsection{From Block-based to End-to-end Structure}
Conventionally, the communication systems are designed based on separate signal processing blocks (i.e., source coding, channel coding and so on) that can be constructed as a chain structure with separate sub-optimal performance. In recent years, with the DL algorithms evolving, the end-to-end system that can utilize the advantages of DL for global optimization has been further investigated. In this subsection, the evolution of the communication structure will be discussed.

\subsubsection{DL in Block-based Communications Structure}
As shown in Fig. \ref{fig:blockdiagram}, a typical wireless communication system can be summarized as a chain diagram with multiple independent blocks as the block-based structure. Each block plays a vital role in executing an independent task, for example, source coding, channel encoding, modulation, channel estimation, demodulation, channel decoding, and source decoding.
\begin{itemize}
    \item \textbf{Signal compression:} In physical layer communications, downlink CSI feedback is one of the determinants to achieve performance gain at the BS. However, the practical challenge is that the large number of antenna elements leads to excessive transmission overhead. Although the sparse spatial and temporal correlations of the CSI have been studied to reduce the heavy feedback overhead, the sparse structure is not practically guaranteed. In \cite{8322184}, the authors proposed a CsiNet as an encoder to compress feedback and reconstruct the CSI. Specifically, the compression task is done at the user side by inputting the angular-delay domain channel matrix to a CNN layer. Then two feature maps at the output are vectorized as real-valued compressed information for feedback.
    \item \textbf{Modulation classification:} Automatic modulation recognition has been studied for many decades. In \cite{664294}, the neural network architecture has been designed as the modulation classifier after the feature extraction step to distinguish signals from both digital and analog modulation schemes. To omit the feature extraction step, the automatic learning CNN-based method is proposed in \cite{o2018over} to learn the modulation schemes directly from time-sequence raw data in the radio domain.  
    \item \textbf{Channel decoding:} The fundamental purpose of channel coding and channel decoding is to detect and correct errors in the noisy channels. In \cite{7852251}, the authors proposed the special structure, DNN-based belief propagation algorithm, which contains odd hidden layers transmit output from variable nodes to check nodes and even layers transmit output from check nodes to variable nodes. Through such a structure, the performance of decoding high-density parity-check codes can be improved. Alternatively, the authors in \cite{7926071,8254811}  proposed a plain DNN structure-based decoder, named as neural network decoder, to achieve a competitive result and high-level parallelization.
    \item  \textbf{Signal detection:} With the increasingly complex applications emerging in communication systems,  the information detection becomes harder due to the complex time-varying channel model. DL-based detectors have been designed in \cite{8227772,8052521}. However, in \cite{8227772}, the authors only considered the received signal and channel matrix as inputs to reconstruct the transmitted signal. In \cite{8052521}, the authors treated the channel as a black box and designed a five-layer fully connected DNN for Orthogonal Frequency-Division Multiplexing (OFDM) signal detection.
\end{itemize}


\subsubsection{DL for End-to-end Communications}
The structure of the end-to-end communication system can be found at the lower part of Fig. \ref{fig:blockdiagram}. All of the individual blocks at the transmitter (receiver) side are treated as a whole which is called the transmitter (receiver) end. Particularly, this structure can take the advantage of data-driven DL, as both transmitter and receiver ends can learn to automatically encode and decode source data. The DL model embedded at both ends is optimized by minimizing the loss function which consists of the difference between the true value and the estimation value. Compared to independent block optimization in block-based communications, the end-to-end optimization can guarantee a global solution \cite{8663966}. In \cite{8262721},  the DL-based auto-encoder end-to-end communication system in MIMO channels for both closed-loop and open-loop systems was proposed. Specifically, closed-loop and open-loop are distinguished by whether to consider a CSI feedback system. However, these end-to-end model training mechanisms incur a practical problem, and the back-propagation stage during model training has to pass through the unknown wireless channel. Hence, more practically, the authors in \cite{9360873} separately designed a DNN-based transmitter and a DNN-based receiver. Explicitly, the transmitter that is robust to various channel conditions learns to transform the input data. Apart from this, the receiver consists of two respective DNN modules used for channel information extraction and data recovery.

\subsection{DL for Wireless Communications Technologies}
In the past decade, with the explosive demands of wireless communications wireless technologies such as mmWave, massive MIMO, NOMA, and IRS have been developed to improve the communication performance from spatial-efficiency, spectral-efficiency, and energy-efficiency perspectives. However, with the advanced technologies implemented in the communication systems, it is even challenging to acquire precise complex mathematical model to realize robust communications. Therefore, DL is a reliable candidate to support practical implementations of the aforementioned advanced technologies. In this subsection, the works focused on DL-based frameworks in advanced technology-assisted wireless communications will be introduced.
%

\subsubsection{DL for MmWave Massive MIMO System}
MmWave band has been recognized as the spectrum that can bring magnitude improvement of speed and capacity for future wireless communications. To mitigate the poor diffraction ability of mmWave, it is widely studied to implement massive MIMO in mmWave systems and apply hybrid precoding techniques to achieve multiplex data streams, thus enhance the system throughput. Although compressive sensing related algorithms have been broadly deployed to reduce the computation complexity caused by the massive number of antenna elements in the mmWave massive MIMO systems for precoding design, the inadequate leverage of the structural characteristics of mmWave systems brings the urgent needs of developing more advanced methods. Therefore, a DL-based mmWave massive MIMO framework for effective hybrid precoding design has been proposed\cite{8618345}. Specifically, the selection of the optimized hybrid precoders is designed as the mapping relation in the DNN. Similarly, \cite{9204436} explored the DNN-based beam training schemes to deal with the nonlinear and nonmonotomic properties of channel power leakage in mmWave.
\subsubsection{DL for NOMA Scheme}
Apart from exploring under-utilized spectrum in the ultra-high-frequency bands (i.e., mmWave), as the spectral efficient technology that enables each user to operate in the same frequency band at the same time through assigning different power levels, NOMA has also been drawn significant attention. Conventional methods for sum data rate and reliability optimization in NOMA systems require high computation complexity to solve the nonlinear optimization power allocation problems with known channel state information. However, in practice, acquiring fast time-varying channel information is very challenging.
Conventional methods are not efficient and reliable enough to capture the complicated channel characteristics. To overcome such difficulty, DL-aided NOMA system has been proposed in \cite{8387468}. To be specific, a LSTM network has been established to detect the channel information automatically through offline training and online learning process.

\subsubsection{DL for IRS-assisted System}
With the flexible feature that can control and reflect the electromagnetic signal by changing the phase of the impinging signals, IRSs has been recognized as a promising technique to broaden the communication coverage for future wireless communication systems. Although the implementation of the IRSs with almost-passive elements is inexpensive, the challenges have been raised at the receiver on estimating the CSI and the signal phase angles. \cite{9625398} modelled the IRS-assisted communication systems as the end-to-end systems through the auto-encoder DL technique. Explicitly, the cascaded channels, which are the channels reflected from IRSs, have been designed as a DNN that can reduce the environment impairments effect. Moreover, \cite{9743298} proposed LSTM-based algorithm to track the constantly change CSI in IRS-assisted UAV communication network.
\subsection{Discussion and Outlook}

\begin{table}[t]
\centering
\caption{DL for Wireless Communications from Block-based to End-to-end Structure}
\label{table:dlcommunication}
\resizebox{\columnwidth}{!}{
\begin{tabular}{|m{3cm}|m{3cm}|m{2cm}|m{1.5cm}|m{3cm}|m{3cm}|m{2cm}|m{1.5cm}|}
\hline
\multicolumn{4}{|c|}{\textbf{DL in block-based structure}}                            & \multicolumn{4}{c|}{\textbf{DL in end-to-end structure}}                      \\
\hline
\textbf{Block}  & \textbf{Purpose}            & \textbf{DL Algorithm} & \textbf{Ref.} & \textbf{System} & \textbf{Purpose}   & \textbf{DL Algorithm} & \textbf{Ref.}  \\
\hline
Source encoder  & Signal compression          & CNN                   & \cite{8322184}             & MIMO    & Transceiver design & DNN                   & \cite{8262721,9360873}              \\
\hline
Modulation      & Modulation
  classification &  NN,CNN           & \cite{664294,o2018over}             & mmWave massive MIMO     & Precoder design    & DNN                   & \cite{8618345,9204436}              \\
\hline
Channel decoder & Channel decoding            & DNN                   & \cite{7852251,7926071,8254811}             & NOMA            & Channel estimation & LSTM                  & \cite{8387468}              \\
\hline
Source decoder  & Signal detection            & DNN                   & \cite{8227772,8052521}             & IRS             & Symbol recovery, channel tracking    & DNN, LSTM             & \cite{9625398,9743298}              \\
\hline
\end{tabular}}
\end{table}

The conventional signal processing algorithms with tractable information theory mathematical models have become unable to model the imperfection and non-linearity of the complex and time-varying wireless communication systems. Therefore, the model-free characteristic of the DL algorithm motivates researchers to deploy it in physical layer communications.
Table \ref{table:dlcommunication} summarizes the aforementioned research works that focus on implementing DL algorithms for different purposes in wireless communication systems from block-based to end-to-end structures.
When applying DL algorithms in the block-based structure, it only can provide local sub-optimal in each individual block (i.e., source encoder, modulation, and so on), while the global optimal can be achieved when applying it in end-to-end structure.

Therefore, with the development of more advanced communication technologies such as MIMO, mmWave massive MIMO, NOMA, and IRS, DL has been widely studied and deployed in the end-to-end communications. As the fundamental supportive for model updates transmission in distributed learning, wireless communications is expected to evolve with the combination of these advanced technologies for superior performance. Additionally, with the time-series property of the dynamic environment, LSTM that can extract the time relationships has draw increasingly attention to be applied in physical layer. Most of the above studies trained the DL model through supervised learning (i.e., with labelled data), however, it is not practical in the real-world to obtain the accurate labelled data in advance for model training. Therefore, it is necessary to design the dynamic loss function for unsupervised learning to remain solid performance for both DL model training and execution in physical layer wireless communications.

\section{Challenges and Future Opportunities}
With the sustained success of applying AI techniques and edge computing in wireless networks, AI is envisioned to become native and ubiquitous in the next generation communications, i.e., 6G and beyond. The research on distributed intelligence in wireless networks makes the fundamental step of native AI wireless networks, which has brought many new research opportunities that we have reviewed in previous sections, but there are still many challenges that need thoughtful exploration. In this section, we outline the challenges and future opportunities separately related to each topic we discussed above.

\subsection{Distributed Computation Offloading}
Researchers are now moving the focus to designing efficient offloading schemes and resource allocation methods for more practical multi-user computation offloading problems. The multi-agent RL framework has drawn significant attention from academia to model the multi-user computation offloading problem, and a few approaches, including independent learning, information sharing, conjecture-based and prediction-based algorithms, have been proposed to address the formulated multi-agent computation offloading problem. Specifically, the conjecture-based and prediction-based algorithms addressed the non-stationary issue in independent learning and large communication overhead in information sharing algorithms and became potential solutions to the multi-agent problem. However, the conjecture-based algorithm needs the training data collected from online interactions with the network elements, which slows down the training process. The potential solution is to develop off-policy learning algorithms that utilize the pre-collected online interaction data for offline training. Another approach is the prediction-based algorithm that exploits the LSTM to predict the global state with the past side information, but this approach relies on centralized offline training and that is challenging when supporting large-scale networks.



\subsection{Customized Distributed Learning}
Most existing works have studied distributed learning to enable
training the traditional ML models in the distributed manner. However, due to the increasing user-centric applications and the heterogeneity of the local dataset and wireless environment, it is necessary to ensure that the learned model can capture users' individual characteristics. Thus, designing the customized distributed learning model is an inevitable direction in wireless networks.
\subsubsection{From Zero-shot Learning to Meta-learning}
Recently, multiple learning schemes,  such as zero-shot, one-shot, few-shot, and meta-learning,  have been designed based on personalized fewer sample datasets  to train the ML models and save the wireless communication resources \cite{paz2020zest,zhou2020distilled,chen2021shot,huisman2021survey}. In \cite{paz2020zest}, the authors proposed a learning framework called zero-shot learning which firstly distinguished the features of the input without any learning and then trained the ML model based on these features. 
Similarly, the authors in \cite{zhou2020distilled} proposed one-shot federated learning which firstly distills the client's private dataset and sends the synthetic data to the server to train the global model. Moreover, the few-shot learning framework refers to learning from a few labelled datasets \cite{chen2021shot}. As one of the special categories of few-shot learning, meta-learning attempts to reduce human intervention and let the system learn by itself \cite{huisman2021survey}. As the aforementioned learning frameworks can be customized to certain applications and save the communication resources at the same time, it is worthy to extend these learning schemes to the applications in the distributed wireless network that has limited communication resources \cite{fallah2020personalized}.

\subsubsection{Personalized Distributed Learning}
The primary purpose of involving distributed users in distributed learning is that a global model can be trained by benefiting from collaborative training of these users and their decentralized computational resources. However, the heterogeneity of the users, including user heterogeneity (e.g., diverse storage hardware, computational capacities, network conditions, battery power) \cite{wu2020personalized}, data heterogeneity (e.g., non-IID and imbalanced data distribution), and model heterogeneity (e.g., hetero-modal data), will affect the convergence performance of the model training.
For instance, when the users have sufficient personalized data, joining the global model training can hurt the model's ability for personalization \cite{kulkarni2020survey}. With non-IID data, the local model updates of each user are of different significance to the global model training \cite{zhao2018federated}. Moreover, distributed learning with hetero-modal data is challenging, thus the multi-model fusion of RF and image data is considered to train a global model for received power prediction in mmWave networks \cite{koda2020distributed}.
Therefore, personalized distributed learning taking into account the diversity of users and the hetero-modal data is a practical issue and full of challenges.

\subsection{Contribution-dependent Incentive Mechanisms}
Designing proper incentive mechanisms for active participants in distributed learning is an emerging research topic, since clients that hold useful data sources may not want to actively provide local updates without rewards. To design an optimal incentive mechanism that can motivate clients to participate in distributed learning, there are some key characteristics, such as information unsharing and contribution evaluation, that can be considered as the metrics to develop incentive decisions for users \cite{Zhan2020Incentive}. However, it becomes more challenging if taking the uncertainty of the dynamic conditions into consideration (e.g., the unpredictable decisions of the participants, unfixed training periods, time-varying data source, diverse data quality caused by communication environment and so on).

\subsection{Asynchronous Distributed Learning}
Most of the existing works focus on synchronous federated learning assuming synchronous model aggregation, but it is not practical since the users do not always complete local gradient calculation and model parameters transmission at the same time due to the heterogeneity of devices and their individual datasets. Asynchronous federated learning has been studied intensively to address this challenge using dynamic learning rates, weight aggregation and a regularized loss function at local users. However, the fully asynchronous federated learning with sequential updating can face the problem of high communication costs caused by frequent model updating and transmission of local updates. A few approaches, such as cluster FL and periodic model aggregation,
have been proposed to tackle those concerns by managing the update frequency of the local users. Thus, a trade-off between convergence performance and communication costs needs to be carefully considered by designing proper update strategies.

The existing strategies, such as user selection, weight aggregation and cluster FL, are effective to improve convergence performance for asynchronous federated learning with heterogeneous users. However, different performance improvement strategies are suitable for different application scenarios. For example, a semi-asynchronous FL with suitable weighted aggregation strategies could be an optimal solution to the scenario when the disparity in computing capabilities among heterogeneous devices is extremely high. Hence, several performance improvement strategies could be developed together to improve the efficiency of asynchronous federated learning, but this could result in a decline in efficiency to a certain extent. In \cite{Xu2021}, the authors pointed out a potential research direction on the comprehensive analysis of the balance between multiple performance improvement strategies and time consumption.

When considering asynchronous learning in the hybrid distributed learning architectures discussed in section IV,
asynchronous distributed learning needs to be redesigned to adapt to the specific learning architecture to improve learning performance. For instance,
with the hybrid learning architecture,
the clustering approach can be used to group the users into different clusters according to the learning method they choose, and then weighted aggregation could be used to aggregate updates from different clusters in an asynchronous way.

\begin{figure}[t]
  \centering
  \includegraphics[width=3.8in]{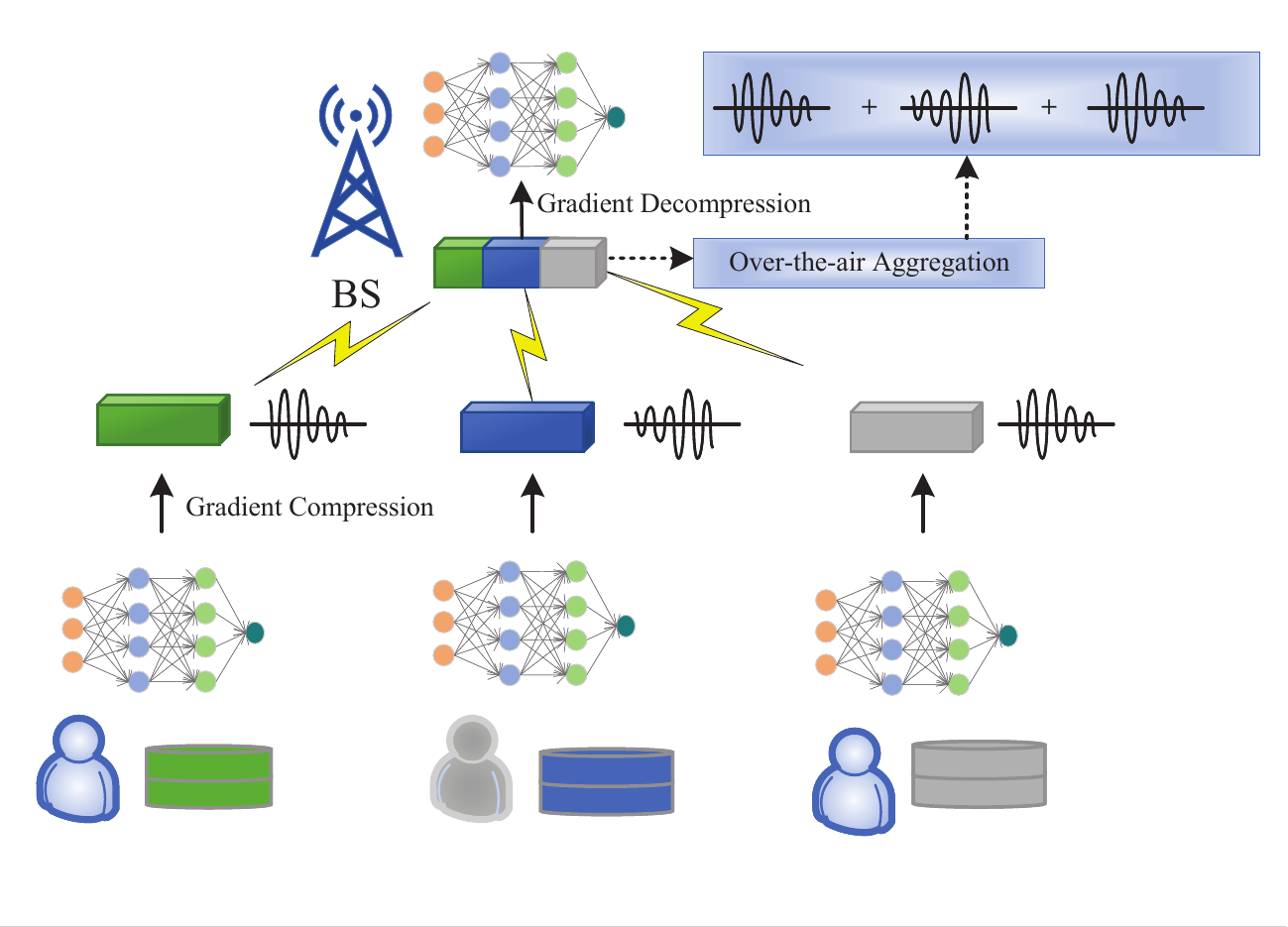}\\
  \caption{The architecture of joint gradient compression and over-the-air aggregation.}\label{overtheair}
\end{figure}

\subsection{Communication Efficiency for Distributed Learning}
In distributed learning, the communications between the central server and the local users constantly exchange information during the training stage, which consumes a huge amount of communication resources. To improve the network efficiency, the communication-efficient technologies, including AirComp and gradient compression, have been proposed but each of them still faces some challenges. Moreover, the assisted wireless technologies, such as IRS, could also be a potential option to improve the communication performance for distributed learning.

\subsubsection{Gradient Compression and Over-the-air Computation}
Gradient compression and AirComp are two main technologies that have been exploited to save communication resources for deploying FL in wireless communications. With gradient compression, although the quantification \cite{seide20141,alistarh2016qsgd,wen2017terngrad,zhou2016dorefa} and sparsification \cite{strom2015scalable,aji2017sparse,dryden2016communication,chen2018adacomp} were studied and could achieve solid compression performance, some useful gradient  information can still be lost. Hence, a  specific gradient compression method should be designed in certain learning models for various wireless applications that can tolerate some degree of information loss. In AirComp, analog-based transmission is designed, which allows the weighted aggregation to be obtained directly over the air without aggregating individual parameters acquired from distributed users. However, if a large number of users participating in training, it is more challenging to realize reliable aggregation using the wireless multiple access channel in the complex systems \cite{8952884}. Therefore, it is essential to design effective approaches that can mitigate the channel distortion in the network and interference among users to provide robust and efficient transmission that can support AirComp. Moreover, collectively considering appropriate gradient compression and dependable over-the-air aggregation to stack the communication benefits can be another promising research direction. As shown in Fig. \ref{overtheair}, the architecture of joint gradient compression and AirComp aggregation is presented.

\subsubsection{IRS-assisted Distributed Learning}
Due to the vulnerability of mmWave transmission, IRS is an emerging low-cost technology that can reconfigure the wireless propagation directions to improve both spectrum and energy efficiencies in wireless networks. Particularly, the phase shifts of the signal can be adjusted through a large amount of passive reflecting elements to steer the signal in specific directions. Hence, IRS can be leveraged to enhance the received signal strength and this is beneficial to both gradient transmission and AirComp \cite{9199786,9502547}. However, jointly designing the aggregation beamformers at the BS and the phase shifts at the IRS can be a very challenging task.

\subsection{Privacy and Security}
Although distributed learning is capable of preventing direct raw data leakage from the local devices, private information can still be extracted through intercepted gradient updates that are exchanged between the distributed devices and central server. Moreover, during the gradients transmission, attacks and data poisoning can also threaten the security of the distributed system \cite{9048613}.

\subsubsection{Privacy Leakage Protection}
Distorting \cite{geyer2017differentially} and dummy \cite{7373646} are two techniques that can protect data privacy from the user side. In \cite{geyer2017differentially}, the authors proposed the  randomized mechanism that consists of random sub-sampling and distorting steps to approximate the average and hide the individual client's contributions. However, the trade-off between privacy-preserving and model convergence performance should be further studied. In \cite{7373646}, the authors designed a method that transmits the original information together with probabilistically dummy packets. Since dummy parameters are sent as redundancy, extra communication resources, such as bandwidth and transmission energy, are required.  The encryption-based technique to prevent data inspection at the central server side was proposed \cite{8241854}. However, additional overhead is needed for encryption in this case. Therefore, for privacy protection, it is essential to find the balance among privacy, model performance, and communication efficiency, at both clients and central server sides.

\subsubsection{Anomaly Detection}
When training the distributed learning model, the model parameters are transmitted through the wireless network. However, abnormal data samples can greatly influence the overall model training. Anomaly detection which can distinguish abnormal data can not only be used to detect data poisoning and attacks from adversaries \cite{9424138} but also can monitor the abnormal operations in the wireless network, such as traffic load, computation resources usage and etc. Therefore, embedding anomaly detection into appropriate distributed learning techniques can provide extensive contributions to both secure model updating and system inspection.

\section{Conclusions}
In this article, the recent literature of distributed intelligence in wireless networks has been surveyed with an emphasis on the following aspects: the basic concepts of native-AI networks, ML techniques enabled edge computing, distributed learning architectures for heterogeneous networks, communication-efficient technologies for distributed learning, as well as DL-enabled end-to-end communication structure and DL-assisted advanced communication technologies. Specifically, we highlighted the comparisons of different ML algorithms enabled edge computing in section III, the advantages of different distributed learning architectures in section IV. Investigating ML-assisted
communication technologies and  structures is particularly important, since they can provide more reliable and ultra low latency communication performance. It is worth pointing that when designing efficient communication technologies, the convergence-based metrics are proposed to investigate user scheduling and resource allocation, and the special technology with direct ML model aggregation, namely over-the-air computation, is also presented in section V. Finally, the challenges of existing research works on distributed intelligence in wireless networks have been
identified, and also the future opportunities were discussed.

\bibliographystyle{IEEEtran}
\bibliography{IEEEabrv,my_references}

\begin{thebibliography}{100}
\providecommand{\url}[1]{#1}
\csname url@samestyle\endcsname
\providecommand{\newblock}{\relax}
\providecommand{\bibinfo}[2]{#2}
\providecommand{\BIBentrySTDinterwordspacing}{\spaceskip=0pt\relax}
\providecommand{\BIBentryALTinterwordstretchfactor}{4}
\providecommand{\BIBentryALTinterwordspacing}{\spaceskip=\fontdimen2\font plus
\BIBentryALTinterwordstretchfactor\fontdimen3\font minus
  \fontdimen4\font\relax}
\providecommand{\BIBforeignlanguage}[2]{{%
\expandafter\ifx\csname l@#1\endcsname\relax
\typeout{** WARNING: IEEEtran.bst: No hyphenation pattern has been}%
\typeout{** loaded for the language `#1'. Using the pattern for}%
\typeout{** the default language instead.}%
\else
\language=\csname l@#1\endcsname
\fi
#2}}
\providecommand{\BIBdecl}{\relax}
\BIBdecl

\bibitem{letaief2019roadmap}
K.~B. Letaief, W.~Chen, Y.~Shi, J.~Zhang, and Y.-J.~A. Zhang, ``{The roadmap to
  6G: AI empowered wireless networks},'' \emph{IEEE Commun. Mag.}, vol.~57,
  no.~8, pp. 84--90, Aug. 2019.

\bibitem{wu2021toward}
J.~Wu, R.~Li, X.~An, C.~Peng, Z.~Liu, J.~Crowcroft, and H.~Zhang, ``Toward
  native artificial intelligence in {6G} networks: System design,
  architectures, and paradigms,'' \emph{arXiv preprint arXiv:2103.02823}, Mar.
  2021.

\bibitem{Knud2020}
K.~L. Lueth, ``State of the {IoT} 2020: 12 billion {IoT} connections,
  surpassing {non-IoT} for the first time,'' \emph{IoT Analytics}, Nov. 2020.

\bibitem{Radicati2021}
T.~R. Group, ``Mobile statistics report, 2021-2025,'' \emph{online access
  available,
  $https://www.radicati.com/wp/wp-content/uploads/2021/Mobile_Statistics_Report,_2021-2025_Executive_Summary.pdf$},
  Apr. 2021.

\bibitem{GSMA2021}
D.~Anne and B.~Kalvin, ``The state of mobile internet connectivity 2021,''
  \emph{online access available,
  https://www.gsma.com/r/wp-content/uploads/2021/09/The-State-of-Mobile-Internet-Connectivity-Report-2021.pdf},
  Sep. 2021.

\bibitem{Levitate2020}
L.~Capital, ``The future of thedrone economy,'' \emph{online access available,
  https://levitatecap.com/levitate/wp-content/uploads/2020/12/White-Paper-v4.pdf},
  Dec. 2020.

\bibitem{Mach2017}
P.~Mach and Z.~Becvar, ``Mobile edge computing: A survey on architecture and
  computation offloading,'' \emph{IEEE Commun. Surveys Tuts.}, vol.~19, no.~3,
  pp. 1628--1656, Mar. 2017.

\bibitem{Taleb2017}
T.~Taleb, K.~Samdanis, B.~Mada, H.~Flinck, S.~Dutta, and D.~Sabella, ``{On
  multi-access edge computing: A survey of the emerging 5G network edge cloud
  architecture and orchestration},'' \emph{IEEE Commun. Surveys Tuts.},
  vol.~19, no.~3, pp. 1657--1681, May. 2017.

\bibitem{verbraeken2020survey}
\BIBentryALTinterwordspacing
J.~Verbraeken, M.~Wolting, J.~Katzy, J.~Kloppenburg, T.~Verbelen, and J.~S.
  Rellermeyer, ``A survey on distributed machine learning,'' \emph{ACM Comput.
  Surv.}, vol.~53, no.~2, Mar. 2020. [Online]. Available:
  \url{https://doi.org/10.1145/3377454}
\BIBentrySTDinterwordspacing

\bibitem{mcmahan2017}
B.~McMahan, E.~Moore, D.~Ramage, S.~Hampson, and B.~A. y~Arcas, ``Applied
  federated learning: Improving google keyboard query suggestions,'' in
  \emph{Artificial Intelligence and Statistics}.\hskip 1em plus 0.5em minus
  0.4em\relax PMLR, Apr. 2017, pp. 1273--1282.

\bibitem{Elbir2020}
A.~M. Elbir, S.~Coleri, and K.~V. Mishra, ``Hybrid federated and centralized
  learning,'' in \emph{Proc. 29th European Signal Processing Conference
  ({EUSIPCO})}.\hskip 1em plus 0.5em minus 0.4em\relax {IEEE}, Aug. 2021.

\bibitem{Liu2021}
X.~Liu, Y.~Deng, and T.~Mohammoodi, ``A novel hybrid split and federated
  learning architecture in wireless {UAV} networks,'' \emph{Proc. IEEE Int.
  Conf. Commun. (ICC)}, early access, May 2022.

\bibitem{Liu2022_energy}
------, ``Energy efficient user scheduling for hybrid split and federated
  learning in wireless {UAV} networks,'' \emph{Proc. IEEE Int. Conf. Commun.
  (ICC)}, early access, May 2022.

\bibitem{Zhan2021}
Y.~Zhan, J.~Zhang, Z.~Hong, L.~Wu, P.~Li, and S.~Guo, ``A survey of incentive
  mechanism design for federated learning,'' \emph{IEEE Trans. Emerg. Topics
  Comput.}, pp. 1--1, Mar. 2021.

\bibitem{Lee2021}
H.-S. Lee and J.-W. Lee, ``Adaptive transmission scheduling in wireless
  networks for asynchronous federated learning,'' \emph{IEEE J. Sel. Areas
  Commun.}, vol.~39, pp. 3673 -- 3687, Oct. 2021.

\bibitem{Xu2021}
J.~Xu and H.~Wang, ``Client selection and bandwidth allocation in wireless
  federated learning networks: A long-term perspective,'' \emph{{IEEE} Trans.
  Wireless Commun.}, vol.~20, no.~2, pp. 1188--1200, Feb. 2021.

\bibitem{Kato2020}
N.~Kato, B.~Mao, F.~Tang, Y.~Kawamoto, and J.~Liu, ``Ten challenges in
  advancing machine learning technologies toward {6G},'' \emph{{IEEE} Wireless
  Commun.}, vol.~27, no.~3, pp. 96--103, Jun. 2020.

\bibitem{8052521}
H.~Ye, G.~Y. Li, and B.-H. Juang, ``{Power of deep learning for channel
  estimation and signal detection in OFDM systems},'' \emph{IEEE Wireless
  Commun. Lett.}, vol.~7, no.~1, pp. 114--117, Sep. 2018.

\bibitem{8382166}
Q.~Mao, F.~Hu, and Q.~Hao, ``Deep learning for intelligent wireless networks: A
  comprehensive survey,'' \emph{IEEE Commun. Surveys Tuts.}, vol.~20, no.~4,
  pp. 2595--2621, Jun. 2018.

\bibitem{wang2021joint}
C.~Wang, D.~Deng, L.~Xu, W.~Wang, and F.~Gao, ``Joint interference alignment
  and power control for dense networks via deep reinforcement learning,''
  \emph{{IEEE} Wireless Commun. Lett.}, vol.~10, no.~5, pp. 966--970, May 2021.

\bibitem{o2018over}
T.~J. O’Shea, T.~Roy, and T.~C. Clancy, ``Over-the-air deep learning based
  radio signal classification,'' \emph{IEEE J. Sel. Topics Signal Process.},
  vol.~12, no.~1, pp. 168--179, Feb. 2018.

\bibitem{zhang2021countermeasures}
L.~Zhang, S.~Lambotharan, G.~Zheng, B.~AsSadhan, and F.~Roli, ``Countermeasures
  against adversarial examples in radio signal classification,'' \emph{{IEEE}
  Wireless Commun. Lett.}, vol.~10, no.~8, pp. 1830--1834, Aug. 2021.

\bibitem{liu2021distributed}
J.~Liu, J.~Huang, Y.~Zhou, X.~Li, S.~Ji, H.~Xiong, and D.~Dou, ``From
  distributed machine learning to federated learning: A survey,''
  \emph{Knowledge and Information Systems}, pp. 1--33, Mar. 2022.

\bibitem{9060868}
W.~Y.~B. Lim, N.~C. Luong, D.~T. Hoang, Y.~Jiao, Y.-C. Liang, Q.~Yang,
  D.~Niyato, and C.~Miao, ``Federated learning in mobile edge networks: A
  comprehensive survey,'' \emph{IEEE Commun. Surveys Tuts.}, vol.~22, no.~3,
  pp. 2031--2063, Apr. 2020.

\bibitem{9205981}
Y.~Liu, X.~Yuan, Z.~Xiong, J.~Kang, X.~Wang, and D.~Niyato, ``{Federated
  learning for 6G communications: challenges, methods, and future
  directions},'' \emph{China Communications}, vol.~17, no.~9, pp. 105--118,
  Sep. 2020.

\bibitem{9530714}
Z.~Qin, G.~Y. Li, and H.~Ye, ``Federated learning and wireless
  communications,'' \emph{IEEE Wireless Commun.}, vol.~28, no.~5, pp. 134--140,
  Sep. 2021.

\bibitem{9446488}
S.~Hu, X.~Chen, W.~Ni, E.~Hossain, and X.~Wang, ``Distributed machine learning
  for wireless communication networks: Techniques, architectures, and
  applications,'' \emph{IEEE Commun. Surveys Tuts.}, vol.~23, no.~3, pp.
  1458--1493, Jun. 2021.

\bibitem{Park2021}
J.~Park, S.~Samarakoon, A.~Elgabli, J.~Kim, M.~Bennis, S.-L. Kim, and
  M.~Debbah, ``Communication-efficient and distributed learning over wireless
  networks: Principles and applications,'' \emph{Proc. IEEE}, vol. 109, no.~5,
  pp. 796--819, Feb. 2021.

\bibitem{9562559}
M.~Chen, D.~Gündüz, K.~Huang, W.~Saad, M.~Bennis, A.~V. Feljan, and H.~V.
  Poor, ``Distributed learning in wireless networks: Recent progress and future
  challenges,'' \emph{IEEE J. Sel. Areas Commun.}, vol.~39, no.~12, pp.
  3579--3605, Apr. 2021.

\bibitem{chen2019deep}
J.~Chen and X.~Ran, ``Deep learning with edge computing: A review,''
  \emph{Proc. IEEE}, vol. 107, no.~8, pp. 1655--1674, Jul. 2019.

\bibitem{Xu2020}
D.~Xu, T.~Li, Y.~Li, X.~Su, S.~Tarkoma, T.~Jiang, J.~Crowcroft, and P.~Hui,
  ``Edge intelligence: Architectures, challenges, and applications,''
  \emph{arXiv preprint arXiv:2003.12172}, Mar. 2020.

\bibitem{li2017scalable}
S.~Li, Q.~Yu, M.~A. Maddah-Ali, and A.~S. Avestimehr, ``A scalable framework
  for wireless distributed computing,'' \emph{IEEE/ACM Trans. Netw.}, vol.~25,
  no.~5, pp. 2643--2654, May 2017.

\bibitem{Li2016a}
------, ``Edge-facilitated wireless distributed computing,'' in \emph{Proc.
  IEEE Global Commun. Conf. (GLOBECOM)}, Dec. 2016.

\bibitem{Qualcomm2020}
Qualcomm, ``Mobile future of extended reality,'' \emph{online access,
  https://www.qualcomm.com/media/documents/files/the-mobile-future-of-extended-reality-xr.pdf},
  Nov. 2020.

\bibitem{Tara2017}
\BIBentryALTinterwordspacing
T.~J. Brigham, ``Reality check: Basics of augmented, virtual, and mixed
  reality,'' \emph{Medical Reference Services Quarterly}, vol.~36, no.~2, pp.
  171--178, Apr. 2017, pMID: 28453428. [Online]. Available:
  \url{https://doi.org/10.1080/02763869.2017.1293987}
\BIBentrySTDinterwordspacing

\bibitem{Google2008}
P.~McDonald, ``Google developer blog,'' \emph{online access,
  http://googleappengine.blogspot.com/2008/04/introducing-google-app-engine-our-new.html},
  Apr. 2008.

\bibitem{Cisco2020}
Cisco, ``Cisco annual internet report (2018-2023),'' \emph{online access,
  https://www.cisco.com/c/en/us/solutions/collateral/executive-perspectives/annual-internet-report/white-paper-c11-741490.pdf},
  Feb. 2020.

\bibitem{Satyanarayanan2011}
M.~Satyanarayanan, V.~Bahl, R.~Caceres, and N.~Davies, ``{The case for VM-based
  cloudlets in mobile computing},'' \emph{IEEE Pervasive Comput.}, Oct. 2009.

\bibitem{bonomi2012fog}
F.~Bonomi, R.~Milito, J.~Zhu, and S.~Addepalli, ``Fog computing and its role in
  the internet of things,'' in \emph{Proc. 1st Edition MCC Workshop Mobile
  Cloud Comput.}, Aug. 2012, pp. 13--16.

\bibitem{hu2015mobile}
Y.~C. Hu, M.~Patel, D.~Sabella, N.~Sprecher, and V.~Young, ``{Mobile edge
  computing-a key technology towards 5G},'' \emph{ETSI white paper}, vol.~11,
  no.~11, pp. 1--16, Sep. 2015.

\bibitem{MEC2019}
A.~Filali, A.~Abouaomar, S.~Cherkaoui, A.~Kobbane, and M.~Guizani,
  ``Multi-access edge computing: A survey,'' \emph{{IEEE} Access}, vol.~8, pp.
  197\,017--197\,046, Oct. 2020.

\bibitem{Baktir2017}
A.~C. Baktir, A.~Ozgovde, and C.~Ersoy, ``How can edge computing benefit from
  software-defined networking: A survey, use cases, and future directions,''
  \emph{IEEE Commun. Surveys Tuts.}, vol.~19, no.~4, pp. 2359--2391, Jun. 2017.

\bibitem{Abdellatif2019}
A.~A. Abdellatif, A.~Mohamed, C.~F. Chiasserini, M.~Tlili, and A.~Erbad, ``Edge
  computing for smart health: Context-aware approaches, opportunities, and
  challenges,'' \emph{{IEEE} Netw.}, vol.~33, no.~3, pp. 196--203, May 2019.

\bibitem{Siriwardhana2021}
Y.~Siriwardhana, P.~Porambage, M.~Liyanage, and M.~Ylianttila, ``{A survey on
  mobile augmented reality with 5G mobile edge computing: Architectures,
  applications, and technical aspects},'' \emph{IEEE Commun. Surveys Tuts.},
  vol.~23, no.~2, pp. 1160--1192, Feb. 2021.

\bibitem{Li2014Exploring}
Y.~Li, L.~Sun, and W.~Wang, ``Exploring device-to-device communication for
  mobile cloud computing,'' in \emph{Proc. IEEE Int. Conf. Commun. (ICC)}, Jun.
  2014.

\bibitem{Nir2014}
M.~Nir, A.~Matrawy, and M.~St-Hilaire, ``An energy optimizing scheduler for
  mobile cloud computing environments,'' in \emph{Proc. IEEE Conf. Comput.
  Commun. Workshops (INFOCOM WKSHPS)}, Apr. 2014.

\bibitem{Tao2017}
X.~Tao, K.~Ota, M.~Dong, H.~Qi, and K.~Li, ``Performance guaranteed computation
  offloading for mobile-edge cloud computing,'' \emph{{IEEE} Wireless Commun.
  Lett.}, vol.~6, no.~6, pp. 774--777, Dec. 2017.

\bibitem{cloud10key}
I.~M. Cloud, ``Key marketing trends for 2017 and ideas for exceeding customer
  expectations,'' \emph{IBM Marketing Cloud}, 2017.

\bibitem{andalibi2021making}
V.~Andalibi, J.~Dev, D.~Kim, E.~Lear, and L.~J. Camp, ``{Making access control
  easy in IoT},'' in \emph{Int. Symposium on Human Aspects of Information
  Security and Assurance}.\hskip 1em plus 0.5em minus 0.4em\relax Springer,
  Jul. 2021, pp. 127--137.

\bibitem{Sardellitti2015}
S.~Sardellitti, G.~Scutari, and S.~Barbarossa, ``Joint optimization of radio
  and computational resources for multicell mobile-edge computing,'' \emph{IEEE
  Trans. Signal Inf. Process. Over Netw.}, vol.~1, no.~2, pp. 89--103, Jun.
  2015.

\bibitem{You2017}
C.~You, K.~Huang, H.~Chae, and B.-H. Kim, ``Energy-efficient resource
  allocation for mobile-edge computation offloading,'' \emph{{IEEE} Trans.
  Wireless Commun.}, vol.~16, no.~3, pp. 1397--1411, Mar. 2017.

\bibitem{Mao2017}
Y.~Mao, J.~Zhang, and K.~B. Letaief, ``Joint task offloading scheduling and
  transmit power allocation for mobile-edge computing systems,'' in \emph{Proc.
  {IEEE} Wireless Communs. and Netw. Conf. ({WCNC})}, Mar. 2017.

\bibitem{Chen2018}
M.~Chen and Y.~Hao, ``Task offloading for mobile edge computing in software
  defined ultra-dense network,'' \emph{IEEE J. Sel. Areas Commun.}, vol.~36,
  no.~3, pp. 587--597, Mar. 2018.

\bibitem{Tran2019}
T.~X. Tran and D.~Pompili, ``Joint task offloading and resource allocation for
  multi-server mobile-edge computing networks,'' \emph{{IEEE} Trans. Veh.
  Technol.}, vol.~68, no.~1, pp. 856--868, Jan. 2019.

\bibitem{Mao2016}
Y.~Mao, J.~Zhang, and K.~B. Letaief, ``Dynamic computation offloading for
  mobile-edge computing with energy harvesting devices,'' \emph{IEEE J. Sel.
  Areas Commun.}, vol.~34, no.~12, pp. 3590--3605, Dec. 2016.

\bibitem{Mao2017a}
Y.~Mao, J.~Zhang, S.~H. Song, and K.~B. Letaief, ``Stochastic joint radio and
  computational resource management for multi-user mobile-edge computing
  systems,'' \emph{{IEEE} Trans. Wireless Commun.}, vol.~16, no.~9, pp.
  5994--6009, Sep. 2017.

\bibitem{Liu2017}
C.-F. Liu, M.~Bennis, and H.~V. Poor, ``Latency and reliability-aware task
  offloading and resource allocation for mobile edge computing,'' in
  \emph{Proc. {IEEE} Globecom Workshops ({GC} Wkshps)}, Dec. 2017.

\bibitem{Zhang2018}
J.~Zhang, W.~Xia, F.~Yan, and L.~Shen, ``Joint computation offloading and
  resource allocation optimization in heterogeneous networks with mobile edge
  computing,'' \emph{{IEEE} Access}, vol.~6, pp. 19\,324--19\,337, Mar. 2018.

\bibitem{Pham2018}
Q.-V. Pham, T.~Leanh, N.~H. Tran, B.~J. Park, and C.~S. Hong, ``Decentralized
  computation offloading and resource allocation for mobile-edge computing: A
  matching game approach,'' \emph{{IEEE} Access}, vol.~6, pp. 75\,868--75\,885,
  Nov. 2018.

\bibitem{Li2018}
J.~Li, H.~Gao, T.~Lv, and Y.~Lu, ``Deep reinforcement learning based
  computation offloading and resource allocation for {MEC},'' in \emph{Proc.
  {IEEE} Wireless Communs. and Netw. Conf. ({WCNC})}, Apr. 2018.

\bibitem{Chen2019}
X.~Chen, H.~Zhang, C.~Wu, S.~Mao, Y.~Ji, and M.~Bennis, ``Optimized computation
  offloading performance in virtual edge computing systems via deep
  reinforcement learning,'' \emph{IEEE Internet Things J.}, vol.~6, no.~3, pp.
  4005--4018, Jun 2019.

\bibitem{hornik1989multilayer}
K.~Hornik, M.~Stinchcombe, and H.~White, ``Multilayer feedforward networks are
  universal approximators,'' \emph{Neural Netw.}, vol.~2, no.~5, pp. 359--366,
  Jan. 1989.

\bibitem{schmidhuber2015deep}
J.~Schmidhuber, ``Deep learning in neural networksn overview,'' \emph{Neural
  Netw.}, vol.~61, pp. 85--117, Jan. 2015.

\bibitem{6638950}
S.~Zhang, C.~Zhang, Z.~You, R.~Zheng, and B.~Xu, ``Asynchronous stochastic
  gradient descent for {DNN} training,'' in \emph{Proc. IEEE Int. Conf.
  Acoustics, Speech and Signal Processing (ICASSP)}, May 2013, pp. 6660--6663.

\bibitem{krizhevsky2017imagenet}
A.~Krizhevsky, I.~Sutskever, and G.~E. Hinton, ``Imagenet classification with
  deep convolutional neural networks,'' \emph{Communications of the ACM},
  vol.~60, no.~6, pp. 84--90, May 2017.

\bibitem{8496892}
R.~C. Gonzalez, ``Deep convolutional neural networks [lecture notes],''
  \emph{IEEE Signal Process. Mag.}, vol.~35, no.~6, pp. 79--87, Nov. 2018.

\bibitem{Mnih2015}
V.~Mnih, K.~Kavukcuoglu, D.~Silver, A.~A. Rusu, J.~Veness, M.~G. Bellemare,
  A.~Graves, M.~Riedmiller, A.~K. Fidjeland, G.~Ostrovski, S.~Petersen,
  C.~Beattie, A.~Sadik, I.~Antonoglou, H.~King, D.~Kumaran, D.~Wierstra,
  S.~Legg, and D.~Hassabis, ``Human-level control through deep reinforcement
  learning,'' \emph{Nature}, vol. 518, no. 7540, pp. 529--533, Feb. 2015.

\bibitem{Hasselt2015}
H.~van Hasselt, A.~Guez, and D.~Silver, ``Deep reinforcement learning with
  double {Q}-learning,'' \emph{Proceedings of the AAAI Conference on Artificial
  Intelligence}, Mar. 2016.

\bibitem{Wang2015Dueling}
Z.~Wang, T.~Schaul, M.~Hessel, H.~van Hasselt, M.~Lanctot, and N.~de~Freitas,
  ``Dueling network architectures for deep reinforcement learning,''
  \emph{Proceedings of the Int. Conf. on Machine Learning}, vol.~48, pp.
  1995--2003, Nov. 2015.

\bibitem{Silver2014Deterministic}
D.~Silver, G.~Lever, N.~Heess, T.~Degris, D.~Wierstra, and M.~Riedmiller,
  ``Deterministic policy gradient algorithms,'' in \emph{Proceedings of the
  31st Int. Conf. on Machine Learning}, ser. ICML'14, vol.~32.\hskip 1em plus
  0.5em minus 0.4em\relax JMLR.org, Jan. 2014, pp. 387--395.

\bibitem{Lillicrap2015}
T.~P. Lillicrap, J.~J. Hunt, A.~Pritzel, N.~Heess, T.~Erez, Y.~Tassa,
  D.~Silver, and D.~Wierstra, ``Continuous control with deep reinforcement
  learning,'' \emph{arXiv preprint arXiv:1509.02971}, Sep. 2015.

\bibitem{Mnih2016}
V.~Mnih, A.~P. Badia, M.~Mirza, A.~Graves, T.~P. Lillicrap, T.~Harley,
  D.~Silver, and K.~Kavukcuoglu, ``Asynchronous methods for deep reinforcement
  learning,'' \emph{Proceedings of the 33rd Int. Conf. on Machine Learning},
  vol.~48, pp. 1928--1937, Feb. 2016.

\bibitem{Schulman2015}
J.~Schulman, S.~Levine, P.~Moritz, M.~I. Jordan, and P.~Abbeel, ``Trust region
  policy optimization,'' \emph{Proceedings of the 32nd Int. Conf. on Machine
  Learning}, vol.~37, pp. 1889--1897, Feb. 2015.

\bibitem{Schulman2017}
J.~Schulman, F.~Wolski, P.~Dhariwal, A.~Radford, and O.~Klimov, ``Proximal
  policy optimization algorithms,'' \emph{arXiv preprint arXiv:1707.06347},
  Jul. 2017.

\bibitem{Zhang2019Multi}
K.~Zhang, Z.~Yang, T.~BaÅŸar, K.~Zhang, Z.~Yang, and T.~Ba{\c{s}}ar,
  ``Multi-agent reinforcement learning: A selective overview of theories and
  algorithms,'' \emph{Handbook of Reinforcement Learning and Control}, pp.
  321--384, Jun. 2021.

\bibitem{Tan1993}
M.~Tan, ``Multi-agent reinforcement learning: Independent vs. cooperative
  agents,'' in \emph{Machine Learning Proceedings 1993}.\hskip 1em plus 0.5em
  minus 0.4em\relax Elsevier, 1993, pp. 330--337.

\bibitem{Littman1994}
M.~L. Littman, ``Markov games as a framework for multi-agent reinforcement
  learning,'' in \emph{Machine Learning Proceedings 1994}.\hskip 1em plus 0.5em
  minus 0.4em\relax Elsevier, Jan. 1994, pp. 157--163.

\bibitem{2007a}
C.~tao Chu, S.~Kim, Y.~an~Lin, Y.~Yu, G.~Bradski, K.~Olukotun, and A.~Ng,
  ``Map-reduce for machine learning on multicore,'' in \emph{Proc. Adv. Neural
  Inf. Process. Syst}.\hskip 1em plus 0.5em minus 0.4em\relax The {MIT} Press,
  Dec. 2007.

\bibitem{Shanahan2015}
J.~G. Shanahan and L.~Dai, ``Large scale distributed data science using apache
  spark,'' in \emph{Proceedings of the 21th {ACM} {SIGKDD} Int. Conf. on
  Knowledge Discovery and Data Mining}.\hskip 1em plus 0.5em minus 0.4em\relax
  {ACM}, Aug. 2015.

\bibitem{li2013parameter}
M.~Li, L.~Zhou, Z.~Yang, A.~Li, F.~Xia, D.~G. Andersen, and A.~Smola,
  ``Parameter server for distributed machine learning,'' in \emph{Proc. Big
  Learning NIPS Workshop}, vol.~6, Dec. 2013, p.~2.

\bibitem{Hu2021}
C.-H. Hu, Z.~Chen, and E.~G. Larsson, ``Device scheduling and update
  aggregation policies for asynchronous federated learning,'' \emph{Proc. IEEE
  22nd Int. Workshop on Signal Processing Advances in Wireless Communs.
  (SPAWC)}, Jul. 2021.

\bibitem{yang2018applied}
T.~Yang, G.~Andrew, H.~Eichner, H.~Sun, W.~Li, N.~Kong, D.~Ramage, and
  F.~Beaufays, ``Applied federated learning: Improving google keyboard query
  suggestions,'' \emph{arXiv preprint arXiv:1812.02903}, Dec. 2018.

\bibitem{Gupta2018}
O.~Gupta and R.~Raskar, ``Distributed learning of deep neural network over
  multiple agents,'' \emph{Journal of Network and Computer Applications}, vol.
  116, pp. 1--8, Aug. 2018.

\bibitem{Vepakomma2018}
P.~Vepakomma, O.~Gupta, T.~Swedish, and R.~Raskar, ``Split learning for health:
  Distributed deep learning without sharing raw patient data,'' \emph{arXiv
  preprint arXiv:1812.00564}, Dec. 2018.

\bibitem{vepakomma2019reducing}
P.~Vepakomma, O.~Gupta, A.~Dubey, and R.~Raskar, ``Reducing leakage in
  distributed deep learning for sensitive health data,'' \emph{arXiv preprint
  arXiv:1812.00564}, May 2019.

\bibitem{vepakomma2018no}
P.~Vepakomma, T.~Swedish, R.~Raskar, O.~Gupta, and A.~Dubey, ``No peek: A
  survey of private distributed deep learning,'' \emph{arXiv preprint
  arXiv:1812.03288}, Dec. 2018.

\bibitem{Yousefpour2018}
A.~Yousefpour, C.~Fung, T.~Nguyen, K.~Kadiyala, F.~Jalali, A.~Niakanlahiji,
  J.~Kong, and J.~P. Jue, ``All one needs to know about fog computing and
  related edge computing paradigms: A complete survey,'' \emph{Journal of
  Systems Architecture}, vol.~98, pp. 289--330, Aug. 2018.

\bibitem{bittencourt2018internet}
L.~Bittencourt, R.~Immich, R.~Sakellariou, N.~Fonseca, E.~Madeira, M.~Curado,
  L.~Villas, L.~DaSilva, C.~Lee, and O.~Rana, ``The internet of things, fog and
  cloud continuum: Integration and challenges,'' \emph{Internet of Things},
  vol.~3, pp. 134--155, Oct. 2018.

\bibitem{hong2019resource}
C.-H. Hong and B.~Varghese, ``Resource management in fog/edge computing: A
  survey on architectures, infrastructure, and algorithms,'' \emph{ACM Comput.
  Surv.}, vol.~52, no.~5, pp. 1--37, Sep. 2019.

\bibitem{nguyen2019market}
D.~T. Nguyen, L.~B. Le, and V.~K. Bhargava, ``A market-based framework for
  multi-resource allocation in fog computing,'' \emph{IEEE/ACM Trans. Netw.},
  vol.~27, no.~3, pp. 1151--1164, Apr. 2019.

\bibitem{Liu2016}
J.~Liu, Y.~Mao, J.~Zhang, and K.~B. Letaief, ``Delay-optimal computation task
  scheduling for mobile-edge computing systems,'' in \emph{Proc. {IEEE} Int.
  Symposium Infor. Theory ({ISIT})}, Jul. 2016.

\bibitem{Xu2017}
J.~Xu, L.~Chen, and S.~Ren, ``Online learning for offloading and autoscaling in
  energy harvesting mobile edge computing,'' \emph{{IEEE} Trans. Cognitive
  Communs. and Netw.}, vol.~3, no.~3, pp. 361--373, Sep. 2017.

\bibitem{weiss1999multiagent}
G.~Weiss, \emph{Multiagent systems: A modern approach to distributed artificial
  intelligence}.\hskip 1em plus 0.5em minus 0.4em\relax MIT press, 1999.

\bibitem{Busoniu2008}
L.~Busoniu, R.~Babuska, and B.~D. Schutter, ``A comprehensive survey of
  multiagent reinforcement learning,'' \emph{IEEE Trans. Syst. Man. Cybern.},
  vol.~38, no.~2, pp. 156--172, Mar. 2008.

\bibitem{Yu2017}
S.~Yu, X.~Wang, and R.~Langar, ``Computation offloading for mobile edge
  computing: A deep learning approach,'' in \emph{Proc. {IEEE} 28th Annual
  International Symposium on Personal, Indoor, and Mobile Radio Communications
  ({PIMRC})}.\hskip 1em plus 0.5em minus 0.4em\relax IEEE, Oct. 2017.

\bibitem{Yu2020Intelligent}
S.~Yu, X.~Chen, L.~Yang, D.~Wu, M.~Bennis, and J.~Zhang, ``Intelligent edge:
  Leveraging deep imitation learning for mobile edge computation offloading,''
  \emph{IEEE Wireless Commun.}, vol.~27, no.~1, pp. 92--99, Feb. 2020.

\bibitem{Liu2020}
X.~Liu, J.~Yu, Z.~Feng, and Y.~Gao, ``Multi-agent reinforcement learning for
  resource allocation in {IoT} networks with edge computing,'' \emph{China
  Communications}, vol.~17, no.~9, pp. 220--236, Sep. 2020.

\bibitem{Nasir2019}
Y.~S. Nasir and D.~Guo, ``Multi-agent deep reinforcement learning for dynamic
  power allocation in wireless networks,'' \emph{IEEE J. Sel. Areas Commun.},
  vol.~37, no.~10, pp. 2239--2250, Oct. 2019.

\bibitem{Ali2019}
Z.~Ali, L.~Jiao, T.~Baker, G.~Abbas, Z.~H. Abbas, and S.~Khaf, ``A deep
  learning approach for energy efficient computational offloading in mobile
  edge computing,'' \emph{{IEEE} Access}, vol.~7, pp. 149\,623--149\,633, Oct.
  2019.

\bibitem{Wang2021multiagent}
X.~Wang, Z.~Ning, and S.~Guo, ``Multi-agent imitation learning for pervasive
  edge computing: A decentralized computation offloading algorithm,''
  \emph{{IEEE} Trans. Parallel Distrib. Syst.}, vol.~32, no.~2, pp. 411--425,
  Feb. 2021.

\bibitem{Chen2021}
X.~Chen, C.~Wu, Z.~Liu, N.~Zhang, and Y.~Ji, ``Computation offloading in beyond
  {5G} networks: A distributed learning framework and applications,''
  \emph{{IEEE} Wireless Commun.}, vol.~28, no.~2, pp. 56--62, Apr. 2021.

\bibitem{Zhan2020}
W.~Zhan, C.~Luo, J.~Wang, C.~Wang, G.~Min, H.~Duan, and Q.~Zhu,
  ``Deep-reinforcement-learning-based offloading scheduling for vehicular edge
  computing,'' \emph{IEEE Internet Things J.}, vol.~7, no.~6, pp. 5449--5465,
  Jun. 2020.

\bibitem{Alam2016}
M.~G.~R. Alam, Y.~K. Tun, and C.~S. Hong, ``Multi-agent and reinforcement
  learning based code offloading in mobile fog,'' in \emph{Proc. Int. Conf. on
  Infor. Netw. ({ICOIN})}.\hskip 1em plus 0.5em minus 0.4em\relax {IEEE}, Jan.
  2016.

\bibitem{Ranadheera2017}
S.~Ranadheera, S.~Maghsudi, and E.~Hossain, ``Mobile edge computation
  offloading using game theory and reinforcement learning,'' \emph{arXiv
  preprint arXiv:1711.09012}, Nov. 2017.

\bibitem{Ning2019}
Z.~Ning, P.~Dong, X.~Wang, J.~J. P.~C. Rodrigues, and F.~Xia, ``Deep
  reinforcement learning for vehicular edge computing,'' \emph{{ACM} Trans. on
  Intelligent Systems and Technology}, vol.~10, no.~6, pp. 1--24, Dec. 2019.

\bibitem{Yan2020}
J.~Yan, S.~Bi, and Y.~J.~A. Zhang, ``Offloading and resource allocation with
  general task graph in mobile edge computing: A deep reinforcement learning
  approach,'' \emph{{IEEE} Trans. Wireless Commun.}, vol.~19, no.~8, pp.
  5404--5419, Aug. 2020.

\bibitem{Huang2020a}
L.~Huang, S.~Bi, and Y.-J.~A. Zhang, ``Deep reinforcement learning for online
  computation offloading in wireless powered mobile-edge computing networks,''
  \emph{{IEEE} Trans. Mobile Comput.}, vol.~19, no.~11, pp. 2581--2593, Nov.
  2020.

\bibitem{Chen2019b}
J.~Chen, S.~Chen, Q.~Wang, B.~Cao, G.~Feng, and J.~Hu, ``{iRAF}: A deep
  reinforcement learning approach for collaborative mobile edge computing {IoT}
  networks,'' \emph{IEEE Internet Things J.}, vol.~6, no.~4, pp. 7011--7024,
  Aug. 2019.

\bibitem{Ho2020}
T.~M. Ho and K.-K. Nguyen, ``Joint server selection, cooperative offloading and
  handover in multi-access edge computing wireless network: A deep
  reinforcement learning approach,'' \emph{{IEEE} Trans. Mobile Comput.}, pp.
  1--1, Dec. 2020.

\bibitem{Bi2021}
S.~Bi, L.~Huang, H.~Wang, and Y.-J.~A. Zhang, ``Lyapunov-guided deep
  reinforcement learning for stable online computation offloading in
  mobile-edge computing networks,'' \emph{{IEEE} Trans. Wireless Commun.}, pp.
  1--1, Jun. 2021.

\bibitem{Wang2021b}
J.~Wang, J.~Hu, G.~Min, A.~Y. Zomaya, and N.~Georgalas, ``Fast adaptive task
  offloading in edge computing based on meta reinforcement learning,''
  \emph{IEEE Trans. Parallel Distrib. Syst.}, vol.~32, no.~1, pp. 242--253,
  Jan. 2021.

\bibitem{Qiu2021}
X.~Qiu, W.~Zhang, W.~Chen, and Z.~Zheng, ``Distributed and collective deep
  reinforcement learning for computation offloading: A practical perspective,''
  \emph{IEEE Trans. Parallel Distrib. Syst.}, vol.~32, no.~5, pp. 1085--1101,
  May 2021.

\bibitem{Ong2015}
H.~Y. Ong, K.~Chavez, and A.~Hong, ``Distributed deep {Q}-learning,''
  \emph{arXiv preprint arXiv:1508.04186}, Aug. 2015.

\bibitem{Tang2020}
M.~Tang and V.~W. Wong, ``Deep reinforcement learning for task offloading in
  mobile edge computing systems,'' \emph{{IEEE} Trans. Mobile Comput.}, pp.
  1--1, Nov. 2020.

\bibitem{Xu2021b}
Y.~Xu, A.~Feriani, and E.~Hossain, ``Decentralized multi-agent reinforcement
  learning for task offloading under uncertainty,'' \emph{arXiv preprint
  arXiv:2107.08114}, Jul. 2021.

\bibitem{Amiri2018}
R.~Amiri, H.~Mehrpouyan, L.~Fridman, R.~K. Mallik, A.~Nallanathan, and
  D.~Matolak, ``A machine learning approach for power allocation in {HetNets}
  considering {QoS},'' in \emph{Proc. IEEE Int. Conf. Communs. (ICC)}.\hskip
  1em plus 0.5em minus 0.4em\relax IEEE, May 2018, pp. 1--7.

\bibitem{Matignon2012}
L.~Matignon, G.~J. Laurent, and N.~Le~Fort-Piat, ``Independent reinforcement
  learners in cooperative markov games: a survey regarding coordination
  problems,'' \emph{The Knowledge Engineering Review}, vol.~27, no.~1, pp.
  1--31, Feb. 2012.

\bibitem{Huang2021}
X.~Huang, S.~Leng, S.~Maharjan, and Y.~Zhang, ``Multi-agent deep reinforcement
  learning for computation offloading and interference coordination in small
  cell networks,'' \emph{IEEE Trans. Veh. Technol.}, vol.~70, no.~9, pp.
  9282--9293, Sep. 2021.

\bibitem{Sacco2021}
A.~Sacco, F.~Esposito, G.~Marchetto, and P.~Montuschi, ``Sustainable task
  offloading in {UAV} networks via multi-agent reinforcement learning,''
  \emph{{IEEE} Trans. Veh. Technol.}, vol.~70, no.~5, pp. 5003--5015, May 2021.

\bibitem{Zhang2021a}
Y.~Zhang, B.~Di, Z.~Zheng, J.~Lin, and L.~Song, ``Distributed multi-cloud
  multi-access edge computing by multi-agent reinforcement learning,''
  \emph{IEEE Trans. Wireless Commun.}, vol.~20, no.~4, pp. 2565--2578, Dec.
  2021.

\bibitem{Gao2020}
H.~Gao, X.~Wang, X.~Ma, W.~Wei, and S.~Mumtaz, ``{Com-DDPG}: A multiagent
  reinforcement learning-based offloading strategy for mobile edge computing,''
  \emph{arXiv preprint arXiv:2012.05105}, Dec. 2020.

\bibitem{Chen2020a}
X.~Chen, C.~Wu, T.~Chen, H.~Zhang, Z.~Liu, Y.~Zhang, and M.~Bennis, ``Age of
  information aware radio resource management in vehicular networks: A
  proactive deep reinforcement learning perspective,'' \emph{IEEE Trans.
  Wireless Commun.}, vol.~19, no.~4, pp. 2268--2281, Apr. 2020.

\bibitem{Zhan2020a}
Y.~Zhan, S.~Guo, P.~Li, and J.~Zhang, ``A deep reinforcement learning based
  offloading game in edge computing,'' \emph{IEEE Trans. Comput.}, vol.~69,
  no.~6, pp. 883--893, Jun. 2020.

\bibitem{Zhuo2019}
H.~H. Zhuo, W.~Feng, Y.~Lin, Q.~Xu, and Q.~Yang, ``Federated deep reinforcement
  learning,'' \emph{arXiv preprint arXiv:1901.08277}, Jan. 2019.

\bibitem{qi2021federated}
J.~Qi, Q.~Zhou, L.~Lei, and K.~Zheng, ``Federated reinforcement learning:
  Techniques, applications, and open challenges,'' \emph{arXiv preprint
  arXiv:2108.11887}, Aug. 2021.

\bibitem{Wang2019}
X.~Wang, Y.~Han, C.~Wang, Q.~Zhao, X.~Chen, and M.~Chen, ``In-edge {AI}:
  Intelligentizing mobile edge computing, caching and communication by
  federated learning,'' \emph{{IEEE} Netw.}, vol.~33, no.~5, pp. 156--165, Sep.
  2019.

\bibitem{Shen2020}
S.~Shen, Y.~Han, X.~Wang, and Y.~Wang, ``Computation offloading with multiple
  agents in edge-computing--supported {IoT},'' \emph{ACM Trans. Sens. Netw.},
  vol.~16, no.~1, pp. 1--27, Feb. 2020.

\bibitem{Hou2021}
W.~Hou, H.~Wen, H.~Song, W.~Lei, and W.~Zhang, ``Multiagent deep reinforcement
  learning for task offloading and resource allocation in cybertwin-based
  networks,'' \emph{IEEE Internet Things J.}, vol.~8, no.~22, pp.
  16\,256--16\,268, Jul. 2021.

\bibitem{Prathiba2021}
S.~B. Prathiba, G.~Raja, S.~Anbalagan, K.~Dev, S.~Gurumoorthy, and A.~P.
  Sankaran, ``{Federated learning empowered computation offloading and resource
  management in 6G-V2X},'' \emph{IEEE Trans. Netw. Sci. Eng.}, pp. 1--1, Aug.
  2021.

\bibitem{Wang2022}
Z.~Wang, T.~Lv, and Z.~Chang, ``Computation offloading and resource allocation
  based on distributed deep learning and software defined mobile edge
  computing,'' \emph{Computer Networks}, vol. 205, p. 108732, Mar. 2022.

\bibitem{Singh2019}
A.~Singh, P.~Vepakomma, O.~Gupta, and R.~Raskar, ``Detailed comparison of
  communication efficiency of split learning and federated learning,''
  \emph{arXiv preprint arXiv: 1909.09145}, Sep. 2019.

\bibitem{Thapa2020}
C.~Thapa, M.~A.~P. Chamikara, and S.~Camtepe, ``{SplitFed}: When federated
  learning meets split learning,'' \emph{arXiv preprint arXiv:2004.12088}, Apr.
  2020.

\bibitem{Elbir2021}
A.~M. Elbir, A.~K. Papazafeiropoulos, and S.~Chatzinotas, ``Federated learning
  for physical layer design,'' \emph{{IEEE} Communs. Mag.}, vol.~59, no.~11,
  pp. 81--87, Nov. 2021.

\bibitem{dean2012large}
J.~Dean, G.~Corrado, R.~Monga, K.~Chen, M.~Devin, M.~Mao, M.~Ranzato,
  A.~Senior, P.~Tucker, K.~Yang \emph{et~al.}, ``Large scale distributed deep
  networks,'' \emph{Advances in Neural Infor. Processing Systems 25 (NIPS
  2012)}, vol.~25, pp. 1223--1231, 2012.

\bibitem{li2014scaling}
M.~Li, D.~G. Andersen, J.~W. Park, A.~J. Smola, A.~Ahmed, V.~Josifovski,
  J.~Long, E.~J. Shekita, and B.-Y. Su, ``Scaling distributed machine learning
  with the parameter server,'' in \emph{11th USENIX Symposium on Operating
  Systems Design and Implementation ($\{$OSDI$\}$ 14)}, 2014, pp. 583--598.

\bibitem{Zheng2016}
S.~Zheng, Q.~Meng, T.~Wang, W.~Chen, N.~Yu, Z.-M. Ma, and T.-Y. Liu,
  ``Asynchronous stochastic gradient descent with delay compensation,''
  \emph{Proceedings of the 34th Int. Conf. on Machine Learning}, vol.~70, pp.
  4120--4129, Sep. 2017.

\bibitem{Xie2019}
C.~Xie, S.~Koyejo, and I.~Gupta, ``Asynchronous federated optimization,''
  \emph{arXiv preprint arXiv:1903.03934}, Mar. 2019.

\bibitem{Zhang2021}
H.~Zhang, J.~Bosch, and H.~H. Olsson, ``Real-time end-to-end federated
  learning: An automotive case study,'' \emph{Proc. IEEE 45th Annual Computers,
  Software, and Applications Conference (COMPSAC)}, Mar. 2021.

\bibitem{Wang2021}
Z.~Wang, Z.~Zhang, and J.~Wang, ``Asynchronous federated learning over wireless
  communication networks,'' in \emph{Proc. {IEEE} Int. Conf. Communs. (ICC)},
  Jun. 2021.

\bibitem{Stripelis2021}
D.~Stripelis, P.~M. Thompson, and J.~L. Ambite, ``Semi-synchronous federated
  learning for energy-efficient training and accelerated convergence in
  cross-silo settings,'' \emph{ACM Transactions on Intelligent Systems and
  Technology (TIST)}, Jun. 2022.

\bibitem{Xu2021Asyn}
C.~Xu, Y.~Qu, Y.~Xiang, and L.~Gao, ``Asynchronous federated learning on
  heterogeneous devices: A survey,'' \emph{arXiv preprint arXiv:2109.04269},
  Sep. 2021.

\bibitem{Hao2020}
J.~Hao, Y.~Zhao, and J.~Zhang, ``Time efficient federated learning with
  semi-asynchronous communication,'' in \emph{Proc. {IEEE} 26th International
  Conference on Parallel and Distributed Systems ({ICPADS})}, Dec. 2020.

\bibitem{Wang2021a}
Z.~Wang, H.~Xu, J.~Liu, H.~Huang, C.~Qiao, and Y.~Zhao, ``Resource-efficient
  federated learning with hierarchical aggregation in edge computing,'' in
  \emph{Proc. IEEE Conf. Comput. Commun. (INFOCOM)}, May 2021.

\bibitem{Sun2021}
W.~Sun, S.~Lei, L.~Wang, Z.~Liu, and Y.~Zhang, ``Adaptive federated learning
  and digital twin for industrial internet of things,'' \emph{IEEE Trans. Ind.
  Informat.}, vol.~17, no.~8, pp. 5605--5614, Oct. 2020.

\bibitem{Yang2021}
H.~Yang, J.~Zhao, Z.~Xiong, K.-Y. Lam, S.~Sun, and L.~Xiao,
  ``Privacy-preserving federated learning for {UAV}-enabled networks:
  Learning-based joint scheduling and resource management,'' \emph{IEEE J. Sel.
  Areas Commun.}, pp. 1--1, Jun. 2021.

\bibitem{liu2021blockchain}
Y.~Liu, Y.~Qu, C.~Xu, Z.~Hao, and B.~Gu, ``Blockchain-enabled asynchronous
  federated learning in edge computing,'' \emph{Sensors}, vol.~21, no.~10, p.
  3335, Jan. 2021.

\bibitem{feng2021blockchain}
L.~Feng, Y.~Zhao, S.~Guo, X.~Qiu, W.~Li, and P.~Yu, ``Blockchain-based
  asynchronous federated learning for internet of things,'' \emph{IEEE Trans.
  Comput.}, Apr. 2021.

\bibitem{Cui2021}
L.~Cui, Y.~Qu, G.~Xie, D.~Zeng, R.~Li, S.~Shen, and S.~Yu, ``Security and
  privacy-enhanced federated learning for anomaly detection in {IoT}
  infrastructures,'' \emph{IEEE Trans. Ind. Informat.}, pp. 1--1, Aug. 2021.

\bibitem{Zeng2021}
R.~Zeng, C.~Zeng, X.~Wang, B.~Li, and X.~Chu, ``A comprehensive survey of
  incentive mechanism for federated learning,'' \emph{arXiv preprint
  arXiv:2106.15406}, Jun. 2021.

\bibitem{Liu2020a}
Y.~Liu and J.~Wei, ``Incentives for federated learning: A hypothesis
  elicitation approach,'' \emph{arXiv preprint arXiv:2007.10596}, Jul. 2020.

\bibitem{Pandey2019}
S.~R. Pandey, N.~H. Tran, M.~Bennis, Y.~K. Tun, A.~Manzoor, and C.~S. Hong, ``A
  crowdsourcing framework for on-device federated learning,'' \emph{IEEE Trans.
  Wireless Commun.}, Nov. 2019.

\bibitem{Khan2019}
L.~U. Khan, S.~R. Pandey, N.~H. Tran, W.~Saad, Z.~Han, M.~N.~H. Nguyen, and
  C.~S. Hong, ``Federated learning for edge networks: Resource optimization and
  incentive mechanism,'' \emph{IEEE Communs. Mag.}, Nov. 2019.

\bibitem{Zeng2020FMore}
R.~Zeng, S.~Zhang, J.~Wang, and X.~Chu, ``{FMore: An incentive scheme of
  multi-dimensional auction for federated learning in MEC},'' \emph{Proc. IEEE
  40th Intern. Conf. on Distributed Computing Systems (ICDCS)}, Feb. 2020.

\bibitem{Jiao2019}
Y.~Jiao, P.~Wang, D.~Niyato, B.~Lin, and D.~I. Kim, ``Toward an automated
  auction framework for wireless federated learning services market,''
  \emph{{IEEE} Trans. Mobile Comput.}, vol.~20, no.~10, pp. 3034--3048, Oct.
  2021.

\bibitem{Kang2019}
J.~Kang, Z.~Xiong, D.~Niyato, S.~Xie, and J.~Zhang, ``Incentive mechanism for
  reliable federated learning: A joint optimization approach to combining
  reputation and contract theory,'' \emph{{IEEE} Internet Things J.}, vol.~6,
  no.~6, pp. 10\,700--10\,714, Dec. 2019.

\bibitem{Lim2020}
W.~Y.~B. Lim, J.~Huang, Z.~Xiong, J.~Kang, D.~Niyato, X.-S. Hua, C.~Leung, and
  C.~Miao, ``{Towards federated learning in UAV-enabled internet of vehicles: A
  multi-dimensional contract-matching approach},'' \emph{IEEE Trans. Intell.
  Transp. Syst.}, Apr. 2020.

\bibitem{Wang2020incentive}
T.~Wang, J.~Rausch, C.~Zhang, R.~Jia, and D.~Song, ``A principled approach to
  data valuation for federated learning,'' \emph{Chapter, pp 153–167,
  Federated Learning}, Sep. 2020.

\bibitem{Zhan2020Incentive}
Y.~Zhan, P.~Li, Z.~Qu, D.~Zeng, and S.~Guo, ``A learning-based incentive
  mechanism for federated learning,'' \emph{{IEEE} Internet Things J.}, vol.~7,
  no.~7, pp. 6360--6368, Jul. 2020.

\bibitem{Liu2020Fed}
Y.~Liu, S.~Sun, Z.~Ai, S.~Zhang, Z.~Liu, and H.~Yu, ``{FedCoin}: A peer-to-peer
  payment system for federated learning,'' \emph{Chapter, pp 125-138, Federated
  Learning}, Feb. 2020.

\bibitem{Yu2020}
T.~Yu, E.~Bagdasaryan, and V.~Shmatikov, ``Salvaging federated learning by
  local adaptation,'' \emph{arXiv preprint arXiv:2002.04758}, Feb. 2020.

\bibitem{9187796}
S.~He, X.~Lyu, W.~Ni, H.~Tian, R.~P. Liu, and E.~Hossain, ``Virtual service
  placement for edge computing under finite memory and bandwidth,'' \emph{IEEE
  Trans. Commun.}, vol.~68, no.~12, pp. 7702--7718, Sep. 2020.

\bibitem{8957702}
J.~Wang, C.~Jiang, H.~Zhang, Y.~Ren, K.-C. Chen, and L.~Hanzo, ``Thirty years
  of machine learning: The road to pareto-optimal wireless networks,''
  \emph{IEEE Commun. Surveys Tuts.}, vol.~22, no.~3, pp. 1472--1514, Jan. 2020.

\bibitem{8865093}
J.~Park, S.~Samarakoon, M.~Bennis, and M.~Debbah, ``Wireless network
  intelligence at the edge,'' \emph{Proc. IEEE}, vol. 107, no.~11, pp.
  2204--2239, Oct. 2019.

\bibitem{10.1145/2810103.2813677}
\BIBentryALTinterwordspacing
M.~Fredrikson, S.~Jha, and T.~Ristenpart, ``Model inversion attacks that
  exploit confidence information and basic countermeasures,'' in \emph{Proc. of
  the 22nd ACM SIGSAC Conf. on Computer and Communs. Security}, ser. CCS
  '15.\hskip 1em plus 0.5em minus 0.4em\relax New York, NY, USA: Association
  for Computing Machinery, Oct. 2015, p. 1322–1333. [Online]. Available:
  \url{https://doi.org/10.1145/2810103.2813677}
\BIBentrySTDinterwordspacing

\bibitem{8737416}
Z.~Wang, M.~Song, Z.~Zhang, Y.~Song, Q.~Wang, and H.~Qi, ``Beyond inferring
  class representatives: User-level privacy leakage from federated learning,''
  in \emph{Proc. IEEE Conf. Comput. Commun. (INFOCOM)}, Apr. 2019, pp.
  2512--2520.

\bibitem{deepleakage}
L.~Zhu and S.~Han, ``Deep leakage from gradients,'' in \emph{Federated
  learning}.\hskip 1em plus 0.5em minus 0.4em\relax Springer, 2020, pp. 17--31.

\bibitem{Xia2021}
W.~Xia, W.~Wen, K.-K. Wong, T.~Q. Quek, J.~Zhang, and H.~Zhu,
  ``Federated-learning-based client scheduling for low-latency wireless
  communications,'' \emph{{IEEE} Wireless Commun.}, vol.~28, no.~2, pp. 32--38,
  Apr. 2021.

\bibitem{Amiri2020b}
M.~M. Amiri, D.~Gündüz, S.~R. Kulkarni, and H.~V. Poor, ``Convergence of
  update aware device scheduling for federated learning at the wireless edge,''
  \emph{IEEE Trans. Wireless Commun.}, vol.~20, no.~6, pp. 3643--3658, Jan.
  2021.

\bibitem{Ren2020}
J.~Ren, Y.~He, D.~Wen, G.~Yu, K.~Huang, and D.~Guo, ``Scheduling for cellular
  federated edge learning with importance and channel awareness,'' \emph{{IEEE}
  Trans. Wireless Commun.}, vol.~19, pp. 7690 -- 7703, Aug. 2020.

\bibitem{Cui2021a}
Y.~Cui, K.~Cao, G.~Cao, M.~Qiu, and T.~Wei, ``Client scheduling and resource
  management for efficient training in heterogeneous {IoT}-edge federated
  learning,'' \emph{{IEEE} Trans. Comput.-Aided Des. Integr. Circuits Syst.},
  pp. 1--1, Sep. 2021.

\bibitem{Vu2020}
T.~T. Vu, D.~T. Ngo, H.~Q. Ngo, M.~N. Dao, N.~H. Tran, and R.~H. Middleton,
  ``User selection approaches to mitigate the straggler effect for federated
  learning on cell-free massive {MIMO} networks,'' \emph{arXiv preprint
  arXiv:2009.02031}, Sep. 2020.

\bibitem{Yang2020a}
H.~H. Yang, A.~Arafa, T.~Q.~S. Quek, and H.~V. Poor, ``Age-based scheduling
  policy for federated learning in mobile edge networks,'' in \emph{Proc.
  {IEEE} Int. Conf. Acoustics, Speech and Signal Processing ({ICASSP})}, May
  2020.

\bibitem{Huang2020}
T.~Huang, W.~Lin, W.~Wu, L.~He, K.~Li, and A.~Zomaya, ``An efficiency-boosting
  client selection scheme for federated learning with fairness guarantee,''
  \emph{{IEEE} Trans. Parallel Distrib. Syst.}, pp. 1--1, Jul. 2020.

\bibitem{cho2020client}
Y.~J. Cho, J.~Wang, and G.~Joshi, ``Client selection in federated learning:
  Convergence analysis and power-of-choice selection strategies,'' \emph{arXiv
  preprint arXiv:2010.01243}, Oct. 2020.

\bibitem{Yang2020}
H.~H. Yang, Z.~Liu, T.~Q.~S. Quek, and H.~V. Poor, ``Scheduling policies for
  federated learning in wireless networks,'' \emph{IEEE Trans. Communs.},
  vol.~68, pp. 317 -- 333, Jan. 2020.

\bibitem{Nishio2019}
T.~Nishio and R.~Yonetani, ``Client selection for federated learning with
  heterogeneous resources in mobile edge,'' \emph{Proc. IEEE Int. Conf. Commun.
  (ICC)}, Apr. 2018.

\bibitem{Yang2018}
K.~Yang, T.~Jiang, Y.~Shi, and Z.~Ding, ``Federated learning via over-the-air
  computation,'' \emph{{IEEE} Trans. Wireless Commun.}, vol.~19, pp. 2022 --
  2035, Dec. 2020.

\bibitem{Amiri2020}
M.~M. Amiri and D.~Gunduz, ``Federated learning over wireless fading
  channels,'' \emph{{IEEE} Trans. Wireless Commun.}, vol.~19, no.~5, pp.
  3546--3557, May 2020.

\bibitem{kang2020reliable}
J.~Kang, Z.~Xiong, D.~Niyato, Y.~Zou, Y.~Zhang, and M.~Guizani, ``Reliable
  federated learning for mobile networks,'' \emph{IEEE Wireless Commun.},
  vol.~27, no.~2, pp. 72--80, Feb. 2020.

\bibitem{Ma2021}
X.~Ma, H.~Sun, Q.~Wang, and R.~Q. Hu, ``User scheduling for federated learning
  through over-the-air computation,'' \emph{Proc. IEEE 94th Vehicular
  Technology Conf. (VTC2021-Fall)}, Sep. 2021.

\bibitem{yin2020joint}
B.~Yin, Z.~Chen, and M.~Tao, ``Joint user scheduling and resource allocation
  for federated learning over wireless networks,'' in \emph{Proc. IEEE Global
  Commun. Conf. (GLOBECOM)}.\hskip 1em plus 0.5em minus 0.4em\relax IEEE, Dec.
  2020, pp. 1--6.

\bibitem{Chen2019a}
M.~Chen, Z.~Yang, W.~Saad, C.~Yin, H.~V. Poor, and S.~Cui, ``A joint learning
  and communications framework for federated learning over wireless networks,''
  \emph{{IEEE} Trans. Wireless Commun.}, vol.~20, pp. 269 -- 283, Sep. 2020.

\bibitem{Zeng2020}
T.~Zeng, O.~Semiari, M.~Mozaffari, M.~Chen, W.~Saad, and M.~Bennis, ``Federated
  learning in the sky: Joint power allocation and scheduling with {UAV}
  swarms,'' in \emph{Proc. IEEE Int. Conf. Commun. (ICC)}, Jun. 2020.

\bibitem{Ma2020}
X.~Ma, H.~Sun, and R.~Q. Hu, ``Scheduling policy and power allocation for
  federated learning in {NOMA} based {MEC},'' in \emph{Proc. IEEE Global
  Commun. Conf. (GLOBECOM)}, Dec. 2020.

\bibitem{Shi2021}
W.~Shi, S.~Zhou, Z.~Niu, M.~Jiang, and L.~Geng, ``Joint device scheduling and
  resource allocation for latency constrained wireless federated learning,''
  \emph{{IEEE} Trans. Wireless Commun.}, vol.~20, no.~1, pp. 453--467, Jan.
  2021.

\bibitem{Zeng2020a}
Q.~Zeng, Y.~Du, K.~Huang, and K.~K. Leung, ``Energy-efficient radio resource
  allocation for federated edge learning,'' in \emph{Proc. {IEEE} Int. Conf.
  Communs Workshops ({ICC} Workshops)}, Jun. 2020.

\bibitem{Yoshida2020}
N.~Yoshida, T.~Nishio, M.~Morikura, and K.~Yamamoto, ``Mab-based client
  selection for federated learning with uncertain resources in mobile
  networks,'' \emph{Proc. IEEE Globecom Workshops (GC Wkshps)}, Sep. 2020.

\bibitem{Xia2020}
W.~Xia, T.~Q.~S. Quek, K.~Guo, W.~Wen, H.~H. Yang, and H.~Zhu, ``Multi-armed
  bandit based client scheduling for federated learning,'' \emph{{IEEE} Trans.
  Wireless Commun.}, vol.~19, pp. 7108 -- 7123, Jul. 2020.

\bibitem{Cho2020}
Y.~J. Cho, S.~Gupta, G.~Joshi, and O.~Yağan, ``Bandit-based
  communication-efficient client selection strategies for federated learning,''
  \emph{Proc. IEEE 54th Asilomar Conference on Signals, Systems, and
  Computers}, Nov. 2020.

\bibitem{Xu2021a}
B.~Xu, W.~Xia, J.~Zhang, T.~Q.~S. Quek, and H.~Zhu, ``Online client scheduling
  for fast federated learning,'' \emph{{IEEE} Wireless Commun. Lett.}, vol.~10,
  no.~7, pp. 1434--1438, Jul. 2021.

\bibitem{Sery2019}
T.~Sery and K.~Cohen, ``On analog gradient descent learning over multiple
  access fading channels,'' \emph{IEEE Trans. Signal Processing}, vol.~68, pp.
  2897 -- 2911, Apr. 2020.

\bibitem{Shao2021}
Y.~Shao, D.~Gunduz, and S.~C. Liew, ``Federated edge learning with misaligned
  over-the-air computation,'' \emph{IEEE Trans. Wireless Commun.}, Feb. 2021.

\bibitem{Amiri2019a}
M.~M. Amiri, T.~M. Duman, and D.~Gunduz, ``Collaborative machine learning at
  the wireless edge with blind transmitters,'' \emph{Proc. IEEE Global Conf.
  Signal Inf. Process. (GlobalSIP)}, Nov. 2019.

\bibitem{Amiri2020a}
M.~M. Amiri, T.~M. Duman, D.~Gunduz, S.~R. Kulkarni, and H.~V. Poor, ``Blind
  federated edge learning,'' \emph{{IEEE} Trans. Wireless Commun.}, Oct. 2020.

\bibitem{Zhu2020}
G.~Zhu, Y.~Wang, and K.~Huang, ``Broadband analog aggregation for low-latency
  federated edge learning,'' \emph{{IEEE} Trans. Wireless Commun.}, vol.~19,
  no.~1, pp. 491--506, Jan. 2020.

\bibitem{Zhu2020a}
G.~Zhu, Y.~Du, D.~Gunduz, and K.~Huang, ``One-bit over-the-air aggregation for
  communication-efficient federated edge learning: Design and convergence
  analysis,'' \emph{{IEEE} Trans. Wireless Commun.}, vol.~20, pp. 2120 -- 2135,
  Nov. 2020.

\bibitem{Sery2020}
T.~Sery, N.~Shlezinger, K.~Cohen, and Y.~C. Eldar, ``Over-the-air federated
  learning from heterogeneous data,'' \emph{IEEE Trans. Signal Processing},
  vol.~69, pp. 3796 -- 3811, Jun. 2020.

\bibitem{Sun2019}
Y.~Sun, S.~Zhou, and D.~Gündüz, ``Energy-aware analog aggregation for
  federated learning with redundant data,'' \emph{Proc. IEEE Int. Conf.
  Communs. (ICC)}, Jul. 2020.

\bibitem{7239545}
E.~P. Xing, Q.~Ho, W.~Dai, J.~K. Kim, J.~Wei, S.~Lee, X.~Zheng, P.~Xie,
  A.~Kumar, and Y.~Yu, ``Petuum: A new platform for distributed machine
  learning on big data,'' \emph{IEEE Trans. Big Data}, vol.~1, no.~2, pp.
  49--67, Jun. 2015.

\bibitem{li2014communication}
M.~Li, D.~G. Andersen, A.~J. Smola, and K.~Yu, ``Communication efficient
  distributed machine learning with the parameter server,'' \emph{Proc. Adv.
  Neural Inf. Process. Syst}, vol.~27, pp. 19--27, 2014.

\bibitem{seide20141}
F.~Seide, H.~Fu, J.~Droppo, G.~Li, and D.~Yu, ``{1-bit stochastic gradient
  descent and its application to data-parallel distributed training of speech
  DNNs},'' in \emph{Proc. 15th Annual Conf. of the Int. Speech Commun.
  Association}.\hskip 1em plus 0.5em minus 0.4em\relax Citeseer, 2014.

\bibitem{alistarh2016qsgd}
D.~Alistarh, J.~Li, R.~Tomioka, and M.~Vojnovic, ``{QSGD: Randomized
  quantization for communication-optimal stochastic gradient descent},''
  \emph{arXiv preprint arXiv:1610.02132}, vol.~1, Oct. 2016.

\bibitem{wen2017terngrad}
W.~Wen, C.~Xu, F.~Yan, C.~Wu, Y.~Wang, Y.~Chen, and H.~Li, ``{TernGrad: Ternary
  gradients to reduce communication in distributed deep learning},''
  \emph{arXiv preprint arXiv:1705.07878}, May 2017.

\bibitem{zhou2016dorefa}
S.~Zhou, Y.~Wu, Z.~Ni, X.~Zhou, H.~Wen, and Y.~Zou, ``{DoReFa-Net: Training low
  bitwidth convolutional neural networks with low bitwidth gradients},''
  \emph{arXiv preprint arXiv:1606.06160}, Jun. 2016.

\bibitem{strom2015scalable}
N.~Strom, ``{Scalable distributed DNN training using commodity GPU cloud
  computing},'' in \emph{Proc. 16th Annual Conf. of the Int. Speech Commun.
  Association}, 2015.

\bibitem{aji2017sparse}
A.~F. Aji and K.~Heafield, ``Sparse communication for distributed gradient
  descent,'' \emph{arXiv preprint arXiv:1704.05021}, Apr. 2017.

\bibitem{dryden2016communication}
N.~Dryden, T.~Moon, S.~A. Jacobs, and B.~Van~Essen, ``Communication
  quantization for data-parallel training of deep neural networks,'' in
  \emph{Proc. 2nd Workshop on Machine Learning in HPC Environments
  (MLHPC)}.\hskip 1em plus 0.5em minus 0.4em\relax IEEE, Nov. 2016, pp. 1--8.

\bibitem{chen2018adacomp}
C.-Y. Chen, J.~Choi, D.~Brand, A.~Agrawal, W.~Zhang, and K.~Gopalakrishnan,
  ``Adacomp: Adaptive residual gradient compression for data-parallel
  distributed training,'' in \emph{Proceedings of the AAAI Conference on
  Artificial Intelligence}, vol.~32, no.~1, Apr. 2018.

\bibitem{lin2017deep}
Y.~Lin, S.~Han, H.~Mao, Y.~Wang, and W.~J. Dally, ``Deep gradient compression:
  Reducing the communication bandwidth for distributed training,'' \emph{arXiv
  preprint arXiv:1712.01887}, Dec. 2017.

\bibitem{9349624}
W.~Jiang, B.~Han, M.~A. Habibi, and H.~D. Schotten, ``{The road towards 6G: A
  comprehensive survey},'' \emph{IEEE Open J. Commun. Soc.}, vol.~2, pp.
  334--366, Feb. 2021.

\bibitem{6932503}
T.~Bai and R.~W. Heath, ``Coverage and rate analysis for millimeter-wave
  cellular networks,'' \emph{IEEE Trans. Wireless Commun.}, vol.~14, no.~2, pp.
  1100--1114, Feb. 2015.

\bibitem{6798744}
L.~Lu, G.~Y. Li, A.~L. Swindlehurst, A.~Ashikhmin, and R.~Zhang, ``{An overview
  of massive MIMO: benefits and challenges},'' \emph{IEEE J. Sel. Topics Signal
  Process.}, vol.~8, no.~5, pp. 742--758, Apr. 2014.

\bibitem{8910627}
Q.~Wu and R.~Zhang, ``Towards smart and reconfigurable environment: Intelligent
  reflecting surface aided wireless network,'' \emph{IEEE Commun. Mag.},
  vol.~58, no.~1, pp. 106--112, Nov. 2020.

\bibitem{8786074}
H.~Huang, S.~Guo, G.~Gui, Z.~Yang, J.~Zhang, H.~Sari, and F.~Adachi, ``Deep
  learning for physical-layer {5G} wireless techniques: opportunities,
  challenges and solutions,'' \emph{IEEE Wireless Commun.}, vol.~27, no. Aug.,
  pp. 214--222, 2019.

\bibitem{8715338}
H.~He, S.~Jin, C.-K. Wen, F.~Gao, G.~Y. Li, and Z.~Xu, ``Model-driven deep
  learning for physical layer communications,'' \emph{IEEE Wireless Commun.},
  vol.~26, no.~5, pp. 77--83, May 2019.

\bibitem{7738524}
Y.-H. Chen, T.~Krishna, J.~S. Emer, and V.~Sze, ``Eyeriss: An energy-efficient
  reconfigurable accelerator for deep convolutional neural networks,''
  \emph{IEEE J. Solid-State Circuits}, vol.~52, no.~1, pp. 127--138, Jan. 2017.

\bibitem{8054694}
T.~O’Shea and J.~Hoydis, ``An introduction to deep learning for the physical
  layer,'' \emph{IEEE Trans. Cogn. Commun. Netw.}, vol.~3, no.~4, pp. 563--575,
  Oct. 2017.

\bibitem{DBLP:journals/corr/HersheyRW14}
\BIBentryALTinterwordspacing
J.~R. Hershey, J.~L. Roux, and F.~Weninger, ``Deep unfolding: Model-based
  inspiration of novel deep architectures,'' \emph{CoRR}, vol. abs/1409.2574,
  Sep. 2014. [Online]. Available: \url{http://arxiv.org/abs/1409.2574}
\BIBentrySTDinterwordspacing

\bibitem{7905837}
M.~Borgerding and P.~Schniter, ``Onsager-corrected deep learning for sparse
  linear inverse problems,'' in \emph{Proc. IEEE Global Conference on Signal
  and Information Processing (GlobalSIP)}, Dec. 2016, pp. 227--231.

\bibitem{8322184}
C.-K. Wen, W.-T. Shih, and S.~Jin, ``{Deep learning for massive MIMO CSI
  feedback},'' \emph{IEEE Wireless Commun. Lett.}, vol.~7, no.~5, pp. 748--751,
  Mar. 2018.

\bibitem{664294}
A.~Nandi and E.~Azzouz, ``Algorithms for automatic modulation recognition of
  communication signals,'' \emph{IEEE Trans. Commun.}, vol.~46, no.~4, pp.
  431--436, Apr. 1998.

\bibitem{7852251}
E.~Nachmani, Y.~Be'ery, and D.~Burshtein, ``Learning to decode linear codes
  using deep learning,'' in \emph{Proc. IEEE 54th Annual Allerton Conf. on
  Commun., Control, and Comput. (Allerton)}, Sep. 2016, pp. 341--346.

\bibitem{7926071}
T.~Gruber, S.~Cammerer, J.~Hoydis, and S.~t. Brink, ``On deep learning-based
  channel decoding,'' in \emph{Proc. IEEE 51st Annual Conf. on Infor. Sciences
  and Systems (CISS)}, Mar. 2017, pp. 1--6.

\bibitem{8254811}
S.~Cammerer, T.~Gruber, J.~Hoydis, and S.~ten Brink, ``Scaling deep
  learning-based decoding of polar codes via partitioning,'' in \emph{Proc.
  IEEE Global Commun. Conf. (GLOBECOM)}, Dec. 2017, pp. 1--6.

\bibitem{8227772}
N.~Samuel, T.~Diskin, and A.~Wiesel, ``{Deep MIMO detection},'' in \emph{Proc.
  IEEE 18th Int. Workshop on Signal Process. Advances in Wireless Commun.
  (SPAWC)}, Jul. 2017, pp. 1--5.

\bibitem{8663966}
Z.~Qin, H.~Ye, G.~Y. Li, and B.-H.~F. Juang, ``Deep learning in physical layer
  communications,'' \emph{IEEE Wireless Commun.}, vol.~26, no.~2, pp. 93--99,
  Mar. 2019.

\bibitem{8262721}
T.~J. O'Shea, T.~Erpek, and T.~C. Clancy, ``{Physical layer deep learning of
  encodings for the MIMO fading channel},'' in \emph{55th Annual Allerton Conf.
  on Commun., Control, and Comput. (Allerton)}, Oct. 2017, pp. 76--80.

\bibitem{9360873}
H.~Ye, G.~Y. Li, and B.-H. Juang, ``Deep learning based end-to-end wireless
  communication systems without pilots,'' \emph{IEEE Trans. Cogn. Commun.
  Netw.}, vol.~7, no.~3, pp. 702--714, Feb. 2021.

\bibitem{8618345}
H.~Huang, Y.~Song, J.~Yang, G.~Gui, and F.~Adachi, ``Deep-learning-based
  millimeter-wave massive {MIMO} for hybrid precoding,'' \emph{IEEE Trans. Veh.
  Technol.}, vol.~68, no.~3, pp. 3027--3032, Jan. 2019.

\bibitem{9204436}
C.~Qi, Y.~Wang, and G.~Y. Li, ``Deep learning for beam training in millimeter
  wave massive {MIMO} systems,'' \emph{IEEE Trans. Wireless Commun.}, pp. 1--1,
  Sep. 2020.

\bibitem{8387468}
G.~Gui, H.~Huang, Y.~Song, and H.~Sari, ``Deep learning for an effective
  nonorthogonal multiple access scheme,'' \emph{IEEE Trans. Veh. Technol.},
  vol.~67, no.~9, pp. 8440--8450, Jun. 2018.

\bibitem{9625398}
M.~A. Alawad, M.~Q. Hamdan, and K.~A. Hamdi, ``End-to-end deep learning
  {IRS}-assisted communications systems,'' in \emph{Proc. IEEE 94th Vehicular
  Technology Conference (VTC2021-Fall)}, Sep. 2021, pp. 1--6.

\bibitem{9743298}
J.~Yu, X.~Liu, Y.~Gao, C.~Zhang, and W.~Zhang, ``Deep learning for channel
  tracking in {IRS}-assisted {UAV} communication systems,'' \emph{IEEE Trans.
  Wireless Commun.}, pp. 1--1, Mar. 2022.

\bibitem{paz2020zest}
T.~Paz-Argaman, Y.~Atzmon, G.~Chechik, and R.~Tsarfaty, ``{ZEST}: Zero-shot
  learning from text descriptions using textual similarity and visual
  summarization,'' \emph{arXiv preprint arXiv:2010.03276}, Oct. 2020.

\bibitem{zhou2020distilled}
Y.~Zhou, G.~Pu, X.~Ma, X.~Li, and D.~Wu, ``Distilled one-shot federated
  learning,'' \emph{arXiv preprint arXiv:2009.07999}, Sep. 2020.

\bibitem{chen2021shot}
Z.~Chen, S.~Maji, and E.~Learned-Miller, ``Shot in the dark: Few-shot learning
  with no base-class labels,'' in \emph{Proceedings of the IEEE/CVF Conf. on
  Computer Vision and Pattern Recognition}, Jun. 2021, pp. 2668--2677.

\bibitem{huisman2021survey}
M.~Huisman, J.~N. van Rijn, and A.~Plaat, ``A survey of deep meta-learning,''
  \emph{Artificial Intelligence Review}, pp. 1--59, Apr. 2021.

\bibitem{fallah2020personalized}
A.~Fallah, A.~Mokhtari, and A.~Ozdaglar, ``Personalized federated learning: A
  meta-learning approach,'' \emph{arXiv preprint arXiv:2002.07948}, Feb. 2020.

\bibitem{wu2020personalized}
Q.~Wu, K.~He, and X.~Chen, ``{Personalized federated learning for intelligent
  IoT applications: A cloud-edge based framework},'' \emph{IEEE Open J. Comput.
  Soc.}, vol.~1, pp. 35--44, May 2020.

\bibitem{kulkarni2020survey}
V.~Kulkarni, M.~Kulkarni, and A.~Pant, ``Survey of personalization techniques
  for federated learning,'' in \emph{Proc. 4th World Conf. on Smart Trends in
  Systems, Security and Sustainability (WorldS4)}.\hskip 1em plus 0.5em minus
  0.4em\relax IEEE, Jul. 2020, pp. 794--797.

\bibitem{zhao2018federated}
Y.~Zhao, M.~Li, L.~Lai, N.~Suda, D.~Civin, and V.~Chandra, ``Federated learning
  with non-{IID} data,'' \emph{arXiv preprint arXiv:1806.00582}, Jun. 2018.

\bibitem{koda2020distributed}
Y.~Koda, J.~Park, M.~Bennis, K.~Yamamoto, T.~Nishio, and M.~Morikura,
  ``Distributed heteromodal split learning for vision aided {mmWave} received
  power prediction,'' \emph{arXiv preprint arXiv:2007.08208}, Jul. 2020.

\bibitem{8952884}
K.~Yang, T.~Jiang, Y.~Shi, and Z.~Ding, ``Federated learning via over-the-air
  computation,'' \emph{IEEE Trans. Wireless Commun.}, vol.~19, no.~3, pp.
  2022--2035, Jan. 2020.

\bibitem{9199786}
K.~Yang, Y.~Shi, Y.~Zhou, Z.~Yang, L.~Fu, and W.~Chen, ``{Federated machine
  learning for intelligent IoT via reconfigurable intelligent surface},''
  \emph{{IEEE} Netw.}, vol.~34, no.~5, pp. 16--22, Sep. 2020.

\bibitem{9502547}
Z.~Wang, J.~Qiu, Y.~Zhou, Y.~Shi, L.~Fu, W.~Chen, and K.~B. Letaief,
  ``Federated learning via intelligent reflecting surface,'' \emph{IEEE Trans.
  Wireless Commun.}, pp. 1--1, Jul. 2021.

\bibitem{9048613}
C.~Ma, J.~Li, M.~Ding, H.~H. Yang, F.~Shu, T.~Q.~S. Quek, and H.~V. Poor, ``On
  safeguarding privacy and security in the framework of federated learning,''
  \emph{{IEEE} Netw.}, vol.~34, no.~4, pp. 242--248, Mar. 2020.

\bibitem{geyer2017differentially}
R.~C. Geyer, T.~Klein, and M.~Nabi, ``Differentially private federated
  learning: A client level perspective,'' \emph{arXiv preprint
  arXiv:1712.07557}, Dec. 2017.

\bibitem{7373646}
A.~Diyanat, A.~Khonsari, and S.~P. Shariatpanahi, ``A dummy-based approach for
  preserving source rate privacy,'' \emph{IEEE Trans. Inf. Forensics Security},
  vol.~11, no.~6, pp. 1321--1332, Jan. 2016.

\bibitem{8241854}
L.~T. Phong, Y.~Aono, T.~Hayashi, L.~Wang, and S.~Moriai, ``Privacy-preserving
  deep learning via additively homomorphic encryption,'' \emph{IEEE Trans. Inf.
  Forensics Security}, vol.~13, no.~5, pp. 1333--1345, Dec. 2017.

\bibitem{9424138}
V.~Mothukuri, P.~Khare, R.~M. Parizi, S.~Pouriyeh, A.~Dehghantanha, and
  G.~Srivastava, ``{Federated learning-based anomaly detection for IoT security
  attacks},'' \emph{IEEE Internet Things J.}, pp. 1--1, May. 2021.

\end{thebibliography}

\end{document}